\documentclass[fleqn,usenatbib]{mnras}
\usepackage{dsfont}
\usepackage[T1]{fontenc}
\usepackage{xcolor}
\DeclareRobustCommand{\VAN}[3]{#2}
\let\VANthebibliography\thebibliography
\def\thebibliography{\DeclareRobustCommand{\VAN}[3]{##3}\VANthebibliography}
\usepackage{graphicx}	
\usepackage{amsmath}	
\usepackage{amssymb}	
\usepackage{bm}
\usepackage{ulem}
\usepackage{soul}
\usepackage{cancel}
\usepackage{gensymb}
\usepackage{newtxtext,newtxmath}

\newcommand{\bn}{\bm\nabla}
\newcommand{\ft}{\bm f_\mathrm{t}}
\newcommand{\xnw}{\bm\xi_\mathrm{nw}}
\newcommand{\e}[1]{\mathrm{e}^{#1}}
\newcommand{\uw}{\bm u_\mathrm{w}}
\newcommand{\unw}{\bm u_\mathrm{nw}}
\newcommand{\mI}[1]{I_\mathrm{#1}}
\newcommand{\mmI}[1]{\mathcal{I}_\mathrm{#1}}

\newcommand{\pt}{P_\mathrm{t}}
\newcommand{\Ek}{\mathrm{E}}
\newcommand{\fb}{Paper I}
\newcommand{\Roc}{\mathrm{Ro}_\mathrm{c}}
\renewcommand{\sc}{s_\mathrm{c}}
\newcommand{\Ct}{C_\mathrm{t}}
\definecolor{db}{HTML}{191586}
\definecolor{vi}{HTML}{5E1111}
\definecolor{rouge}{HTML}{b90e1e}

\renewcommand{\cor}[1]{\textcolor{black}{#1}}

\title[Nonlinear tidal inertial waves in spherical shells]{
The effects of nonlinearities on tidal flows in the convective envelopes of rotating stars and planets in exoplanetary systems
}

\author[A. Astoul \& A.~J. Barker]{
A. Astoul,$^{1}$\thanks{E-mail: a.a.v.astoul@leeds.ac.uk}
A. J. Barker,$^{1}$\thanks{E-mail: A.J.Barker@leeds.ac.uk}
\\
$^{1}$Department of Applied Mathematics, School of Mathematics, University of Leeds, Leeds, LS2 9JT, UK\\
}

\date{Accepted XXX. Received YYY; in original form 2022 May 24}

\pubyear{2022}
\begin{document}
\label{firstpage}
\pagerange{\pageref{firstpage}--\pageref{lastpage}}
\maketitle
\begin{abstract} 
In close exoplanetary systems, tidal interactions drive orbital and spin evolution of planets and stars over long timescales. Tidally-forced inertial waves (restored by the Coriolis acceleration) in the convective envelopes of low-mass stars and giant gaseous planets contribute greatly to the tidal dissipation when they are excited and subsequently damped (e.g.~through viscous friction), especially early in the life of a system. These waves are known to be subject to nonlinear effects, including triggering differential rotation in the form of zonal flows. In this study, we use a realistic tidal body forcing to excite inertial waves through the residual action of the equilibrium tide in the momentum equation for the waves. By performing 3D nonlinear hydrodynamical simulations in  adiabatic and incompressible convective shells, we investigate how the addition of nonlinear terms affects the tidal flow properties, and the energy and angular momentum redistribution. In particular, we identify and justify the removal of terms responsible for unphysical angular momentum evolution observed in a previous numerical study. Within our new set-up, we observe the establishment of strong cylindrically-sheared zonal flows, which modify the tidal dissipation rates from prior linear theoretical predictions. We demonstrate that the effects of this differential rotation on the waves neatly explains the discrepancies between linear and nonlinear dissipation rates in many of our simulations. We also highlight the major role of both corotation resonances and parametric instabilities of inertial waves, which are observed for sufficiently high tidal forcing amplitudes or low viscosities, in affecting the tidal flow response. 
\end{abstract} 



\section{Introduction}
Over the last $25$ years or so, nearly $5000$ exoplanets have been discovered around mostly late-type stars from M to F spectral types, thanks to spatial and ground-based observational campaigns \citep[see e.g.][for a review]{P2018}. Several hundred
of these exoplanets\footnote{e.g.~from the database \url{http://exoplanet.eu/}.}, known as Hot Jupiters, 
are giant gaseous planets orbiting in less than $10$ days around their host stars. In such compact systems, tides inside the star and the planet are very likely to have played a major role in modifying their rotational periods, their obliquities (also known as spin-orbit angles) and shaping the orbital architecture of the planet \citep[typically the eccentricity and the semi-major axis; see e.g.][]{O2014, O2020, M2019}. In addition to these secular dynamical modifications, star-planet tidal interactions can also bring about internal structural changes by tidal heating, and are for example suspected to affect the size of the planet by inducing radius inflation \citep[e.g.][for close sub-Neptune planets]{M2019b}.

One component of the tidal response in a body due to the tidal gravitational potential of its perturber (the star or the planet) is the quasi-hydrostatic tidal bulge, and its associated large-scale flow, which is the so-called equilibrium tide \citep{L1911,Z1966a,Z1989,EK1998,H2010,H2012}. Signatures of ellipsoidal deformations and equilibrium tides have for example been detected by photometry at the surfaces of HAT-P 7, WASP-18, and WASP-12, around the latter of which orbits a Hot Jupiter undergoing orbital decay
\citep{WO2010,SW2019,MD2016,MN2020,Yee2020,Turner2021}. 
The inward migration of WASP-12 b is thought to be due to the second type of tides, the dynamical tides, which are the oscillatory responses to the tidal potential that arise when considering the non-hydrostatic (i.e.  inertial) terms in the equations of motion for tidal flows  \citep{Z1966c,Z1970,Z1975,OL2004}. In the case of WASP-12, tidally-excited gravity waves (which are restored by the buoyancy force) are thought to propagate within the radiative zone, and if the star has a radiative core they can geometrically focus towards the stellar centre, and break nonlinearly, thus dissipating their kinetic energy 
\citep[e.g.][]{GD1998,OL2007,BO2010,BO2011,WN2017,B2020}. The exchange of angular momentum with the planet due to the tidal dissipation inside the star is what is causing the planetary infall in this picture. However, this dissipative mechanism is likely to be most efficient in systems featuring old 
(towards the end of the main sequence) and slowly rotating stars hosting a massive planet \citep{B2020,L2021}. 

Another powerful mechanism to transfer angular momentum is the dissipation of inertial waves (restored by the Coriolis acceleration), which can propagate in the convective envelopes of low-mass stars and giant gaseous planets \citep[e.g.][]{OL2004,OL2007,W2005a,W2005b}. 
The dissipation of tidal inertial waves in stars is particularly efficient in the early stages of the life of a star (typically during the pre-main sequence of late-type stars) as was shown for example by \cite{M2015},  \cite{B2020} and \cite{L2021} using the frequency-averaged formalism developed by \cite{O2013}. It can then have major consequences for example on the planetary semi-major axis, hastening the infall of the planet if it orbits inside the corotation radius (where the planetary orbital period equals the rotational period of the star), or contribute significantly to pushing it away
in the opposite case \citep[in comparison with dissipation of the equilibrium tide alone, see also][]{BM2016,GB2017}. It can also explain the circularisation periods of solar-type binary stars, which was a long-standing theoretical problem \citep{B2022}. Inertial waves excited inside planets are also believed to be important in explaining the preferentially circular orbits of the shortest-period hot Jupiters through tidally-driven orbital circularisation \citep{OL2004,B2016}.

However, giving robust predictions for the dynamical evolution of a close star-planet system is particularly challenging. The assessment of the tidal dissipation has been shown to significantly depend on many parameters like the stellar angular velocity and metallicity, and the masses of both the star and the planet  \citep[e.g.][]{M2015,GB2017,BG2017,B2020,L2021}. Moreover, the current formalism used for tidal inertial waves is based on a frequency average of the tidal dissipation, even though the dissipation 
is known to be strongly dependent on the tidal forcing frequency in the linear regime when these waves are excited \citep{OL2004,O2005,O2009,RV2010}.
Although this formalism is an efficient and rather straightforward tool to provide information on the long-term tidal evolution of exoplanetary (and close binary) systems, the frequency-dependent tidal dissipation gives us an instant picture of tidal effects at a given evolutionary stage of the system featuring specific tidal forcing frequencies. The resonance locking mechanism, for instance, relies on the continuous match over time between the forcing frequencies of the perturber and some resonant and highly dissipative modes inside the perturbed body \citep{WS1999,FL2016}. This mechanism has been successfully applied to the Saturnian system to provide an explanation for the high tidal dissipation rate in Saturn inferred from the rapid orbital expansion measured for Titan \citep[using inertial modes, with important consequences for the formation scenario of Saturn's satellites,][]{LC2020} and also applied to extra-solar giant gaseous planets orbiting F-type stars with convective cores \citep[using gravity modes,][]{MF2021}. Furthermore, the nature, propagation, and dissipation properties of tidal flows are also modified by the magnetism and differential rotation both present in the convective envelopes of low-mass stars and giant gaseous planets \citep{W2016,W2018,LO2018,BR2013,GB2016,GM2016}.
Different powerful and frequency-dependent tidal dissipation mechanisms can then emerge, like the Ohmic dissipation of magnetic energy overtaking the viscous dissipation of kinetic energy in various low-mass stars throughout their lives \citep[see][]{AM2019}, or enhanced dissipation at corotation resonances (or critical layers, where the phase speed of the wave matches the local velocity of the background flow) in differentially-rotating bodies \citep[see][for an analytical characterisation of this phenomenon]{AP2021}. 

Inertial wave propagation also depends strongly on the geometry of the container. While in the special case of a uniformly-rotating homogeneous full sphere (or spheroid) the governing Poincaré equation for inertial oscillations is separable in the coordinates, and allows for a complete set of regular solutions \citep[i.e. normal modes,][]{B1889,OL2004,W2005a,W2005b,L2020}, the problem is in general ill-posed in other geometries due to the conflicting boundary conditions \citep{RG2000,RG2001}.
In uniformly-rotating spherical shells (relevant for models of the convective envelopes of stars and planets), singular modes show up in the inviscid limit, taking the form of limit cycles, which are periodic and straight paths of characteristics (also called attractors), reflecting from the shell boundaries. In the presence of viscosity, these singular solutions are regularised and both their energy and viscous dissipation are focused along attractors, developing oscillating shear layers, which are centred around them with a certain width depending on the viscosity \citep{K1995,RV1997,RV2018}. The activation of these shear layers is determined by the size of the shell compared to the core  and by the forcing frequency. They are often associated with intense peaks of tidal dissipation when integrating over the shell \citep[e.g.][]{O2009}, and these resonant peaks could be associated with the presence of hidden global modes, by analogy with those observed in full sphere geometry \citep{LO2021}.

On top of the effects listed above, non-linear fluid effects should start to play a role in (at least) the most compact exoplanetary systems. For instance, the nonlinear interactions between the large-scale equilibrium tide and smaller-scale inertial waves gives rise to the elliptical instability. This is thought to be a highly dissipative mechanism able to synchronise the planet and damp its obliquity (but not excite it) and eccentricity, but probably only for the very-shortest Hot Jupiters, and which has been extensively studied in experiments and numerical simulations \citep[e.g.][]{K2002,LL2010,CL2013,BL2013,BL2014,B2016}. Interestingly, non-linear effects are likely to affect small-scale tidal waves for even smaller tidal amplitudes than for the
large-scale equilibrium tidal flows, as discussed for gravity modes in e.g.~\citet{BO2010}, \citet{B2011}, and see also the astrophysical discussion of Sect. \ref{sec:astro} later in this paper. In his pioneering work, \cite{T2007} demonstrated in numerical computations that the nonlinear self-interaction of inertial waves in shear layers in spherical shells can generate strong axisymmetric zonal (or geostrophic) flows in spherical shells. Along these lines, 
\cite{ML2010} and \citet[hereafter Paper I]{FB2014} verified through experiments and numerical simulations, respectively, that tidal forcing is able to trigger this nonlinear phenomenon, through the non-uniform deposition of angular momentum inside a convective tidally-deformed sphere or shell, akin to the convective regions of stars and giant planets. In such a previously poorly-studied and complicated differentially-rotating environment, the dissipative properties of tidal inertial waves can be deeply affected \citep[as was also shown in the linear study of][with cylindrical differential rotation, which depends only on the distance from the axis of rotation]{BR2013}. Moreover, approaching astrophysically-relevant regimes for compact exoplanetary systems, which have low viscosities and high tidal forcing amplitudes, favours such nonlinear effects and is also conducive to the occurrence of various kind of fluid instabilities, like parametric or shear instabilities \citep[Paper I]{JO2014}. 

In this paper, we revisit the effects of nonlinearities on tidally-forced inertial waves in direct numerical simulations
of a convective shell, modelling the convective envelopes of a low-mass star or a giant gaseous planet subjected to an imposed tidal potential. In this study, we continue along the lines of Paper I by simulating an incompressible shell, relegating a study of realistic density profiles to future work. In contrast to previous numerical studies, we use a more realistic tidal forcing to excite inertial waves, through the residual action of the equilibrium tide  acting as an effective body force \citep[e.g.][]{O2013}. In this way, it is possible to analytically and numerically unveil the role of the nonlinear terms in redistributing energy between the different scales and components of the tidal flows, uncovering the origin of some unexpected evolution of angular momentum observed for some tidal frequencies in Paper I. In addition, more robust predictions (in the astrophysical context) can be made for the impact of nonlinearities on tidal dissipation rates and on scaling laws for the emerging zonal flows.

In Section \ref{sec:model}, we describe the analytical model governing the excitation, propagation, and dissipation of tidal inertial waves in an adiabatic and incompressible convective shell. We also derive the energy and angular momentum balances of tidal flows in this framework. Then, in Section \ref{sec:num}, we perform and describe new results obtained from direct numerical simulations of tidal waves in spherical shells, comparing our results with the prior study of 
Paper I. Therein, we also identify the formation and effects of differential rotation, including corotation resonances, and the occurrence of parametric instabilities of inertial waves. Finally, we evaluate the relevance of our work for exoplanetary systems in Section \ref{sec:astro}, and we discuss the limitations of our model and conclude in Sect. \ref{sec:end}.
\section{Model for tidally-forced inertial waves}
\label{sec:model}
\subsection{Tidal flow decomposition and governing equations}
We consider the convective shell of an undistorted spherical body, an idealised representation of the convective envelope of a low-mass star or a giant gaseous planet, which is subjected to the tidal potential $\Psi$ due to an orbiting companion. The thickness of the convective shell is determined by the radial aspect ratio (the ratio of the inner to outer radii) which is  set to $\alpha=0.5$ in the rest of this study. This choice is directly relevant for the envelopes of certain low-mass stars and giant planets with extended dilute cores (e.g.~as observed for Jupiter), and is made so that the peculiar inertial wave behaviour in shells is captured without adopting a very thin shell. The body is assumed to have an initial uniform rotation $\bm\Omega=\Omega\bm e_z$ along the vertical axis with unit vector $\bm e_z$, with constant $\Omega$ that is small compared to the critical angular velocity of the body in order to neglect centrifugal effects. 
Inertial waves, which are restored by the Coriolis pseudo-force, can thus be excited between the cut-off frequencies $\pm2\Omega$ in the fluid frame. For their treatment, the convection inside the shell is treated only insofar as it is assumed to have led to an adiabatic stratification (and it possibly provides a turbulent viscosity), 
which is, for instance, a reasonable assumption for most of the solar convective envelope \citep[e.g.][]{C2021}, and probably also for the outermost convective envelope of Jupiter and Saturn \citep{DC2019,mh2019}. Indeed, the squared Brunt-Väisälä frequency\footnote{The characteristic frequency associated with buoyancy forces.}, is in reality probably negative in the bulk of the convective envelope, thus driving convective motions, but is mostly negligible in absolute value compared to $\Omega^2$, except possibly in a small region near the surface where the density is very low and the convection is no longer efficient. The convective shell is also considered incompressible as a first approach, with a uniform density $\rho$.

In our model, tidal flows are treated as Eulerian perturbations of the hydrostatic equilibrium of the body using  the non-wavelike/wavelike decomposition (equivalent to the
equilibrium/dynamical tide\footnote{Due to the incompressibility, the definitions of the equilibrium tide made by \cite{Z1966a} and \cite{GN1989} are equivalent, since the non-wavelike flow as defined by Eq. (\ref{eq:X}) hereafter is both divergence-free and curl-free.}) introduced by \cite{O2013}.
Accordingly, we use in the rest of the paper the subscripts $_\mathrm{w}$ and $_\mathrm{nw}$ to designate the tidal response related to the wavelike (dynamical) and non-wavelike (equilibrium) tides, respectively. In the frame co-rotating with angular frequency $\boldsymbol{\Omega}$ and spherical polar coordinates $(r,\theta,\varphi)$ centred on the body at time $t$, the tidal potential can be written as
\begin{equation}
    \Psi(r,\theta,\varphi,t)=A\,\mathrm{Re}\left[\left(\frac{r}{R}\right)^lY^m_l(\theta,\varphi)\e{-i\omega t}\right],
\end{equation}
where we introduce the tidal amplitude $A=G M_2R^2/a^3$, which depends on the mass $M_2$ and semi-major axis $a$ of the perturber (with $G$ the gravitational constant), $R$ the radius of the tidally-perturbed body, $Y^m_l$ the orthonormalised spherical harmonic of degree $l$ and order $m$ (the azimuthal wavenumber) with $m\leq l$, and $\omega$ the tidal forcing frequency in the rotating frame. Note that we only consider the dominant quadrupolar tidal component of $\Psi$ for $l=m=2$. This approximation is valid for quasi-coplanar and quasi-circular systems \citep[i.e. for asynchronous tides, e.g.][]{O2014}, but is also the dominant component for an eccentric orbit. The tidal potential satisfies Laplace's equation ($\Delta\Psi=0$) inside the perturbed body \cor{whose} tidally-modified gravitational potential \cor{is} $\Phi_\mathrm{nw}$.

As its name suggests, the non-wavelike tide is the non-oscillatory
, instantaneous and hydrostatic, fluid response to the tidal potential, and its associated flow. By defining its displacement $\bm\xi_\mathrm{nw}$, flow velocity $\unw=\partial_t\bm\xi_\mathrm{nw}$,  pressure perturbation $p_\mathrm{nw}$, and density perturbation $\rho_\mathrm{nw}$, the non-wavelike tide thus satisfies the closed set of linearised equations:
\begin{equation}
\begin{aligned}
    &\partial_t\unw=-\nabla\left(\frac{p_\mathrm{nw}}{\rho}+\Phi_\mathrm{nw}+\Psi\right),\ \ 
   & p_\mathrm{nw}=-\rho\left(\Phi_\mathrm{nw}+\Psi\right),\\
    &\rho_\mathrm{nw}=-\rho\bn\cdot\bm\xi_\mathrm{nw},\hspace{-5ex}
    &\Delta\Phi_\mathrm{nw}=4\pi G\rho_\mathrm{nw},
    \end{aligned}
    \label{eq:nw}
\end{equation}
where the equations correspond respectively to the momentum, hydrostatic equilibrium, continuity, and Poisson equations. These are completed by the inner and outer boundary conditions:
\begin{equation}
\left\{\begin{aligned}
    &\xnw\cdot\bm n=0 &\text{at}~~ r=\alpha R,\\
    &\xnw\cdot\bm n=-\frac{\Phi_\mathrm{nw}+\Psi}{g} &\text{at}~~ r=R,
\end{aligned}\right.
\label{eq:bnw}
\end{equation}
 with $\bm n$ the unit vector normal to the boundaries and $g$ the constant surface gravitational acceleration of the body. Note that here we adopt an impermeable inner boundary, which more appropriately describes tides in the envelopes of giant planets containing a solid core than in a low-mass star, where we must match the non-wavelike tidal displacement to the equilibrium tide in the radiative core \citep[e.g.][]{B2020}. 
 
 From Eqs. (\ref{eq:nw}) and (\ref{eq:bnw}) the non-wavelike displacement can be derived from a potential such that $\xnw=-\bn X$ \citep{O2013} where
 \begin{equation}
    X(r,\theta,\varphi,t)=\frac{\Ct R^2}{2(1-\alpha^5)}\left[\left(\frac{r}{R}\right)^2+\frac{2}{3}\alpha^5\left(\frac{R}{r}\right)^3\right]\e{-i\omega t}Y_2^2(\theta,\varphi),
    \label{eq:X}
\end{equation}
with $\Ct$ a constant embodying the amplitude of the tidal forcing. The dimensionless parameter $\Ct$ can be expressed as 
\begin{equation}
\Ct=(1+k_2)\epsilon,   
\label{eq:CT}
\end{equation}
introducing two useful astrophysical quantities: the usual dimensionless tidal amplitude parameter $ \epsilon=(R/a)^3M_2/M_1$ with $M_1$ the mass of the perturbed body \citep[][]{O2014,B2016}, along with the real part of the quadrupolar Love number $k_2=\frac{3\rho}{5\overline\rho-3\rho}$ with $\overline\rho$ the mean density in the whole body (core plus envelope). The quantity $\epsilon$ quantifies the ratio of the amplitude of the tidal potential over the self-gravity of the perturbed body at its surface, while $k_2$ accounts for its non-dissipative tidal deformation and is determined by calculating the real part (in-phase component) of the ratio of the amplitude of the perturbed gravitational potential to the tidal potential, $\Phi_\mathrm{nw}/\Psi$, for $l=m=2$ \citep{L1911,O2013}. Since $k_2$ is constrained such that $0<k_2\leq3/2$, with its maximum reached for a full homogeneous sphere (when $\rho=\overline\rho$), the value of $\Ct$ using $\alpha=0.5$ is also a measure of the magnitude of the tidal forcing  with a similar value to $\epsilon$ (their astrophysical values will be discussed further in Sect. \ref{sec:astro}).

With the non-wavelike perturbation now being defined as above, the remaining vortical action of the rotation on this tide $\ft = -2\bm\Omega\wedge\bm u_\mathrm{nw}$ is used to excite the non-zero frequency (i.e. wavelike) tidal response \citep{OL2004,O2005,O2013}. The nonlinear equation of motion for the wavelike (dynamical) tides in the fluid frame is thus:
\begin{equation}
    \partial_t\uw+2\bm\Omega\wedge\uw+(\bm u\cdot\bn)\bm u =-\frac{\bn p_\mathrm{w}}{\rho}+\nu\Delta\bm u_\mathrm{w} +\ft,
    \label{eq:w}
\end{equation}
where $\bm u=\uw+\unw$ is the total perturbed flow and $\nu$ is an effective viscosity accounting for the sum of the molecular kinematic viscosity and an eddy viscosity modelling the action of turbulent convective motions on tidal flows. We point out that the molecular  viscosity in convective regions may be negligible compared to the turbulent one. 
While realistic values of the kinematic Ekman number
$\Ek=\nu/(\Omega R^2)$ usually range between\footnote{For example $\Ek=10^{-11}-10^{-12}$ for the Sun \citep[e.g.][]{C2013}, while $\Ek=10^{-18}-10^{-19}$ for Jupiter \citep[e.g.][]{GW2014}.}  $\Ek=10^{-10}-10^{-20}$ in convective envelopes, the turbulent one is often taken to be $10^{-4}-10^{-6}$ when calculated from mixing-length theory in stellar evolution models \citep[e.g. as in][]{AM2019,FG2022}.
It should also be stressed that while the action of turbulent convective motions on the equilibrium tides has long been modelled as a turbulent eddy viscosity\footnote{It has been introduced in the analytical studies of \cite{Z1966a,Z1966b}, \cite{GN1977} and \cite{Z1989} using mixing length theory, and supported by the numerical analysis of \citet[in local box models]{DB2020a,DB2020b} and \citet[in global spherical models]{VB2020b,VB2020a}.}, whether or not this is fully justified, 
the modelling of the action of turbulent motions on the wavelike tides is still (even more) highly speculative.

In Eq. (\ref{eq:w}), we neglect the dissipative term associated with the non-wavelike tidal flow ($\nu\Delta \unw$) since this term is estimated to be small relative to those retained when inertial waves are excited (see also the last paragraph of Sect.~\ref{sec:bal}). In addition, we assume that the non-wavelike tide is perfectly maintained on the timescale of our simulations, which are much shorter than tidal evolutionary timescales and thus probe the ``instantaneous" tidal response for a given system \citep[similar reasoning has been used in][]{BA2021}. However, it is worth noting that we have so far kept the non-wavelike tidal flow in the nonlinear advection term $(\bm  u\cdot\bm\nabla)\bm u$, which thus consists of four terms, and we will discuss this choice in Sect. \ref{sec:simu_nl}. The equation of motion Eq. (\ref{eq:w}) is also combined with the continuity equation for the wavelike perturbations $\bn\cdot\uw=0$, assuming incompressibility. To complete this model, we adopt stress-free and impenetrable boundary conditions for the wavelike flow at both the inner $r=\alpha R$ and outer $r=R$ boundaries. Namely, no tangential stress ($\bm n\wedge([\sigma]\bm n)=\bm 0$ with $[\sigma]$ the viscous stress tensor) 
and no radial velocity $\uw\cdot\bm n=0$ at the boundaries \citep[e.g.][]{R2015}. Though a no-slip boundary condition might be preferred at the inner boundary in giant gaseous planets if they possess a solid core, this choice of boundary condition does not seem to significantly affect the propagation of inertial modes in a spherical shell \citep[Paper I]{RV2010}, and our choice is also more appropriate for modelling stars.

We non-dimensionalise our simulations using the length-scale $R$, the time-scale $\Omega^{-1}$ and the density value $\rho$.
\subsection{Kinetic and angular momentum balances}
\label{sec:bal}
We derive in this section the energy and angular momentum balances in our model, and analyse the energetic redistribution and exchanges between the wavelike and non-wavelike tidal flows, which are of prime importance for analysing our non-linear simulations (Sect. \ref{sec:num}), and motivate our choices for neglecting certain non-linear terms driving unrealistic fluxes through the boundary (Sect. \ref{sec:simu_nl}).

From the scalar product between $\rho\uw$ and the momentum equation Eq. (\ref{eq:w}) and after spatial integration
$\langle\cdot\rangle$ over the shell's volume $V$, one can get the kinetic energy $K=\langle\rho\uw^2/2\rangle$ balance for tidal inertial waves:
\begin{equation}
    \partial_t K= \mI{nw-w}+\mI{w-nw}-D_\nu+\pt,
    \label{eq:bal}
\end{equation}
where we have introduced  the nonlinear energy transfer terms  $\mI{i-j}=-\langle\rho\uw\cdot(\bm u_i\cdot\bn)\bm u_j\rangle$ with $i,~j\in\{\mathrm{w,nw}\}$ between the wavelike and non-wavelike flows, the tidal power
\begin{equation}
    \pt=\langle \rho\uw\cdot\ft\rangle,
\end{equation}
reflecting the energy injected into the wavelike flows by the tidal work, which is damped by the tidal viscous dissipation 
\begin{equation}
    D_\nu=-\langle\rho \nu\uw\cdot\Delta\uw \rangle,
\end{equation}
of the wavelike flow.

In the derivation of Eq. (\ref{eq:bal}), several terms have vanished due to the assumptions of our model. The energy transfer term involving the wavelike/wavelike nonlinearity does not appear in the energy balance because: 
\begin{equation}
\mI{w-w}=-\langle\rho\uw\cdot(\uw\cdot\bn)\uw\rangle=
-\frac{\rho}{2}\int_{\delta V}\uw^2~\uw\cdot\bm n\,\mathrm{d}S=0,
\label{eq:Iww}
\end{equation}
when applying the divergence theorem with a surface element $\mathrm{d}S$ and a unit vector $\bm n$ normal to the shell  boundaries, together with incompressibility and impenetrability at the boundaries $\delta V$ for the wavelike flow ($\bn\cdot\uw=0$ and $\uw\cdot\bm n=0$, respectively). For the same reasons, there is no term associated with the wavelike pressure. The transfer term associated with the non-wavelike/non-wavelike nonlinear advection also cancels out:
\begin{equation}
    \begin{aligned}
    \mI{nw-nw}&=-\langle\rho\uw\cdot(\unw\cdot\bn)\unw\rangle \\
    &=-\left\langle\rho\uw\cdot\left[\bn(\unw^2/2)-\unw\wedge(\bn\wedge\unw)\right]\right\rangle
    \\
    &=-\frac{\rho}{2}\int_{\delta V} \unw^2~\uw\cdot\bm n\,\mathrm{d}S=0,
    \label{eq:Inwnw}
    \end{aligned}
\end{equation}
where we use in the second line standard vector calculus identities as in \citet[Eq. (7)]{BA2021}, along with the fact that the non-wavelike flow is curl-free and the wavelike flow is impenetrable at the boundaries. 

By using the divergence-free assumption for the non-wavelike flow $\bn\cdot\unw=0$, the first "mixed" energy transfer term $\mI{nw-w}$ in Eq. (\ref{eq:bal}) can be recast as
\begin{equation}
    \mI{nw-w}=-\langle\rho\uw\cdot(\unw\cdot\bn)\uw\rangle=-\frac{\rho}{2}\int_{\delta V}~\uw^2\unw\cdot\bm n\,\mathrm{d}S.
    \label{eq:Inww}
\end{equation}
In a tri-axial ellipsoidal body, which more realistically corresponds to a centrifugally and tidally deformed  star or planet, the streamlines of the non-wavelike flow are supposed to be tangent to the surface boundary in the bulge frame (rotating with the orbit, for asynchronous tides). Thus, we expect that the surface integral in Eq. (\ref{eq:Inww}) also vanishes on average in such a deformed geometry in the bulge frame, since $\bm u_\mathrm{nw}\cdot\bm n=0$ there \citep{BO2016,BA2021}. In the fluid frame, we do not expect this term to vanish identically but we will explain later that its magnitude is expected to be much smaller than $\mI{w-w}$ (in a point-wise sense). This integral does not vanish in our undeformed spherical shell model, due to advection of kinetic energy through the spherical boundaries by the non-wavelike flow. This has important consequences for the evolution of the angular momentum, as will be seen in the following. Finally, the last non-zero energy transfer term gives
\begin{equation}
    \mI{w-nw}=-\langle\rho\uw\cdot(\uw\cdot\bn)\unw\rangle=\langle\rho\unw\cdot(\uw\cdot\bn)\uw\rangle,
    \label{eq:Iwnw}
\end{equation}
using the same arguments as for Eq. (\ref{eq:Iww}). The quantity $\mI{w-nw}$ hence features the Reynolds stresses involving correlations between wavelike flows components and gradients of the non-wavelike flow. 

Note that, if we define $\mathcal{K}_{i}=\rho|\bm{u}_i^2|/2$ where $i\in\{\mathrm{w,nw}\}$,
\begin{equation}
\begin{aligned}
    -\langle\rho\uw\cdot(\bm u\cdot\bn)\bm u\rangle&=\sum_{i,j\in\{\mathrm{w,nw}\}}\mI{i,j}\\
    &=-\int_{\delta V}\mathcal{K}_\mathrm{w}\  \unw\cdot\bm n\,\mathrm{d}S~+\mI{w-nw},
\end{aligned}   
\label{eq:14}
\end{equation}
and similarly
\begin{equation}
\begin{aligned}
    -\langle\rho\unw\cdot(\bm u\cdot\bn)\bm u\rangle&=
    -\int_{\delta V}(\mathcal{K}_\mathrm{nw}+\unw\cdot\uw)\  \unw\cdot\bm n\,\mathrm{d}S-\mI{w-nw}.
\end{aligned} 
\label{eq:15}
\end{equation}
From these two equations, $\mI{w-nw}$ is clearly identified as an energy transfer term between the wavelike and non-wavelike flows. It takes the opposite sign in the equation for the energy of the non-wavelike flow, if we were to allow this component to evolve. 
Eqs. (\ref{eq:14}) and (\ref{eq:15}) together imply
\begin{equation}
\begin{aligned}
    -\langle\rho\bm u\cdot(\bm u\cdot\bn)\bm u\rangle
    =&-\int_{\delta V}\mathcal{K}_\mathrm{tot}\  \unw\cdot\bm n\,\mathrm{d}S,
\end{aligned}   
\end{equation}
where $\mathcal{K}_\mathrm{tot}=\rho|\uw+\unw|^2/2$.
Note also that 
\begin{equation}
\langle\rho \bm{u}_\mathrm{w} \cdot \bm{u}_{\mathrm{nw}}\rangle=-\langle\rho \bm{u}_\mathrm{w} \cdot \nabla \dot{X}\rangle=-\int_{\delta V}\rho \dot{X}\bm{u}_\mathrm{w} \cdot\bm n\,\mathrm{d}S=0,
\end{equation}
upon applying the divergence theorem and the boundary conditions,
implying that 
\begin{equation}
K_{\mathrm{tot}}=\langle \mathcal{K}_\mathrm{tot}\rangle =\langle \rho|\uw+\unw|^2\rangle/2=\langle \mathcal{K}_\mathrm{w}+\mathcal{K}_{\mathrm{nw}}\rangle=K + K_{\mathrm{nw}},
\end{equation}
where $K=\langle \mathcal{K}_\mathrm{w}\rangle$ (as above) and $K_\mathrm{nw}=\langle \mathcal{K}_{\mathrm{nw}}\rangle$.

If we were to impose a radial velocity $u_r$ at the outer boundary, like in \cite{O2009} or Paper I, instead of a tidal body force, the (nonlinear) energy balance would be: 
\begin{equation}
    \partial_t K_\mathrm{tot}=-\int_{\delta V(r=R)}(\mathcal{K}_\mathrm{tot}+p_\mathrm{w})\ u_r\,\mathrm{d}S~~~-D_\nu,
    \label{eq:balfab}
\end{equation}
where the first term inside the integral is omitted in linear calculations (since it comes from nonlinear terms), and the second term represents tidal forcing in their model.

The total angular momentum of the fluid in the shell is
\begin{equation}
\bm L=\int_V \rho \bm r\wedge\bm u\,\mathrm{d}V,  
\end{equation}
where $\bm r$ is the position vector. By taking the vector product between $\bm r$ and the momentum equation Eq. (\ref{eq:w}) and integrating over the volume, we get the angular momentum balance
\begin{equation}
    \partial_t\bm L=-2\int_V\rho \bm r\wedge(\bm\Omega\wedge\bm u)\,\mathrm{d}V-\int_V \rho \bm r\wedge(\bm u\cdot\bn\bm u)\,\mathrm{d}V,
    \label{eq:timL}
\end{equation}
where the pressure torque vanishes, as well as the viscous torque by the  adoption of stress-free boundary conditions \citep[Paper I; with the difference that $\bm u=\uw+\unw$ here]{JB2011}.
By injecting the new variable $\tilde{\bm u}=\bm u+\bm\Omega\wedge\bm r$ similarly as in Paper I, Eq. (\ref{eq:timL}) yields
\begin{equation}
  \partial_t\bm L=-\int_V\bm \rho r\wedge\left[(\tilde{\bm u}\cdot\bn)\tilde{\bm u}\right]\,\mathrm{d}V=\int_{\partial V}\rho (\bm r\wedge\tilde{\bm u})\unw\cdot\bm n\,\mathrm{d}S\neq\bm{0},
  \label{eq:ang}
\end{equation}
where the incompressibility of the total flow $\bm u$ has been used in both equalities along with the impenetrability condition for $\uw$. Therefore,
the total angular momentum is not conserved over time in our spherical model, and its evolution is entirely controlled by the surface flux sustained by the "unrealistic" radial non-wavelike flow through the boundaries in our spherical model (since we look at short intervals of time relative to tidal evolutionary timescales, and thus assume $\unw$ to be perfectly maintained). As we will show partly below and in the next section, this non-vanishing flux term, due to either imposed radial velocity boundary conditions, as in Paper I (see Eq. (\ref{eq:balfab})), or through inclusion of the nonlinearity terms involving the non-wavelike tide in our spherical model, is important and causes unexpected tidal evolution, such as the sustained unrealistic tidal de-synchronisation observed in the aforementioned study. Note that the tidal effective forcing itself does not contribute to changing the total angular momentum (in the context of our model that focuses on an ``instantaneous snapshot" of the system's evolution) since $\langle \bm{r}\wedge\bm{f}_\mathrm{t} \rangle=\bm{0}$ (due to the integral over azimuthal angles).

We can crudely estimate the importance of $\mI{nw-w}$ compared to the other terms in the energy balance in Eq. (\ref{eq:bal}) using physical scaling arguments.
The non-wavelike tide has a typical velocity  magnitude $u_\mathrm{nw}\sim R \omega \Ct$ and lengthscale $L\sim R$, because it represents large-scale flows in the convective envelope, its timescale is the inverse of the tidal frequency $\omega$, and its displacement scales as $\xi_\mathrm{nw}\sim R \Ct$ (Eq. (\ref{eq:X})). In a spherical shell, the scales of the wavelike tide are controlled by the structure of the inertial wave shear layers, whose thicknesses depend on the Ekman number (in linear theory). The wavelike tide has corresponding velocity $u_\mathrm{w} \sim \Ek^{-\beta}R\omega \Ct$  and a lengthscale $\ell \sim \Ek^{\alpha} R$, where $\alpha,\beta>0$ are coefficients depending on the particular (singular) mode in the shell: for shear layers $\alpha=1/3$ (if $\ell$ is the wavelength of the oscillation inside the shear layer) or $\alpha=1/4$ (if $\ell$ is the  width of the shear layer) and $\beta=1/6$; for Ekman boundary layers near the critical latitude\footnote{It is the latitude where  the shear layer emerges from the surface of the inner boundary.} $\alpha=2/5$ and $\beta=1/5$ \citep[these scaling have been analytically and numerically verified for instance in][and Paper I]{K1995,RV2018}. Given the low values of the turbulent or microscopic Ekman number $\Ek\ll 1$, in any of these cases we have $\ell\ll L$ and $u_\mathrm{w}\gg u_\mathrm{nw}$ in spherical shells.
Furthermore, we also have the following point-wise approximations 
\begin{align*}
\mmI{nw-w}&\sim \rho u_\mathrm{w}^2u_\mathrm{nw}/\ell, & 
\mmI{w-nw}&\sim \rho u_\mathrm{w}^2u_\mathrm{nw}/L, \\
\mmI{w-w}&\sim \rho u_\mathrm{w}^3/\ell, & \mmI{nw-nw}&\sim\rho u_\mathrm{w}u_\mathrm{nw}^2/L, \\
\mathcal{D}_\nu&\sim\rho \nu u_\mathrm{w}^2/\ell^2, &
\mathcal{P}_\mathrm{t}&\sim\Omega u_\mathrm{w} u_\mathrm{nw},
\end{align*}
where $\mmI{i-j}$ and $\mathcal{P}_\mathrm{t}$ are the energy transfer terms (and $\mathcal{D}_\nu$ the viscous term) before computing the volume integral.
Hence, we typically expect the local ordering 
\begin{equation}
\begin{aligned}
~~\mmI{w-w} &\gg \mmI{nw-w} \gg \mmI{w-nw}\gg\mmI{nw-nw}, 
\label{eq:order}
\end{aligned}
\end{equation}
 in magnitude for shear layers and critical latitudes, for the typical Ekman numbers and tidal amplitudes in stars and planets. In addition, while these orderings are also expected to hold in our simulations
 for the typical values of the Ekman number $\Ek\sim10^{-4}-10^{-6}$ and tidal forcing parameters $\Ct\sim10^{-2}-10^{-1}$ we use in our simulations, we may not expect $\mmI{nw-w}\ll \mmI{w-w}$.
 By consequence, if no cancellations in the surface integral of $\mI{nw-w}$ occur, it is expected that this term will play a prominent role in the energy and angular momentum balances Eqs. (\ref{eq:bal}) and (\ref{eq:ang}), and could drive unrealistic evolution. This provides a first reason to motivate our neglect of the mixed (non-wavelike/wavelike and wavelike/non-wavelike) nonlinearities over the wavelike/wavelike one in simulations. If we neglect this term, we emphasise that the total angular momentum is conserved over time $\partial_t\bm L=0$, and the energy balance now reads:
\begin{equation}
    \partial_t K=-D_\nu+P_\mathrm{t}.
    \label{eq:balw}
\end{equation}

\begin{figure*}
    \centering
    \includegraphics[width=0.33\textwidth]{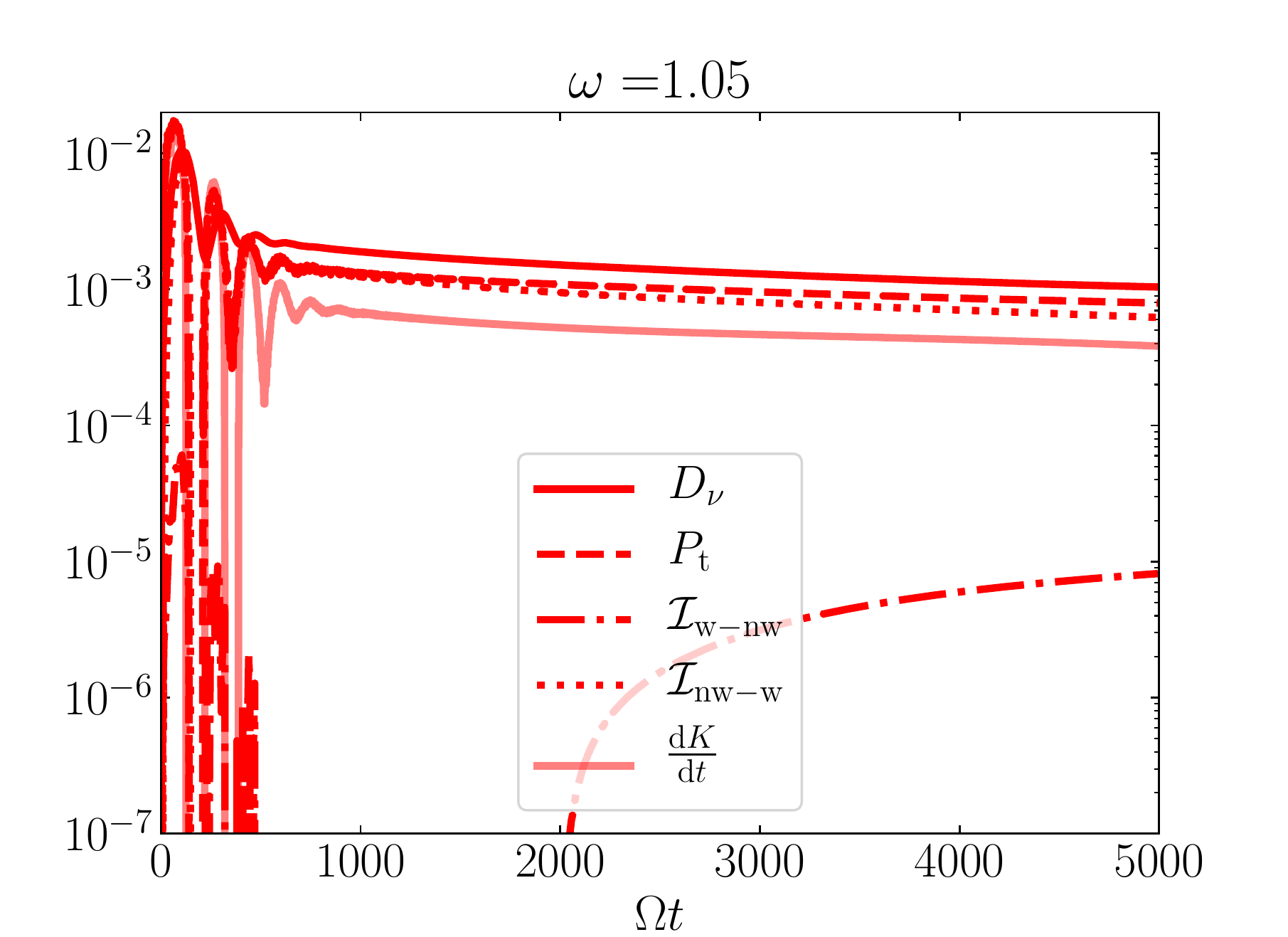}
    \includegraphics[width=0.33\textwidth]{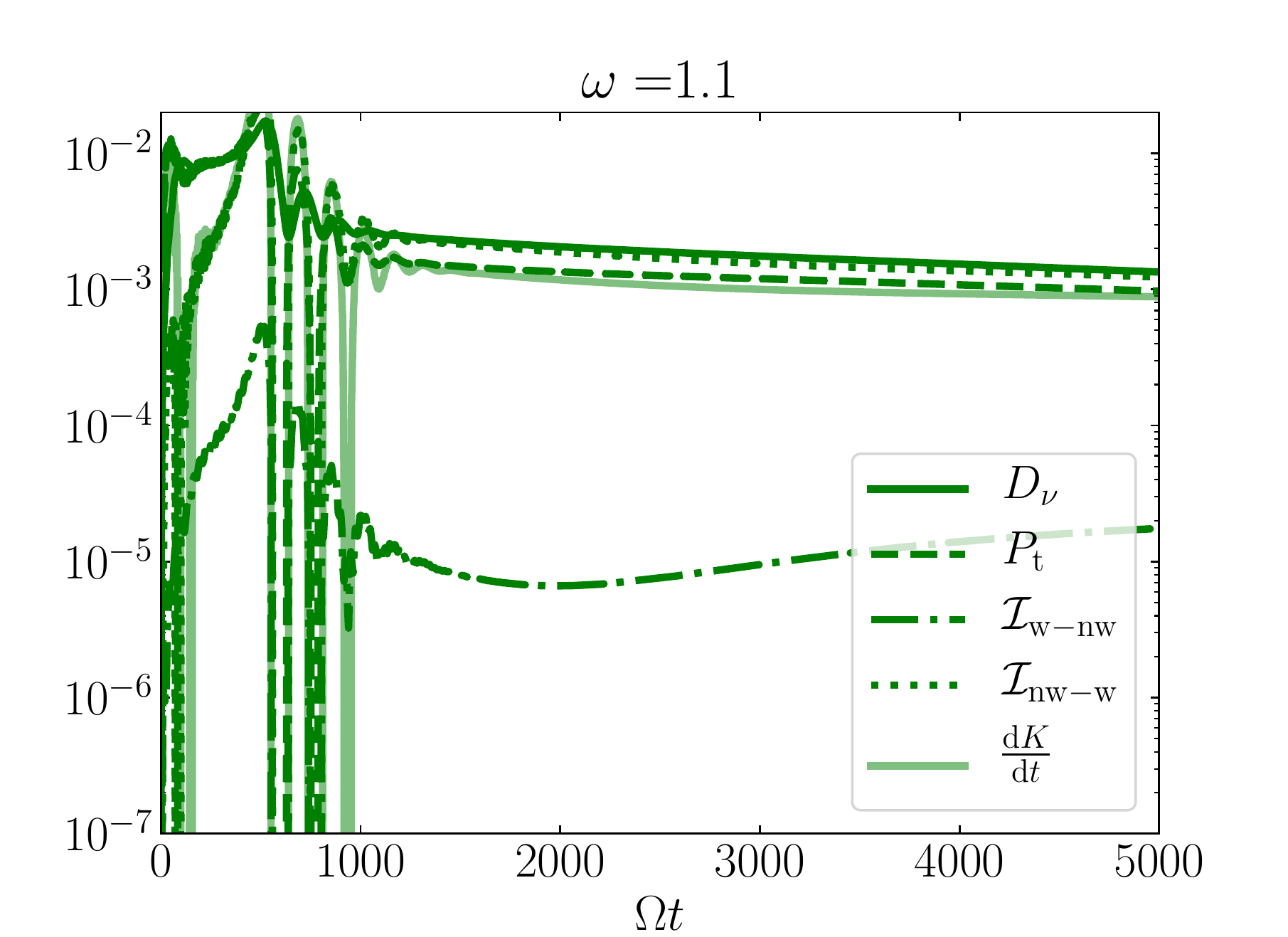}
    \includegraphics[width=0.33\textwidth]{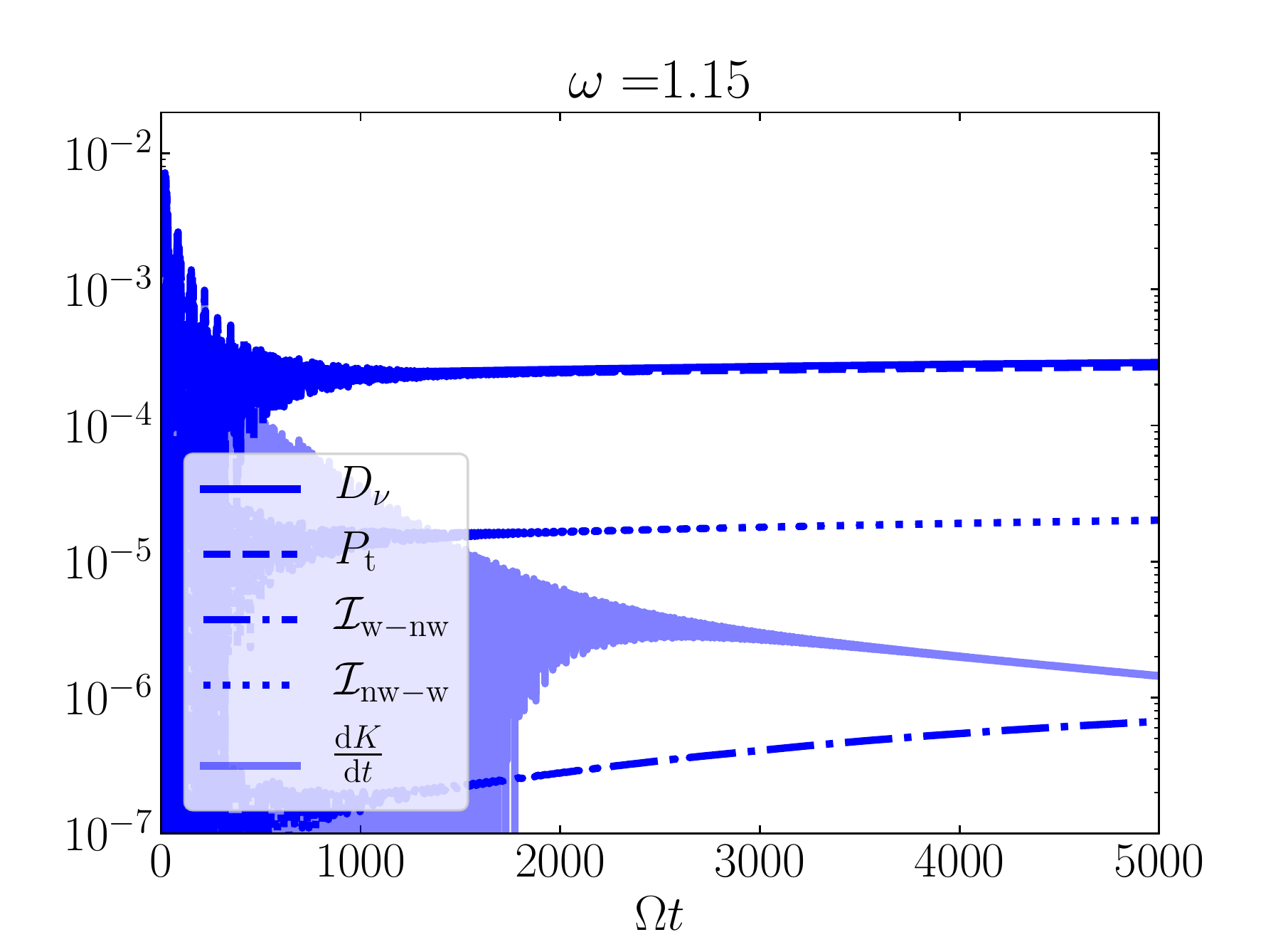}
    \caption{Contributions to the energy balance in Eq. (\ref{eq:bal}) as a function of time for three different forcing frequencies $\omega$, all with $\Ek=10^{-5}$ and $\Ct=10^{-2}/\omega$. A running average with a period $2\pi/\omega$ has been performed to smooth the curves. All y-axis quantities have been re-scaled by a factor $A^2\times32\pi/15\approx10^{-4}$ by taking $A=0.00386$ so as to provide a comparison with Fig. 6 of \fb.}
    \label{fig:rey_om}
\end{figure*}
Morever, given the integral and local relationships Eqs. (\ref{eq:Inwnw}) and (\ref{eq:order}), the non-wavelike/non-wavelike non-linearity contributes negligibly to the energy exchanges in the body. It only modifies the pressure, which is constrained by incompressibility, and therefore this term has no effect in our model with our adopted boundary conditions. 
Finally, assuming the same magnitude of the viscosity holds for both the wavelike and non-wavelike flows, our neglect of the dissipative term for the latter is further justified since $\nu\Delta u_\mathrm{nw}/\nu\Delta u_\mathrm{w}\sim\Ek^{2\alpha+\beta}\ll1$.
\section{Nonlinear simulations}
\label{sec:num}
We now describe and analyse nonlinear simulations of tidally-forced inertial waves governed by Eq. (\ref{eq:w}) for an incompressible flow satisfying stress-free boundary conditions in a convective shell, performed by the open-source code MagIC.
\subsection{Numerical model}
We use version 5.10 of the pseudo-spectral code MagIC\footnote{Available at \url{https://magic-sph.github.io/}.} designed for 3D (magneto-)hydrodynamical simulations of fluid motions in a spherical shell or full sphere \citep[e.g.][for Boussinesq implementation and benchmarks]{CA2001,W2002}.
We employ Chebyshev polynomials in the radial direction and a spherical harmonic decomposition in the azimuthal and latitudinal directions. The code supports a number of mixed implicit/explicit time-stepping schemes, where the nonlinear terms and the Coriolis force are treated explicitly in real (spatial) space to avoid inverting a large matrix (resulting from nonlinear mode coupling), while the remaining linear terms are treated implicitly in spectral space (hence the name pseudo-spectral). The chosen time-stepping is based on an IMEX multistep method, choosing specifically the canonical combination of second-order Crank-Nicolson and Adams-Bashforth schemes to treat the implicit and explicit terms, respectively. The choice of Chebyshev polynomials evaluated at the Gauss-Lobatto collocation gridpoints, instead of low-order finite differences in radius, for example, allows for much better accuracy for smooth solutions for the same number of points, and it is also well adapted to resolve thin (viscous) layers near the boundaries thanks to a denser grid of points close to the inner and outer radii. 
All of our simulations have benefited from the MKL and SHTns \citep{S2013} libraries for fast Fourier, Legendre and spherical harmonics transforms between real and spectral space, and vice versa, along with MPI parallelisation on high performance computing clusters, which significantly improves the speed of our calculations. 
Like in the code PARODY used in Paper I, the wavelike velocity is decomposed using poloidal-toroidal potentials, which automatically ensure that the flow is a solenoidal (i.e. divergence-free) field. Details about this decomposition and the spherical harmonic and Chebyshev representations can be found in the manual of the MagIC code, at  \url{https://magic-sph.github.io/numerics.html#secnumerics}. Spectral projections of the unforced momentum equation for the poloidal and toroidal potentials, and the associated stress-free boundary conditions, in this formalism are also described there.

We have modified MagIC subroutines to implement in real space the tidal forcing $\ft$ and each of the non-linear terms involving the non-wavelike flow which contributes to the energy exchanges according to Sect. \ref{sec:bal}, namely the mixed non-linear terms $(\unw\cdot\bn)\uw$ and $(\uw\cdot\bn)\unw$. 
Since these changes are highly non-trivial (particularly for the mixed nonlinear terms using the code variables), we ensured that we had done this correctly by comparing results obtained independently with an implementation for all of these terms (and subsets of these terms) computed using the spectral element code Nek5000 (which we also used for some of the simulations in \fb). This detailed comparison is omitted here to save space. We have also modified the viscous dissipation routines to calculate the appropriate viscous dissipation and tidal power, and implemented routines to enable volume integrals (of products) of real space quantities to be computed, so that, for example, we can evaluate each of the terms that contribute to Eq.~(\ref{eq:bal}).

In most of our simulations\footnote{Typically for simulations with $\Ek\gtrsim10^{-5}$ and $A\lesssim5\cdot10^{-2}$.}, we adopted a maximum spherical harmonic degree of $l_\mathrm{max}=85$ (meaning  $n_\varphi=256$ longitudinal grid points and $n_\theta=128$ latitudinal grid points), along with a number of radial grid points $n_r=97$ (other spatial resolution choices will be specified later). 
This spatial resolution is sufficient to allow more than three orders of magnitude difference in the kinetic energy spectrum between the low and high order harmonic degrees and azimuthal wavenumbers, while maintaining a fairly good speed for the simulations\footnote{On the order of a few hours with our choice of time step \cor{with 96 cores} to reach several thousand rotation times.}.
We also used a time step of $\delta t=10^{-2}$ in most of our simulations, and checked that this value is low enough to ensure both accuracy (by checking results for convergence using smaller $\delta t$) and stability according to the Courant–Friedrichs–Lewy condition. \cor{Lastly, we checked that strictly enforcing conservation of the total (equatorial and axial) angular momentum at every time step, which is an option in MagIC, does not change our results with only wavelike-wavelike non-linearities.}
\begin{figure*}
    \includegraphics[width=\columnwidth]{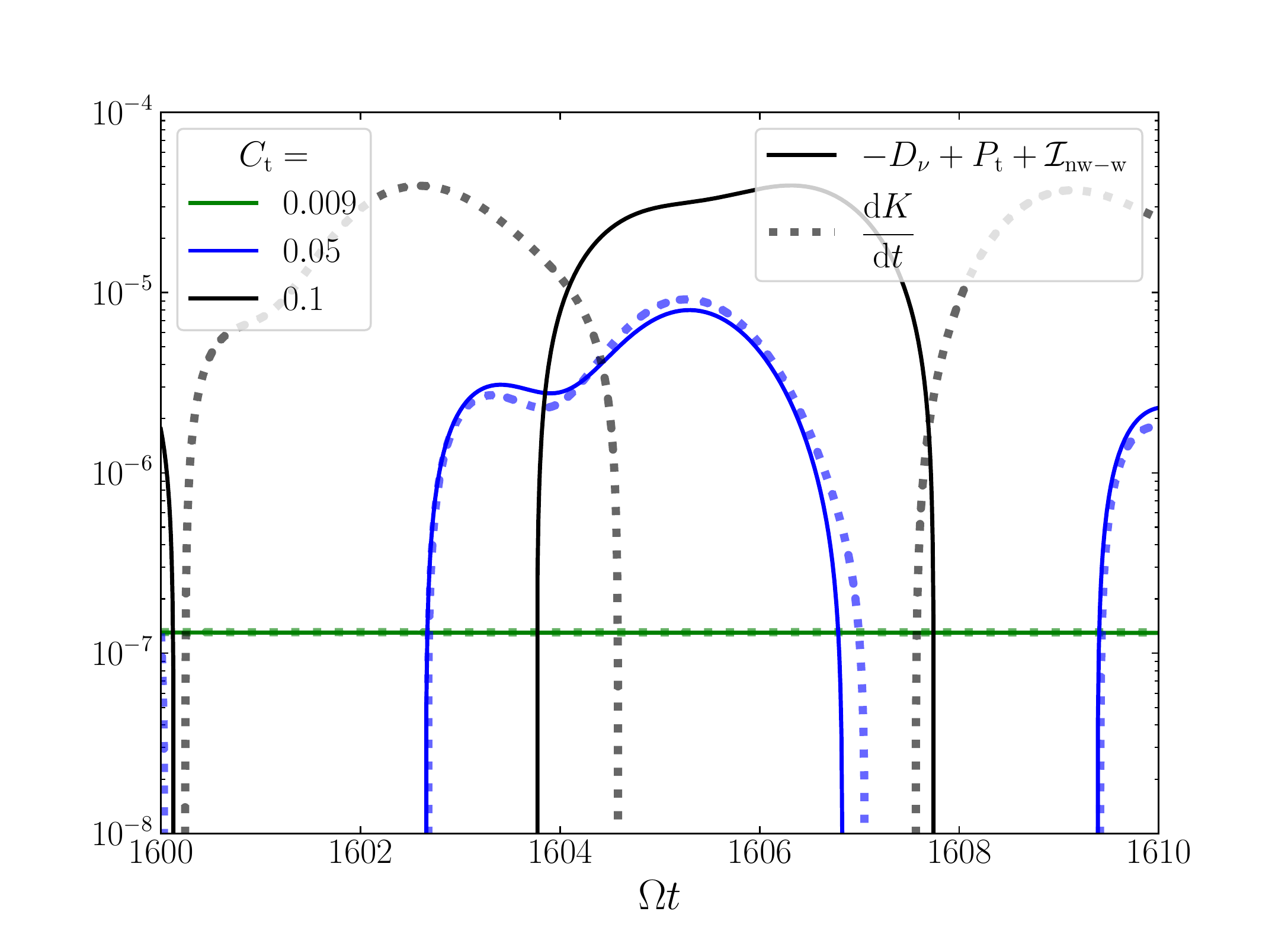}
\includegraphics[width=\columnwidth]{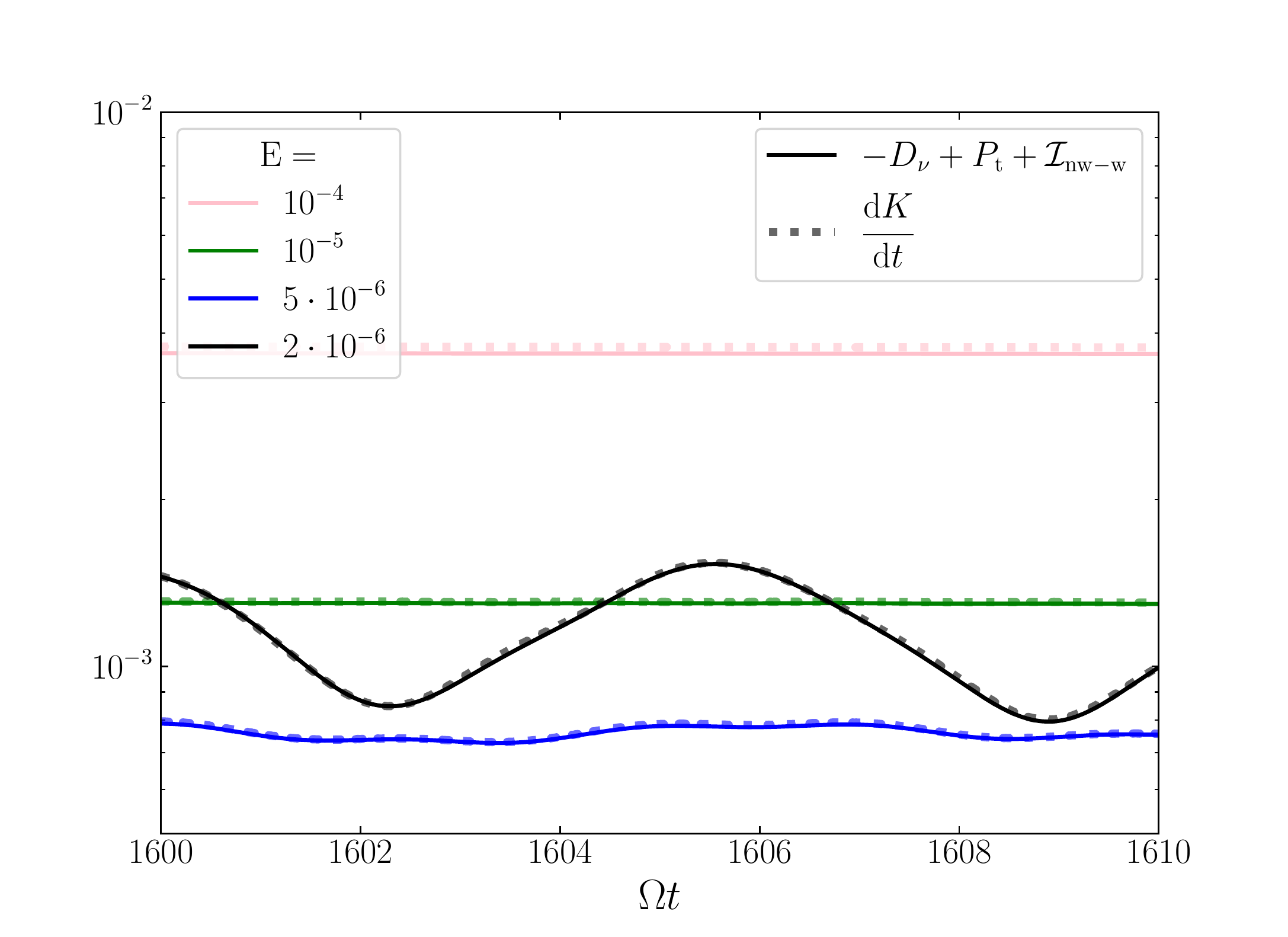}
    \caption{Comparison between the left and right hand sides of the energy balance Eq. (\ref{eq:bal}) for $\omega=1.1$ without the transfer term $\mI{w-nw}$ involving wavelike correlations and gradients of the non-wavelike flow. 
    \textit{Left:} for different amplitudes of the tidal forcing $\Ct$, at fixed Ekman number $\Ek=10^{-5}$ (not re-scaled). \textit{Right:} for different values of the Ekman number $\Ek$ at fixed tidal forcing amplitude $\Ct\approx0.009$ (re-scaled as in Fig. \ref{fig:rey_om}).}
    \label{fig:rey}
\end{figure*}
\subsection{Surface versus body forcing}
\label{sec:simu_nl}
The main difference in our model with \fb, apart from the use of a different numerical code, is the way inertial waves are forced. They imposed an outer radial velocity at the surface $r=R$ to model the 
non-wavelike tidal flow and how it forces inertial waves, which we will refer to as ``surface forcing" in the following. In that case the impenetrability condition only applies at the inner radius $r=\alpha R$. When studying the linear tidal response, surface forcing and the body forcing that we use are equivalent in the asymptotic limit of small tidal forcing frequencies compared to the dynamical frequency $\omega_\mathrm{d}=\sqrt{g/R}$, and they are approximately similar for astrophysically-relevant small but nonzero values of $\omega/\omega_\mathrm{d}$. This regime is valid for slowly-rotating stars or planets for which centrifugal effects can be neglected
\citep[e.g.][]{O2009}. Within this limit, 
both forcing mechanisms can be related by equating the amplitude of the non-wavelike velocity 
in both models, i.e.,
\begin{equation}
\sqrt{\frac{32\pi}{15}}A\approx\omega \Ct,
\label{eq:amp}
\end{equation}
where $A$ is the forcing amplitude in \fb~and $\Ct$ was defined above. For the purposes of recovering the results of Paper I, our tidal forcing amplitude $\Ct$ must be scaled with frequency, since they chose to fix $A$ instead. 

We now discuss nonlinear simulations performed for three initial tidal forcing frequencies $\omega=1.05,~1.1,~1.15$, which were those initially analysed in the linear regime by  \cite{O2009} and in the prior nonlinear simulations of \fb. 
These three cases were chosen because they exhibited drastically different behaviours of the dissipation rate as a function of Ekman number, and had correspondingly different flow structures, despite having similar tidal frequencies.
The tidal forcing amplitude $\Ct$ has been chosen to satisfy Eq. (\ref{eq:amp}) with $A=10^{-2}/\sqrt{32\pi/15}\approx0.00386$ (e.g.~for comparison with Fig. 6 of \fb), but which is slightly different in our simulations for the three tidal frequencies (but $\Ct\approx0.009$). Fig. \ref{fig:rey_om} shows each of the terms appearing in the energy balance Eq. (\ref{eq:bal}) over time, for the three nonlinear simulations with different frequencies and $\Ek=10^{-5}$.

First, we checked that the energy balance as described in Eq. (\ref{eq:bal}) is well satisfied in these simulations. Moreover, in all three cases, the energy transfer term $\mI{w-nw}$ is negligible compared to the viscous dissipation and tidal power, while the term \cor{$\mI{nw-w}$} sustaining the unphysical non-wavelike flux through the boundaries contributes significantly to the energy balance for $\omega=1.05$ and $1.1$. For the $\omega=1.15$ case, the energy injected by the tidal forcing is mainly viscously dissipated, without important transfers of energy between wavelike and non-wavelike flows. These observations are consistent with our discussion of how the terms in the energy balance scale in the last paragraph of Sect. \ref{sec:bal}, by noting that highly dissipative shear layers are excited in linear simulations for $\omega=1.05$ and $1.1$ unlike $\omega=1.15$ \citep[as described in][]{O2009}, making the scaling laws most relevant for the first two frequencies there. However, one may wonder whether the "physical" transfer term $\mI{w-nw}$ is always negligible for our parameters (as seen in Fig. \ref{fig:rey_om}) or not. In Fig. \ref{fig:rey}, we investigate how the omission of $\mI{w-nw}$ impacts the energy balance for different tidal forcing amplitudes $\Ct$ and Ekman numbers $\Ek$ at $\omega=1.1$. In the left panel, one can observe that this term is important for very high tidal forcing amplitudes $\Ct\gtrsim0.1$ (the energy transfer has been checked beforehand), but for moderate to low $\Ct$, typically $\Ct\lesssim 0.05$, the energy balance without $\mI{w-nw}$ remains almost unchanged for these parameters (similar results are found for $\omega=1.05$). Changes in Ekman number (right panel), from $\Ek=10^{-4}$ to $\Ek=2\cdot10^{-6}$, show that $\mI{w-nw}$ is also playing a negligible role in the energy balance for $\Ct\approx0.009$.

Since the non-wavelike flux through the boundaries described by $\mI{nw-w}$ in these simulations is artificial, we wish to remove the contribution from this term from our simulations. To do so we can remove both of the nonlinear terms involving the non-wavelike tide in Eq.~(\ref{eq:w}), which is justified to represent the dominant balance of terms in the astrophysical regime based on the physical scaling arguments we outlined in Sect. \ref{sec:bal}. However, this approach is only justified in our simulations if $\mI{w-nw}$ also does not contribute significantly to the energy balance (since we also remove this term in our approach). From the above discussion, this is likely to be valid for $\Ct\lesssim0.05$ and $\Ek\sim10^{-6}-10^{-4}$.

\begin{figure*}
\includegraphics[width=\columnwidth]{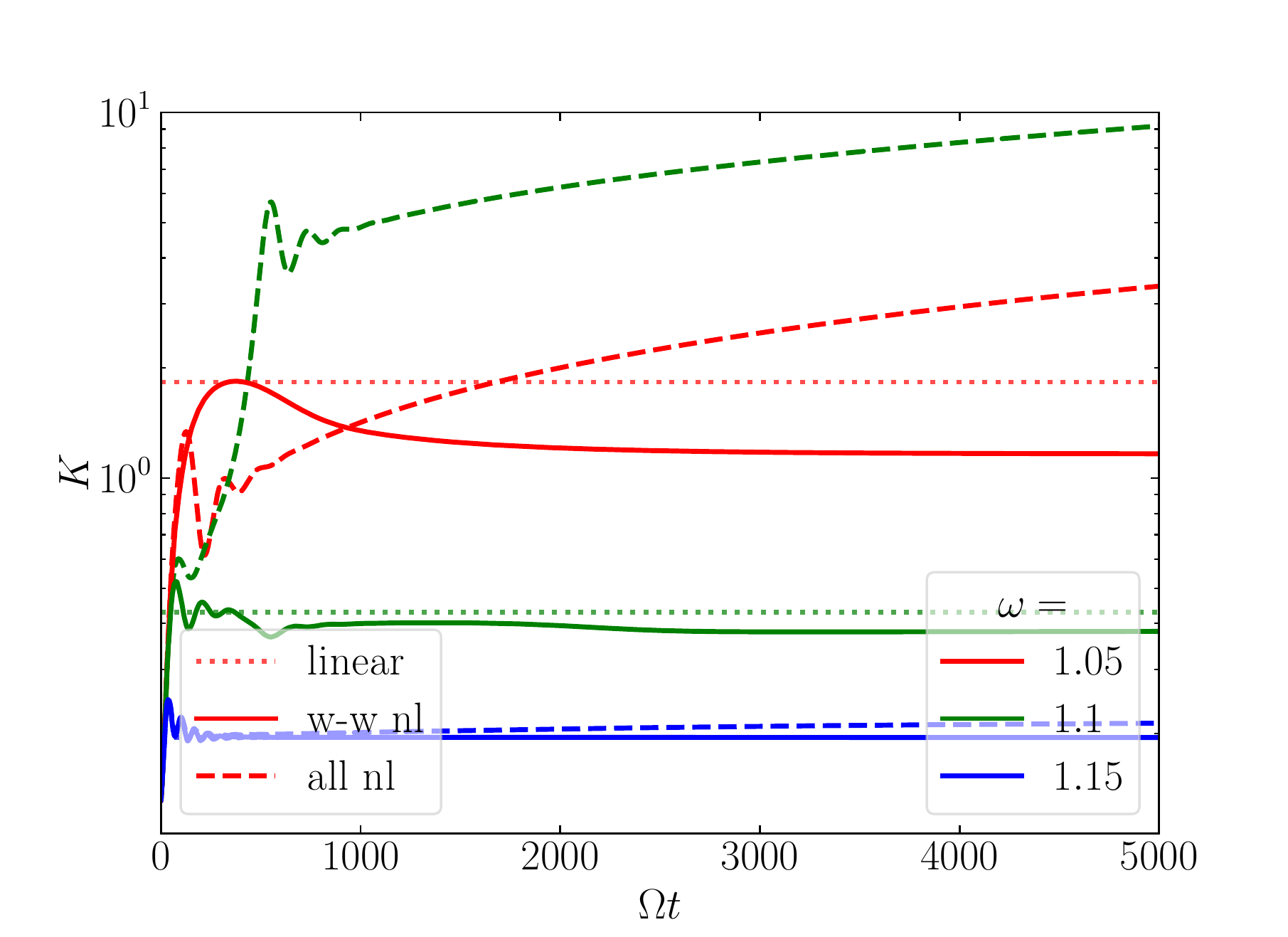}
\includegraphics[width=\columnwidth]{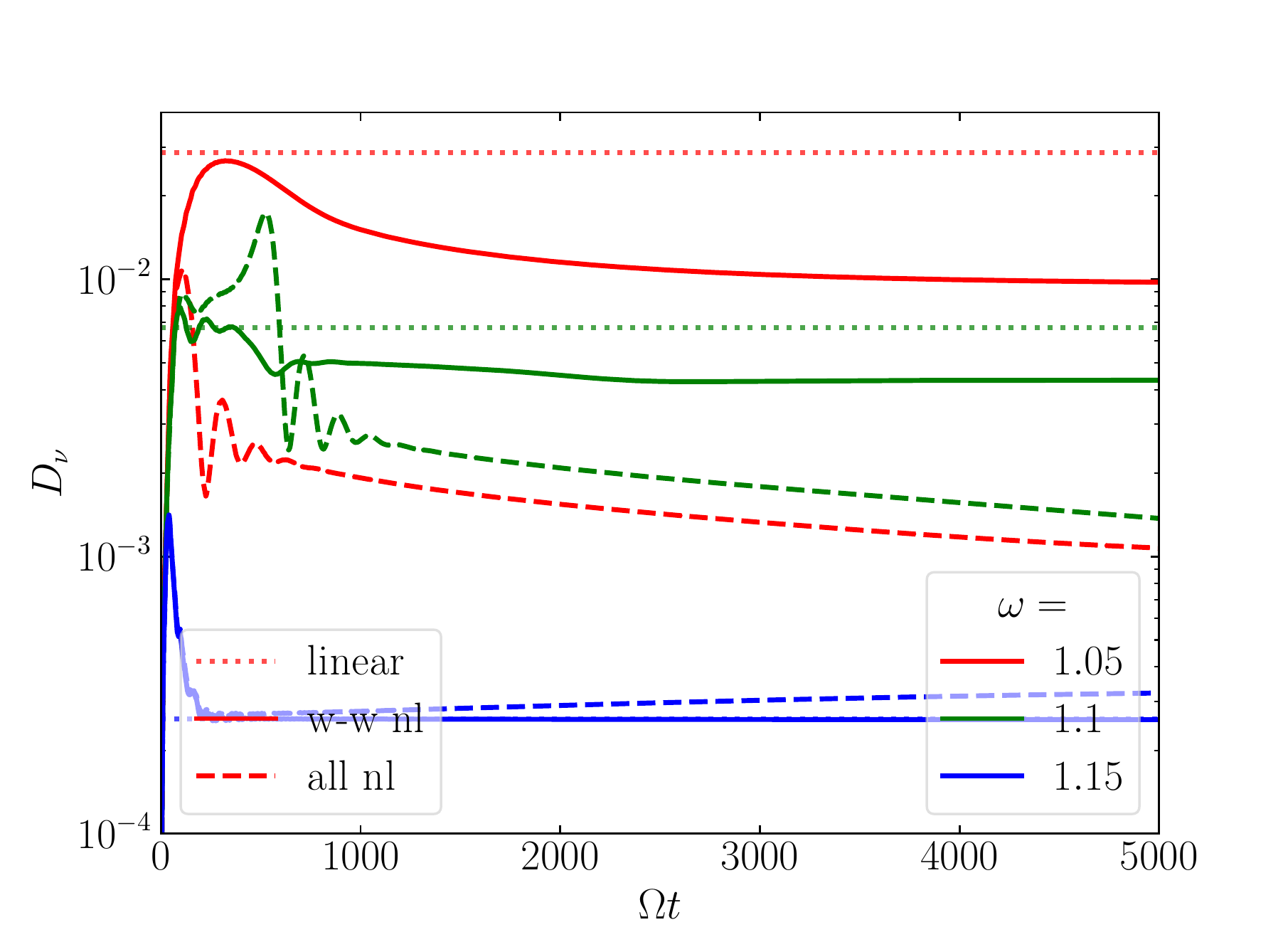}
\caption{Kinetic energy $K$ (\textit{left}) and dissipation $D_\nu$ (\textit{right}) over time for three different forcing frequencies (colour), comparing linear calculations (horizontal dotted lines) with simulations including wavelike/wavelike (w-w, solid lines) nonlinearities, or all nonlinear terms (all nl, dashed lines). 
Kinetic energy and dissipation have been re-scaled as in Fig. \ref{fig:rey_om}. The Ekman number is set to $\Ek=10^{-5}$ and the forcing amplitude $\Ct=10^{-2}/\omega$.
}
\label{fig:kD}
\end{figure*}
\begin{figure*}
    \centering
    \includegraphics[width=\columnwidth]{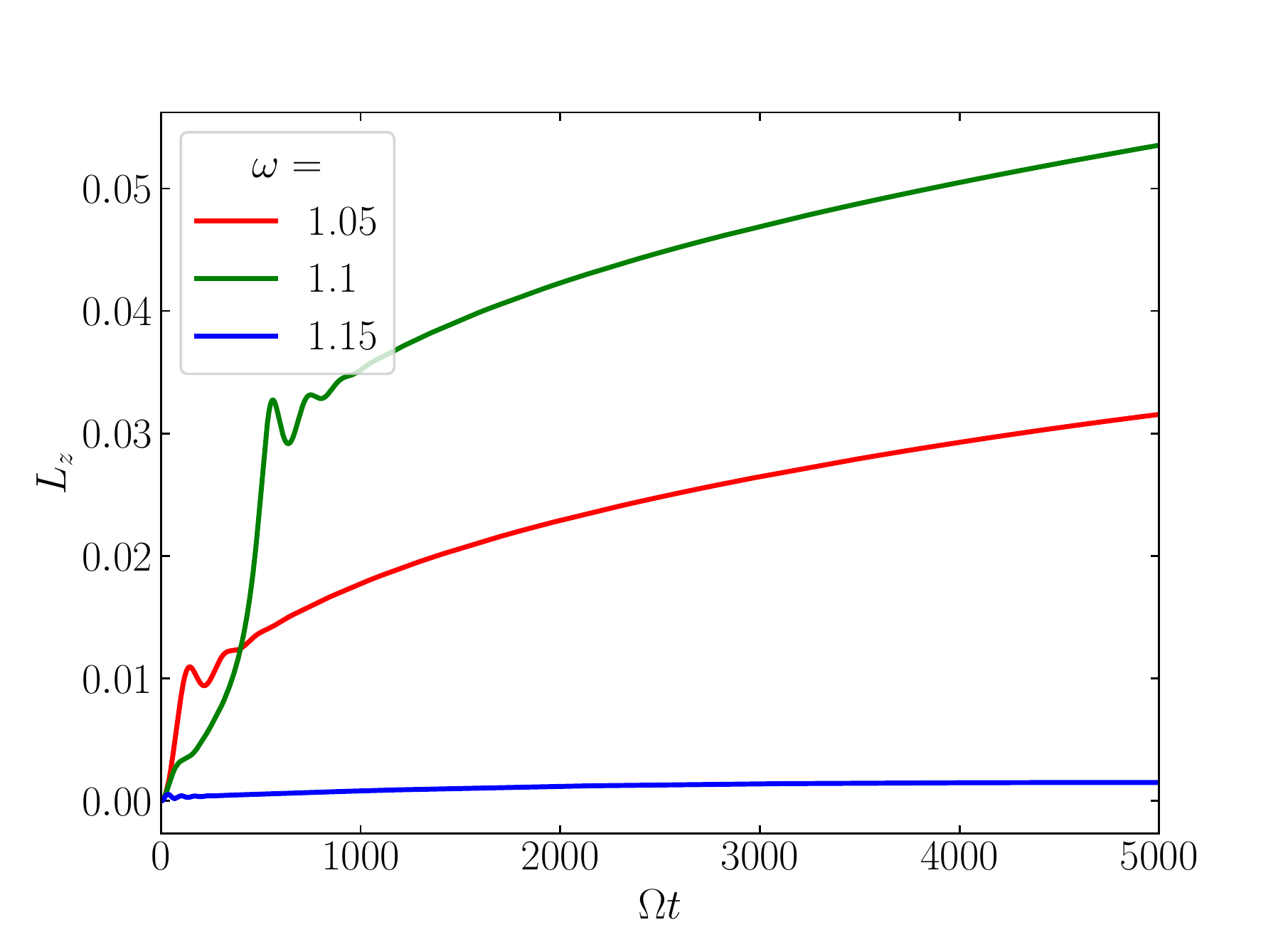}    
\includegraphics[width=\columnwidth]{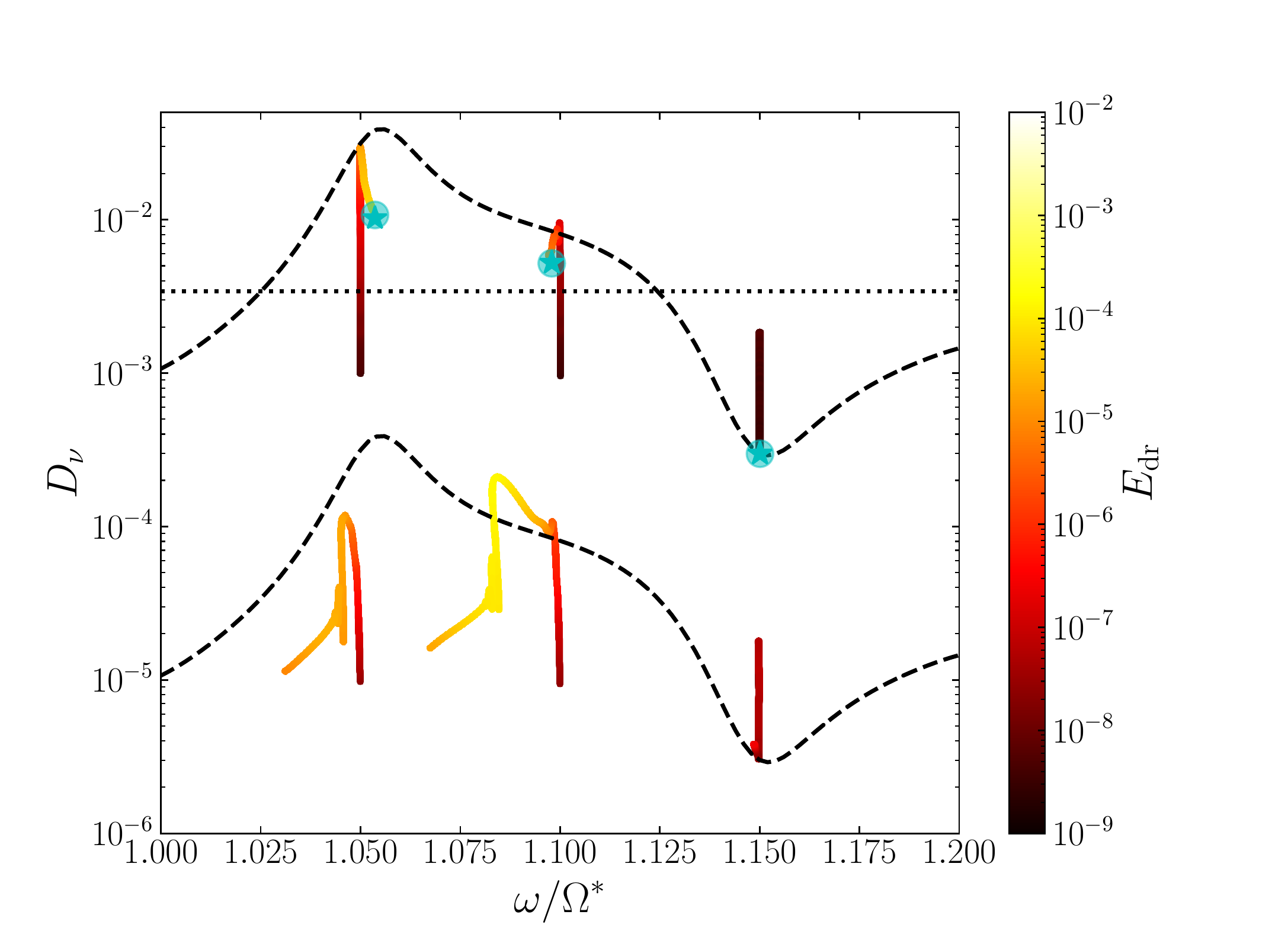}
\caption{\textit{Left:}
Vertical component of the total angular momentum in the rotating frame, $L_z$, as a function of time $\Omega t$ in simulations including all nonlinearities (w-w, nw-w and w-nw) for three initial forcing frequencies $\omega$. \textit{Right:}
Nonlinear dissipation $D_\nu$ (solid lines) as a function of normalised frequency $\omega/\Omega^*$ in simulations from $\Omega t=20$ to $5000$, with all (\textit{bottom}) or only wavelike/wavelike (\textit{top}) nonlinear terms (blue pale bullets indicate the final dissipation when an average steady state is reached). The colour of each line indicates the kinetic energy in the differential rotation $E_\mathrm{dr}$. For comparison, the frequency-dependent linear dissipation for a uniformly rotating body is plotted as the dashed black lines and the impulsive energy injected $\hat{E}$ as the horizontal dotted line. 
All the dissipation rates are re-scaled by a factor $\Ct^2$ (contrary to previous figures showing dissipation where a different scaling is used to match with \fb) and the bottom curves are shifted downwards by a factor $10^{-2}$ for ease of visualisation.
The tidal power term computed from the final steady-state attained in linear simulations with a background cylindrical zonal flow are also added as blue stars.}
    \label{fig:ang}
\end{figure*}
It is thus of interest to compare simulations which feature all of the nonlinear terms with ones that include only the wavelike/wavelike nonlinearity. This is done in Fig. \ref{fig:kD}, displaying the kinetic energy and dissipation over time for the same three tidal forcing frequencies as in Fig.~\ref{fig:rey_om}. Linear predictions computed with an independent spherical spectral code \citep[similar to e.g.][]{O2009}
are also displayed in both panels as the horizontal dashed lines for each frequency. When all nonlinear terms are included, 
we recover the kinetic energy and dissipation shown in Fig. 6 of \fb~ after appropriate re-scaling by a factor $A^2\times32\pi/15$, except that our model doesn't include the non-wavelike tide dissipation $\rho\nu\Delta\unw$ which is included in their model. However, in all three simulations, this term has been checked to be at least one order of magnitude smaller than the term $\rho\nu\Delta\uw$. Our quantitative agreement with \fb~ (for the times shown in their Fig. 6) verifies that our implementation of tidal forcing and nonlinear terms is correct.

The difference in kinetic energy and dissipation between linear and full nonlinear simulations keeps increasing with time, without a steady state being reached at the end of the simulation (here for $\Omega t=5000$) when all nonlinear terms are included, contrary to the case with only  wavelike/wavelike nonlinear terms in which $K$ and $D_\nu$ are closer to the corresponding linear predictions. This is likely to be due to the evolution of the total angular momentum $\bm L$ in the full nonlinear case (primarily due to unphysical non-wavelike fluxes through the outer boundary, as demonstrated in Eq. (\ref{eq:ang})). The vertical component of angular momentum is shown in the left panel of Fig. \ref{fig:ang}, which shows that the angular momentum is increasing with time, as would be expected due to tidal spin-synchronisation, while $\bm L$ is conserved in wavelike/wavelike nonlinear simulations (since we are not then modelling long-term tidal evolution, just ``instantaneous transfers" in a snapshot of the system). 
\begin{figure*}
    \centering
    \includegraphics[width=\columnwidth]{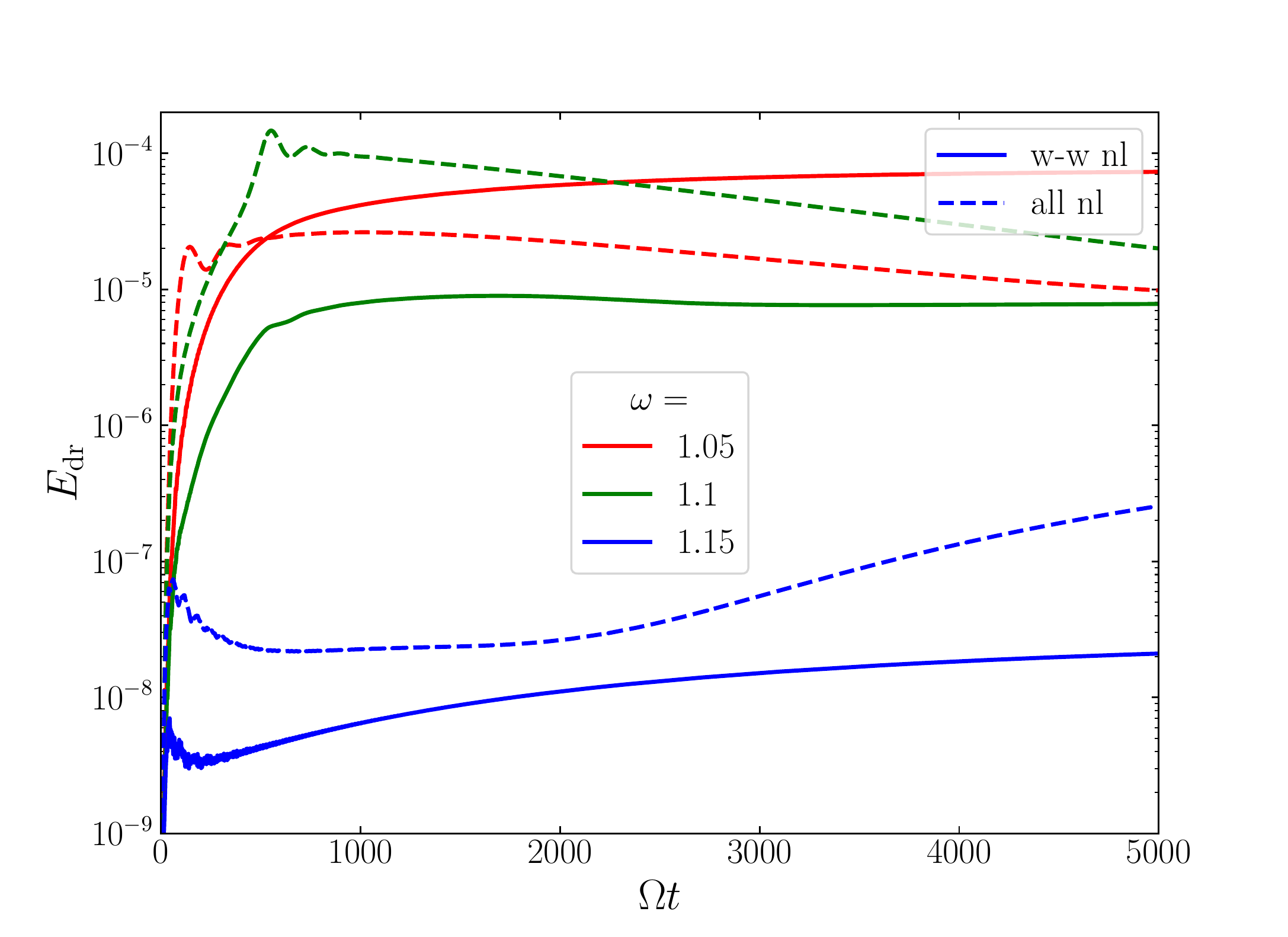}
    \includegraphics[width=\columnwidth]{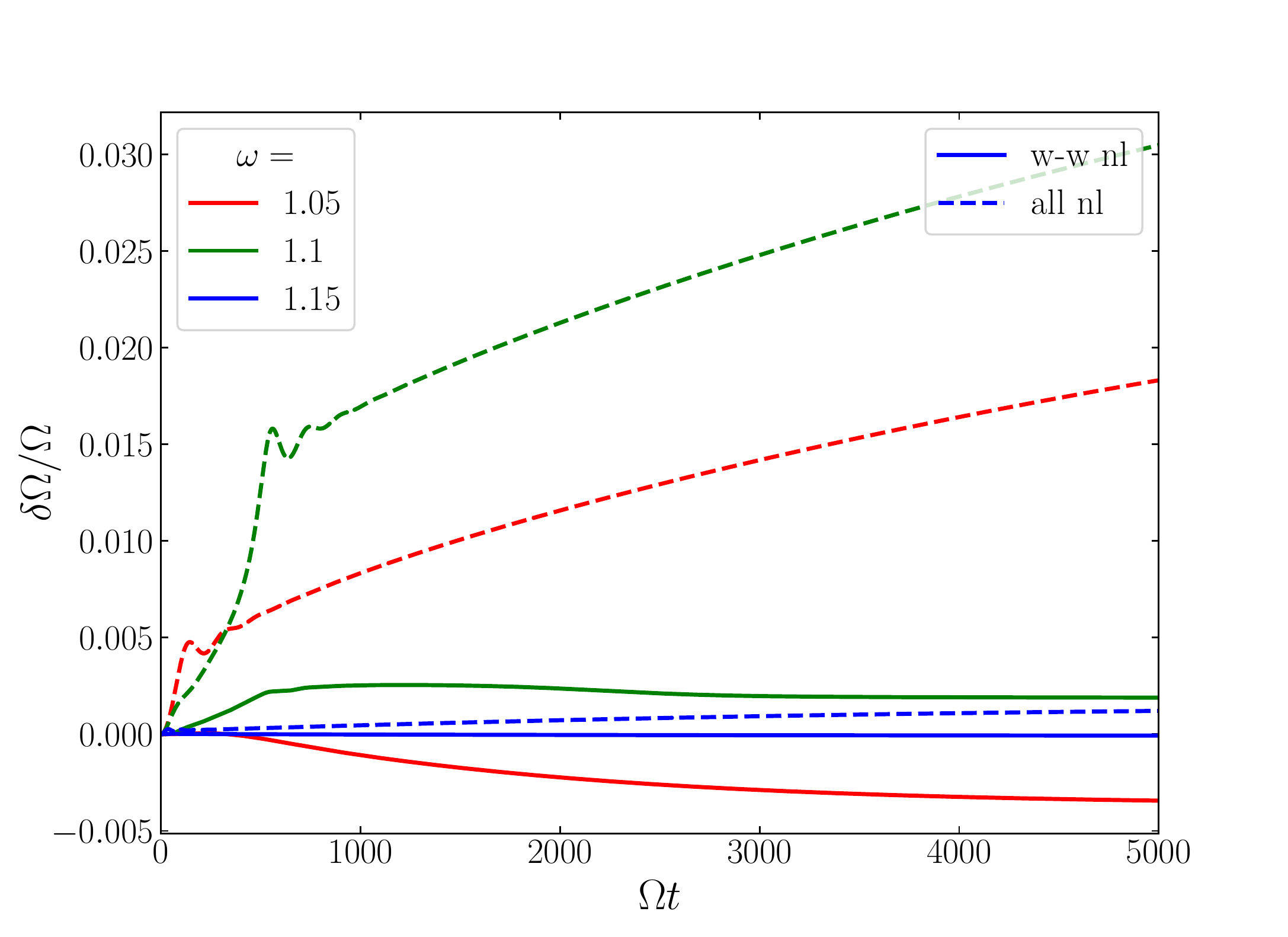}
    \caption{\textit{Left:} Energy in the differential rotation $E_\mathrm{dr}$  versus time for three initial forcing frequencies and for all (dotted lines) or only wavelike/wavelike (w-w) nonlinearities (solid lines). 
    \textit{Right:} Evolution of the mean rotation rate of the fluid normalised by the initial uniform rotation $\delta\Omega/\Omega$ versus time for the same nonlinear simulations.
    }
    \label{fig:edr_dom}
\end{figure*}

Furthermore, the fact that the difference between linear and nonlinear $K$ and $D_\nu$ is more pronounced for $\omega=1.05$ and $1.1$ compared to the $\omega=1.15$ case 
\cor{seems to be linked to the strong focusing of waves into localised shear layers} for the two lowest frequencies, as described in \citet[as opposed to the more diffuse ``space-filling" distribution of inertial wave beams for $\omega=1.15$]{O2009}. 
\cor{While for the linear $\omega=1.1$ case, the excitation of a periodic wave attractor has been identified, the linear $\omega=1.05$ case rather features large-scale structures hidden beneath shear layers \citep[see][and possibly visible in the top left panel of Fig. \ref{fig:vr}]{LO2021}. These have been shown to be potentially related to global modes in full sphere geometry, explaining the resonant peak in linear dissipation rates\footnote{\cor{It is however not clear to what extent this relation holds in our nonlinear simulation for the lowest frequency, where the zonal flow leads to both a detuning effect by shifting the frequency of the wave away from the resonant linear mode, but also presumably changing the location of the resonant peak itself \citep[e.g.][]{GM2016}.}}}.

The nonlinear self-interactions of inertial waves in such localised shear layers favour the emergence of strong zonal flows through the non-uniform deposition of angular momentum in the shell \citep[\fb]{T2007}. These zonal flows can be identified by calculating the energy of the differential rotation
\begin{equation}
    E_\mathrm{dr}=\frac{\rho}{2}\int \left[\langle u_\varphi\rangle_\varphi-\delta\Omega r\sin\theta\right]^2\,\mathrm{d}V,
\end{equation}
where $\langle u_\varphi\rangle_\varphi$ is the azimuthally-averaged azimuthal component of the flow, and 
\begin{equation}
    \delta\Omega=\frac{1}{V}\int\frac{u_\varphi}{r\sin\theta}\,\mathrm{d}V,
\end{equation}
is the newly created volume-averaged rotation rate of the fluid in the rotating frame. The latter is superimposed upon the mean rotation rate $\Omega$ such that the volume-averaged rotation rate in the inertial frame is $\Omega^*=\Omega+\delta\Omega$.
In this new differentially-rotating environment, the propagation and dissipation of inertial waves are significantly modified, as was also shown in the linear studies of \citet[][]{BR2013} and \citet{GB2016,GM2016}, which could be primarily responsible for explaining the discrepancies we find in tidal dissipation rates between linear and nonlinear regimes (this will be further discussed in the next section). 

\begin{figure*}
    \centering
    \includegraphics[trim=0.5cm 0cm 1cm 0.5cm, clip,width=0.33\textwidth]{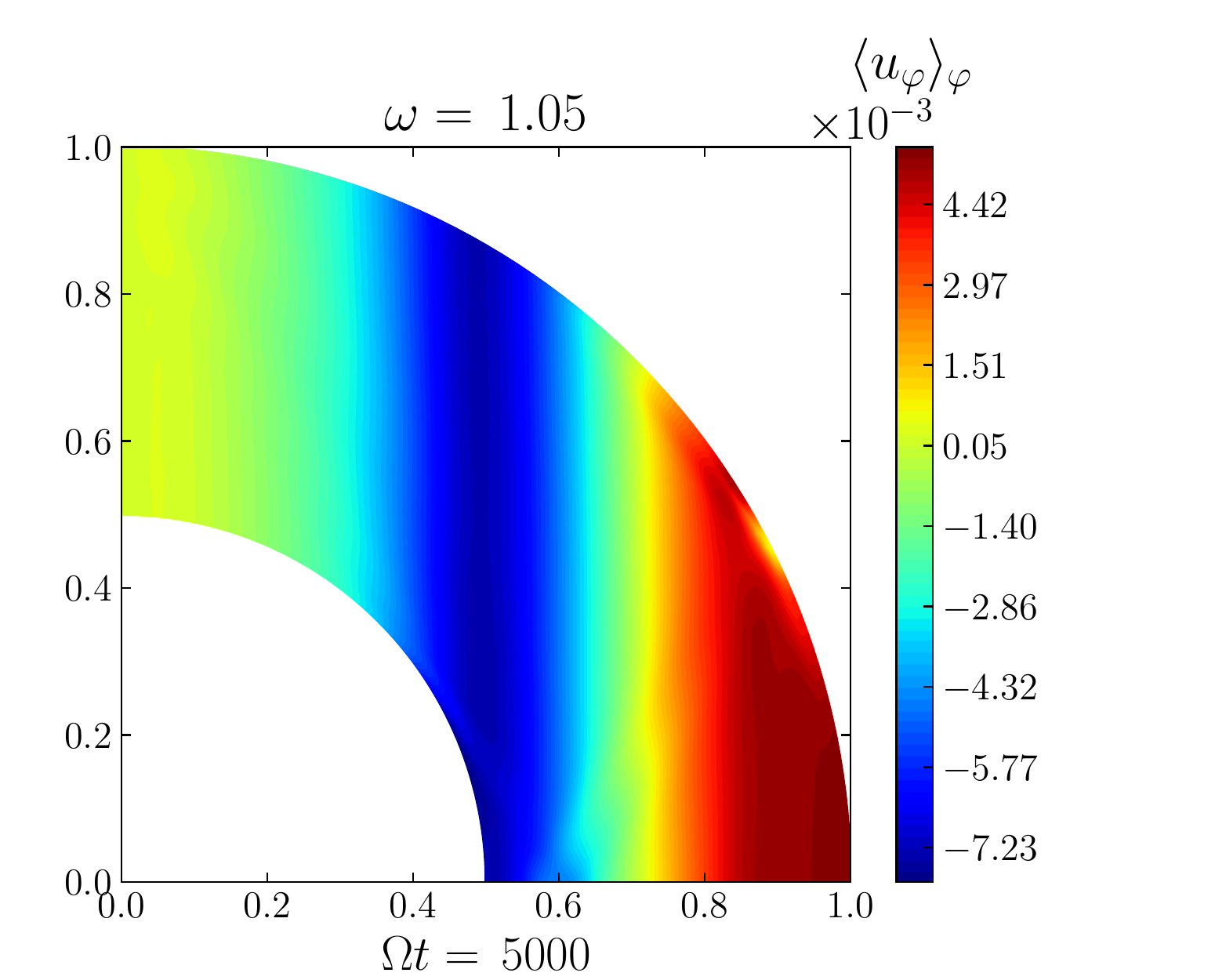}
    \includegraphics[trim=0.5cm 0cm 1cm 0.5cm, clip,width=0.33\textwidth]{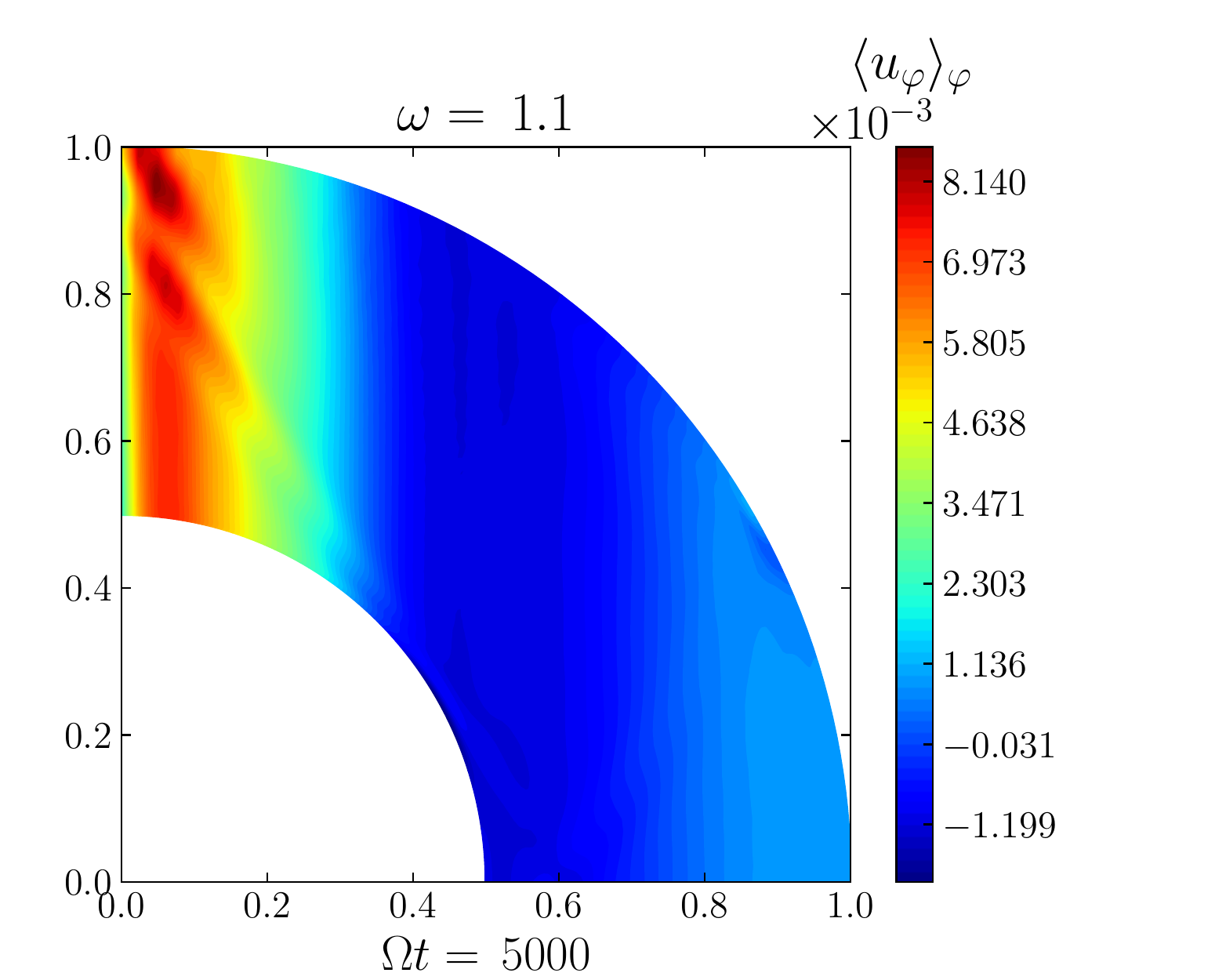}
    \includegraphics[trim=0.5cm 0cm 1cm 0.5cm, clip,width=0.33\textwidth]{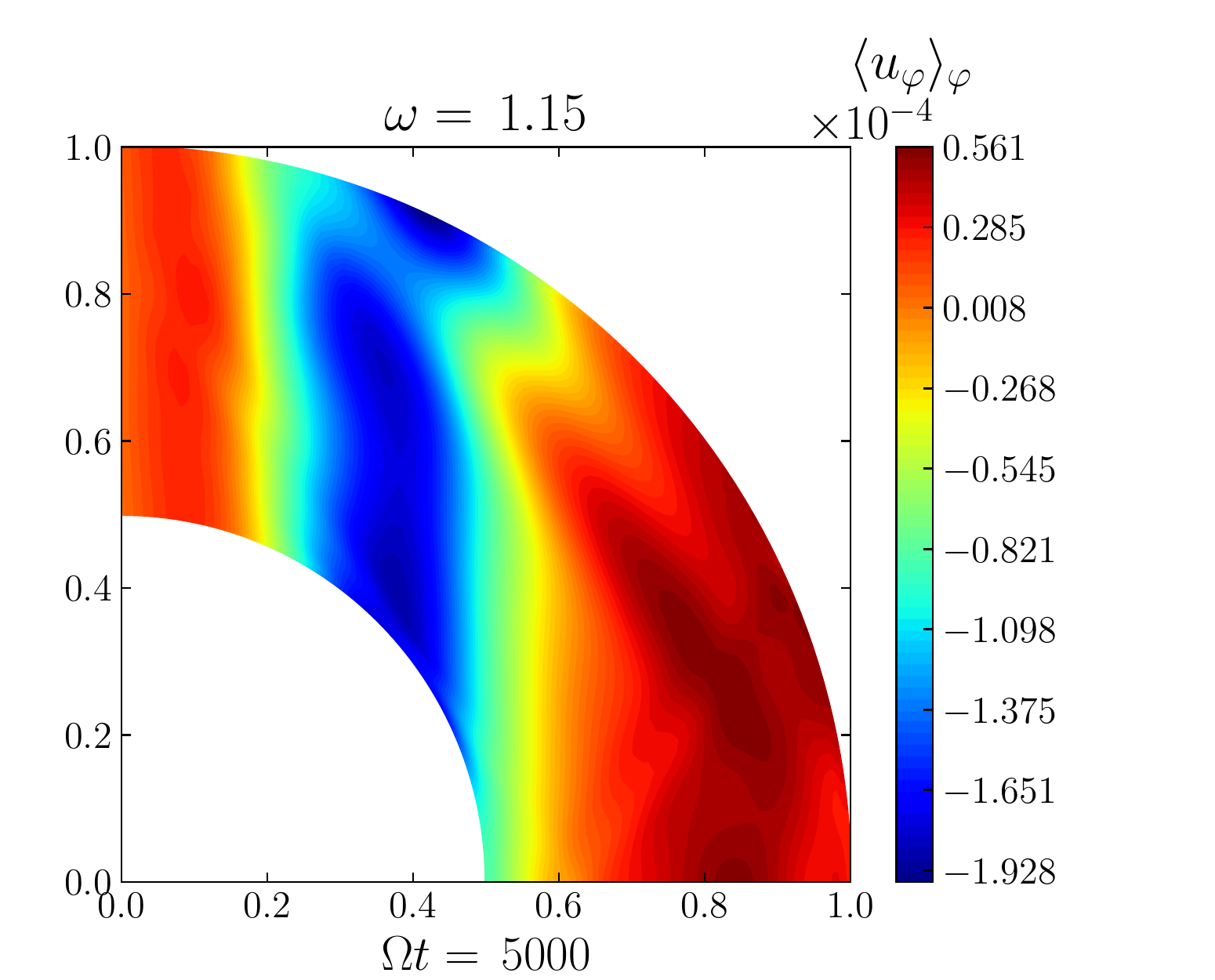}
    \caption{Azimuthally averaged zonal flows $\langle u_\varphi\rangle_\varphi$ for three different frequencies when a steady state is reached in wavelike/wavelike nonlinear simulations. The colormaps are invariant with respect to the equator and the rotation axis in each of these simulations.}
    \label{fig:vp_nonlinear}
\end{figure*}

The right panel of Fig. \ref{fig:ang} shows the tidal dissipation rate as a function of normalised frequency $\omega/\Omega^*$ from nonlinear simulations that include all nonlinear terms (bottom curves, shifted downwards by a factor $10^{-2}$) compared with simulations including only the wavelike/wavelike nonlinearity (upper curves). Since $\delta\Omega$ evolves with time, particularly in the full nonlinear simulations, it means that the simulations also explore a range of tidal frequencies as they progress. We also identify $E_\mathrm{dr}$ by the colour of each line, which shows that differential rotation develops in each of these simulations. Especially, the removal of mixed nonlinear terms does not prevent the development of zonal flows which are only driven by the wavelike/wavelike nonlinearity in that case, as is also shown in Fig. \ref{fig:vp_nonlinear}. In that figure, we display snapshots in the meridional $(r,\theta)$ plane of azimuthally-averaged zonal flows for the three frequencies.
The energy in the differential rotation is similar or even greater on long time scales for $\omega=1.05$ and $1.1$, compared with the cases where all nonlinearities are included, though the change in the rotation rate $\delta\Omega$ is much less important, as shown in both panels of Fig. \ref{fig:edr_dom}. Note that nonzero values of $\delta\Omega$ can still occur in simulations with only wavelike/wavelike nonlinearities despite the conservation of $L_z$ because the flow is differentially rotating. We have checked that simulations including only wavelike/wavelike nonlinearities conserve $L_z$ to an accuracy better than $10^{-10}$.

In the right panel of Fig. \ref{fig:ang}, we have also added the normalised constant $\hat{E}/\Ct^2$ (horizontal dotted lines), which is 
the energy transfer in an impulsive interaction \citep[Eq. (112)]{O2013}
\begin{equation}
    \frac{\hat{E}}{\Ct^2}=\frac{1}{2\pi}\int_{-\infty}^{+\infty}P_\mathrm{t}(\omega)\,\frac{\mathrm{d}\omega}{\omega^2}=\frac{20}{189}\frac{\alpha^5}{1-\alpha^5}\rho R^5\Omega^2,
    \label{eq:hatE}
\end{equation}
with $\rho$, $R$, and $\Omega$ fixed to unity in our simulations. This quantity gives a measure of a typical dissipation rate over one time unit (in units of $\Omega^{-1}$) and is close to a (unweighted) mean value of the linear frequency-dependent tidal dissipation across most of spectrum.
It is related to the frequency-averaged tidal dissipation of inertial waves, which is commonly applied to study dynamical evolution of exoplanetary systems \citep[e.g.][]{M2015,BM2016}, given by\footnote{For a fully homogeneous body, $k_2=3/2$, and one recovers Eq. (113) of \citet{O2013}.}:
\begin{equation}
    \Lambda=\int_{-\infty}^{+\infty}\mathrm{Im}\left[K^2_2(\omega)\right]\,\frac{\mathrm{d}\omega}{\omega}=\frac{16\pi}{63}\epsilon_\Omega^2(1+k_2)^2\frac{\alpha^5}{1-\alpha^5},
    \label{eq:Lambda1}
\end{equation}
where $\epsilon_\Omega^2=\Omega^2/\omega_d^2$. The integrand is the imaginary part of the quadrupolar potential Love number, related to the impulsive energy transfer as $\Lambda=(12\pi\epsilon^2_\Omega/(5\epsilon^2))\,\hat{E}$. 

Final dissipation rates for wavelike/wavelike nonlinear simulations shown in the right panel of Fig. \ref{fig:ang} (bullets) tend to be closer to the mean tidal dissipation proxy given by Eq. (\ref{eq:hatE}), compared to the highly resonant frequency-dependent linear dissipation (dashed lines), which can be one order of magnitude (or more) lower or higher than the mean value.
At the beginning of all nonlinear simulations, the energy injected into the differential rotation increases as the dissipation approaches the value predicted by linear theory. Then, the establishment of zonal flows (in particular) causes the dissipation to deviate from its linear prediction, even more so in the simulations where all nonlinearities are included since $\bm L$ is not conserved (see the left panel of Fig. \ref{fig:ang}). As expected, for $\omega=1.15$ where strong shear layers are not activated and which corresponds to a lower level of dissipation in the linear regime, zonal flows are hardly generated (note the order of magnitude of difference for $\langle u_\varphi\rangle_\varphi$ in Fig. \ref{fig:vp_nonlinear}, right panel, compared to the other frequencies shown in the left and middle panels). There are no appreciable departures in the tidal dissipation rates from the linear regime in this case. From these three cases, it would appear that the closer we are to a ``resonant peak" in which the dissipation is enhanced, the differential rotation excited is stronger and the gap between the linear and wavelike/wavelike nonlinear regimes (upper curves in the right panel of Fig. \ref{fig:ang}) is larger.

\begin{figure*}
    \centering
    \includegraphics[width=0.32\textwidth]{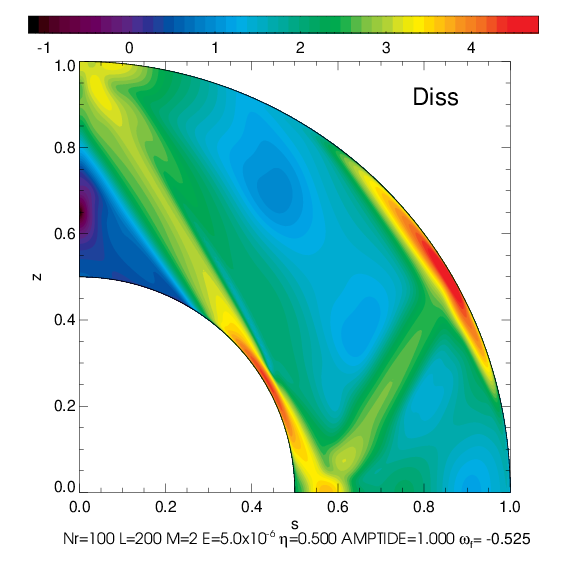}
    \includegraphics[width=0.32\textwidth]{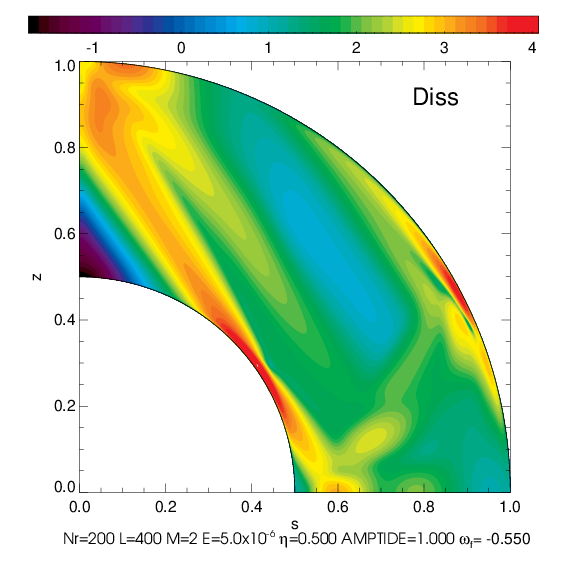}
    \includegraphics[width=0.32\textwidth]{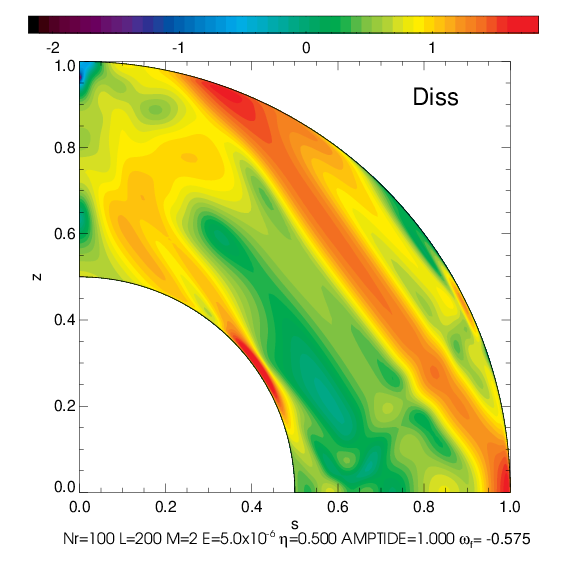}
    \caption{Viscous dissipation in the meridional plane in linear calculations computed with LSB for three different tidal forcing frequencies $\omega=2\omega_\mathrm{f}$. 
    Note the amplitude is arbitrary as this is from a linear calculation.
    }
    \label{fig:linear_diss}
\end{figure*}
As shown in Fig. \ref{fig:vp_nonlinear}, but also in \fb, \citet{T2007}, \citet{ML2010}, or \citet{SS2021}, zonal flows produced by nonlinear self-interaction of non-axisymmetric inertial waves always take the form of nested and axisymmetric cylinders with different angular velocity profiles depending upon cylindrical radius inside the shell. The radial distribution, magnitude and direction of the flow at a given location (prograde or retrograde, i.e.~red or blue in meridional snapshots of azimuthal velocities, respectively) can be very different depending on the tidal frequency. For instance, high-amplitude prograde flows can develop near the equator or near the rotation axis, as we show in the left and middle panels of Fig. \ref{fig:vp_nonlinear}, respectively, even though the tidal forcing frequencies are quite similar. While it is difficult to straightforwardly predict the final magnitude and structure of these columnar flows, it is clear that the strongest zonal flows preferentially develop at the points of reflection between the inertial wave beams and the outer boundary or rotation axis. These are locations characterised by large amplitudes in the velocity components or dissipation, as we can see from the  meridional cuts of the radial component of the linear velocity in Fig. \ref{fig:vr} (top panels) and the dissipation in Fig. \ref{fig:linear_diss} based on complementary linear calculations \citep[computed using the spectral code LSB,][]{VR2007}. Powerful shear layers emerge from the critical latitude at $\theta=\arcsin(\omega/(2\Omega))$ tangent to the inner sphere, approximately at latitudes of $32\degree$, $33\degree$, and $34\degree$, respectively, from the left to the right panels of Figs. \ref{fig:vr} and \ref{fig:linear_diss}. These are recurring features of forced inertial waves in 3D rotating spherical shells \citep[e.g.][]{RV2010,RV2018}. In meridional cuts displaying the nonlinear radial velocity for $\omega=1.05$ and $1.1$ (left and middle bottom panels of Fig.~\ref{fig:vr}), we observe that shear layers are attenuated compared to their linear counterparts (upper panels), presumably due to the development of zonal flows.
Though cylindrical zonal flows are the dominant patterns when looking at the $\varphi-$average of the azimuthal velocity in nonlinear simulations,  the imprints of shear layers are still visible in snapshots 
in the three panels of Fig. \ref{fig:vp_nonlinear} when comparing with the same panels showing linear dissipation in Fig. \ref{fig:linear_diss}. The key role of shear layers and critical latitudes in the establishment of these geostrophic flows have already been pointed out for example in the experimental study of \cite{SS2021}.

From now on and throughout the rest of this paper, we analyse simulations with wavelike/wavelike nonlinearities only.
\begin{figure*}
    \centering
    \includegraphics[trim=1.8cm 0cm 0cm 0cm, clip,width=0.33\textwidth]{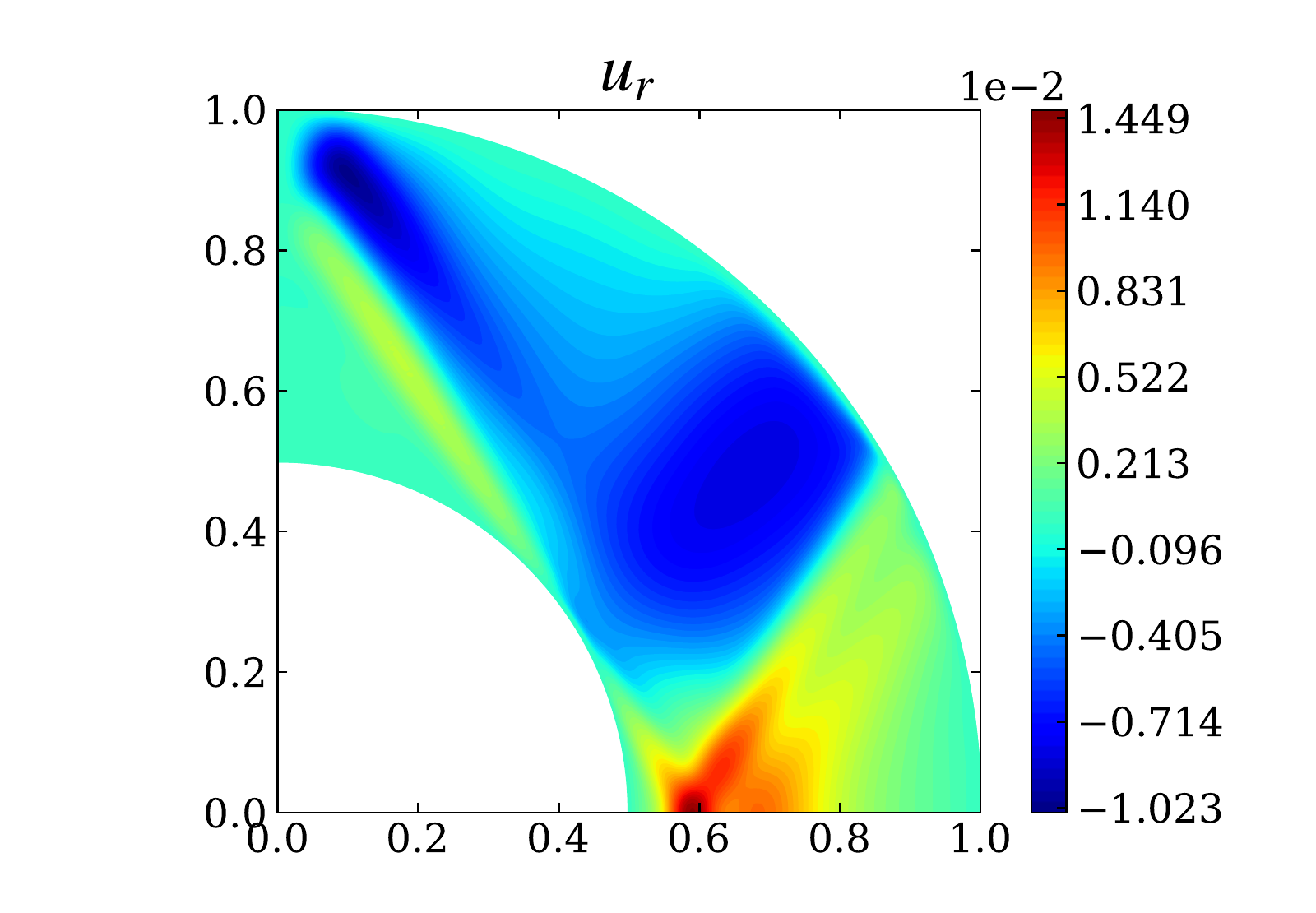}
    \includegraphics[trim=1.8cm 0cm 0cm 0cm, clip,width=0.33\textwidth]{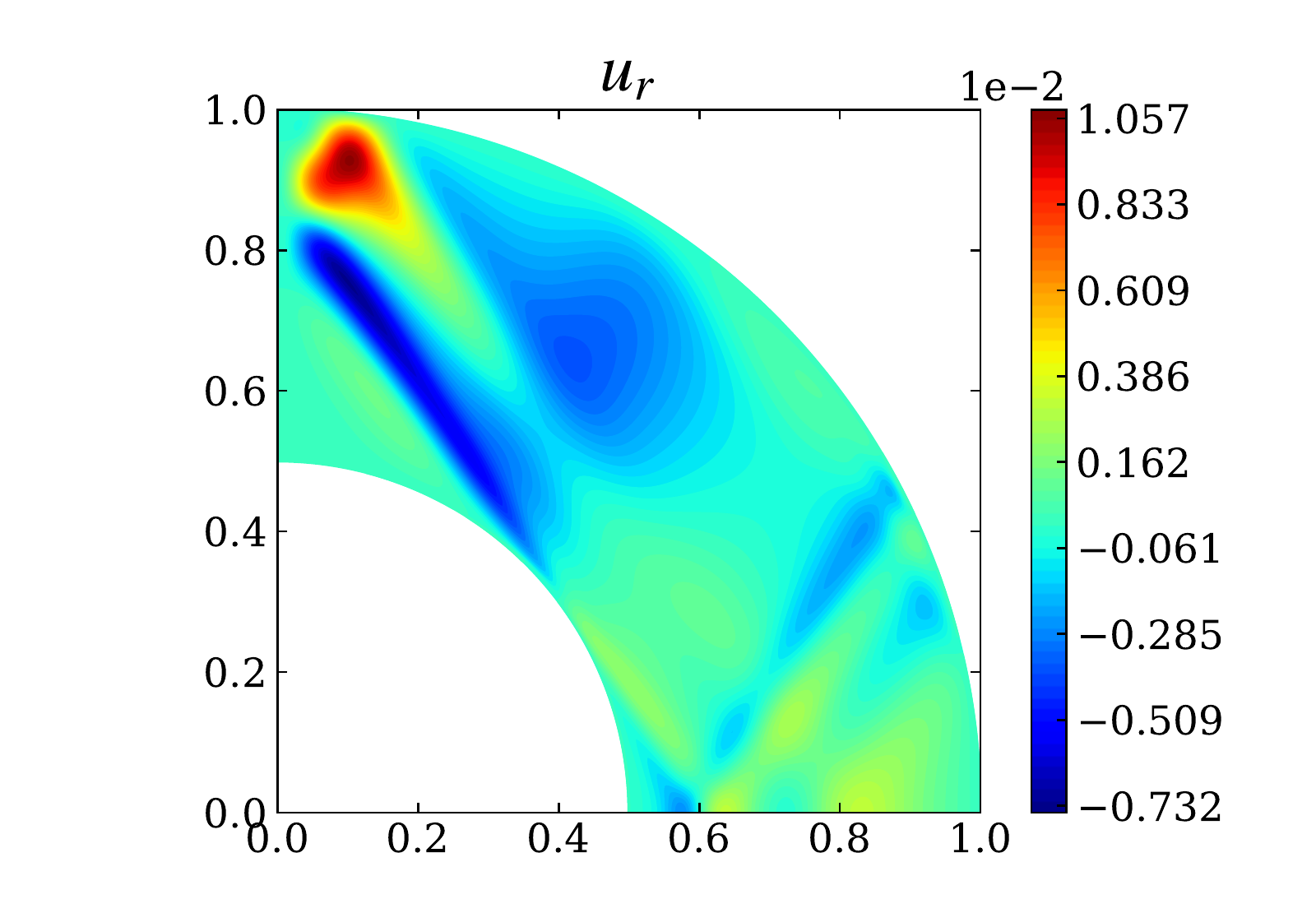}
    \includegraphics[trim=1.8cm 0cm 0cm 0cm, clip,width=0.33\textwidth]{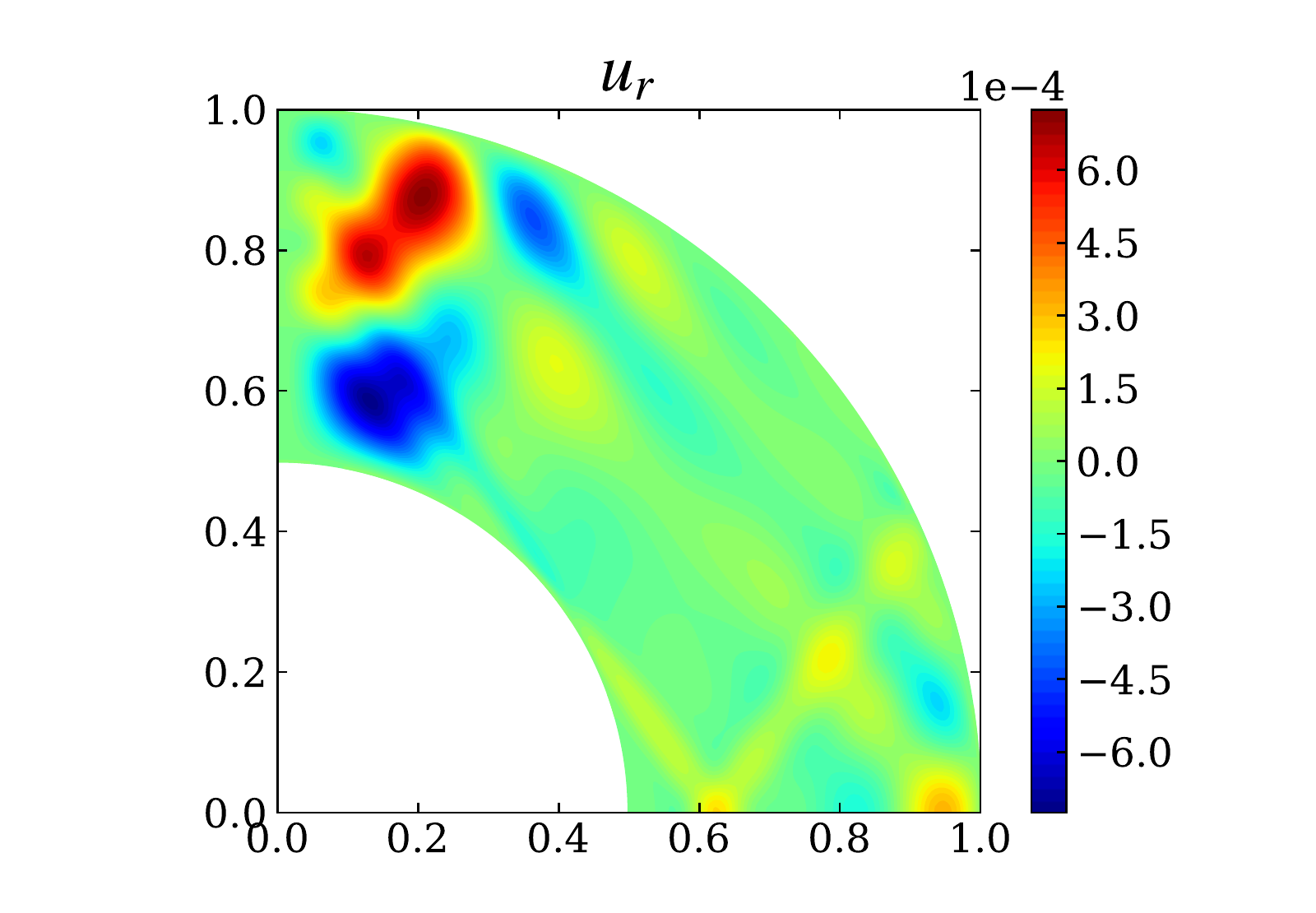}
    \includegraphics[width=0.33\textwidth]{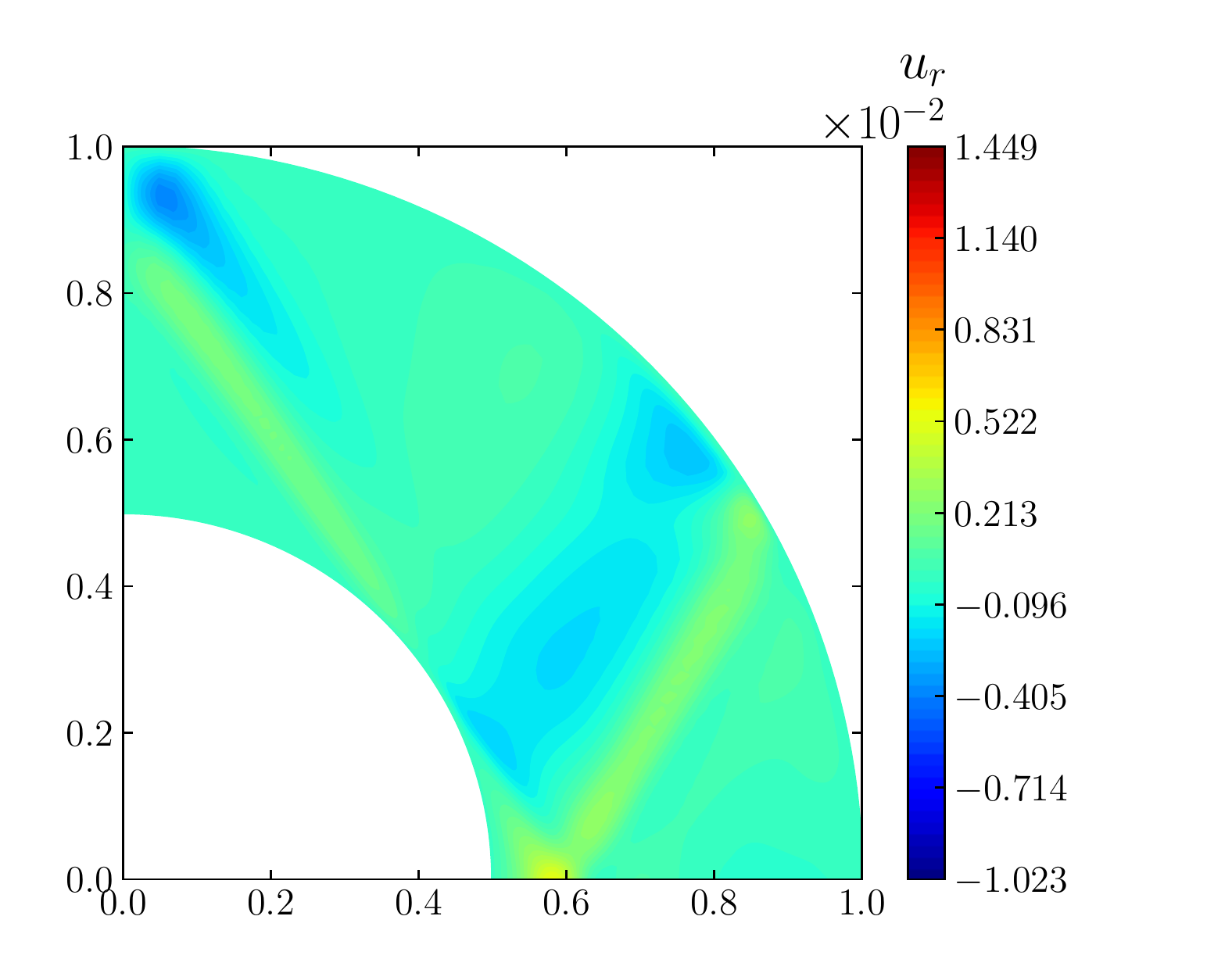}
    \includegraphics[width=0.33\textwidth]{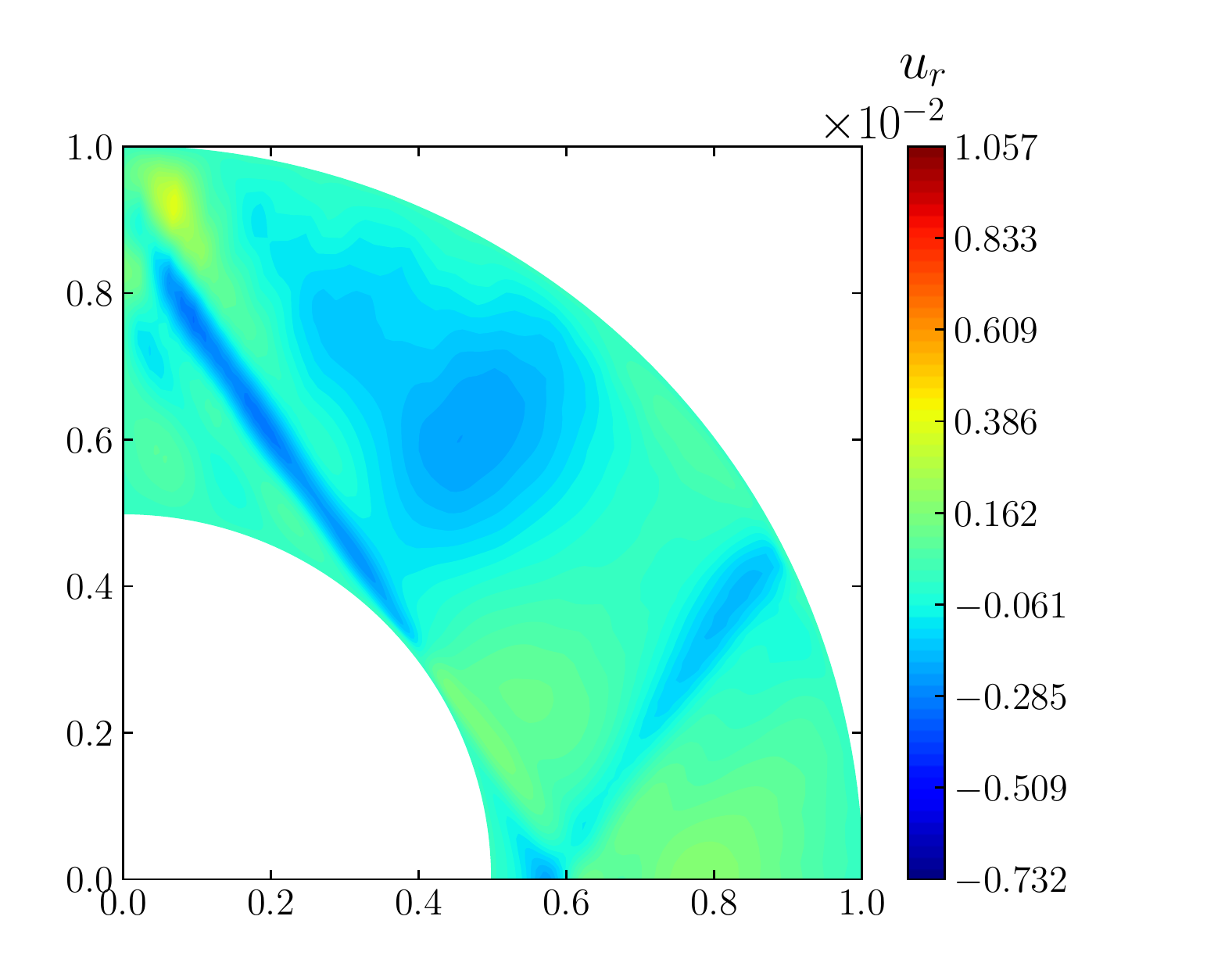}
    \includegraphics[width=0.33\textwidth]{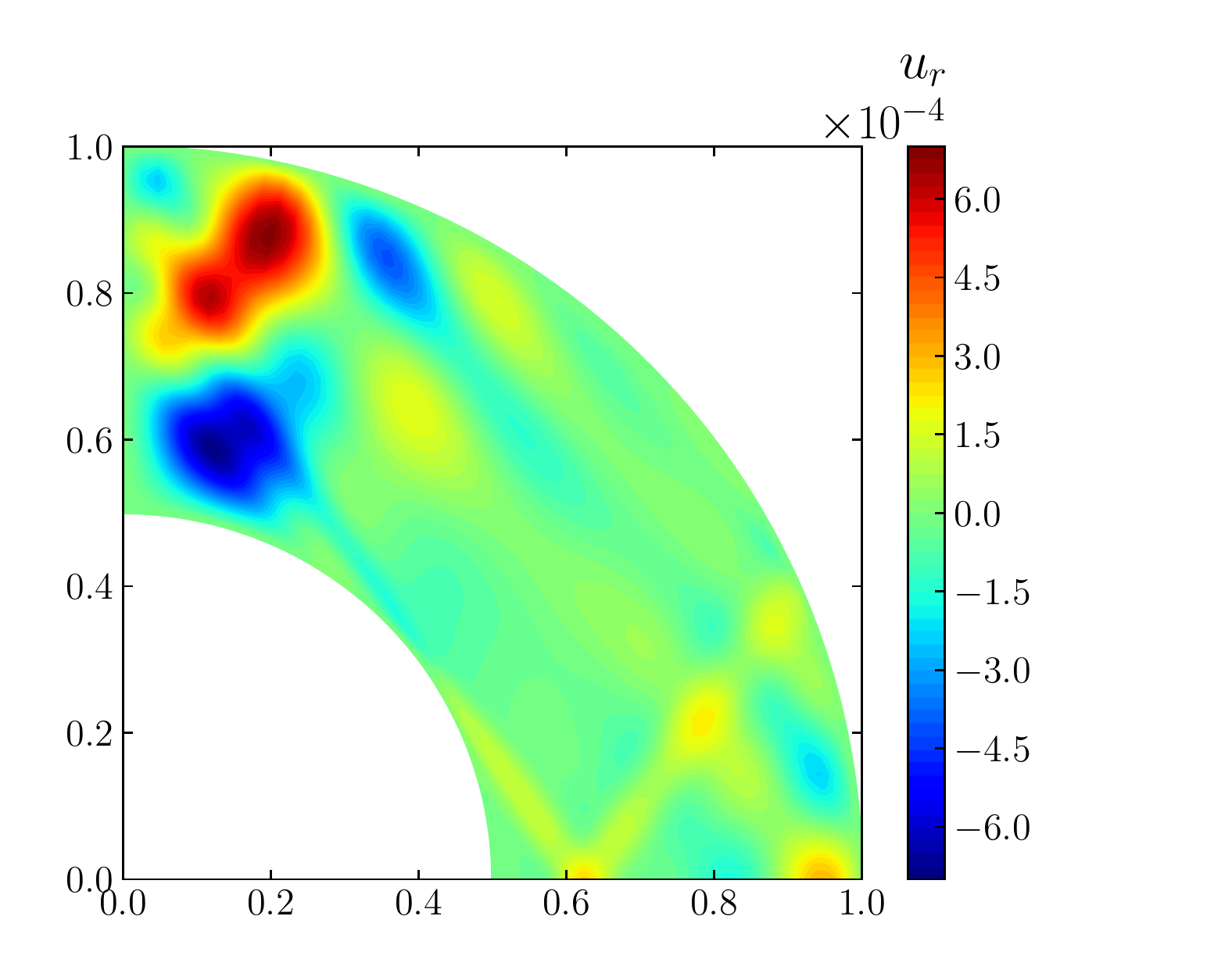}
    \caption{Radial component of the \cor{wavelike} velocity in the meridional plane for the same three frequencies as in Fig. \ref{fig:vp_nonlinear}. \textit{Top:} in linear simulations computed with LSB.
    \textit{Bottom:} in nonlinear simulations computed with MagIC where $\varphi\equiv\omega t/m\,[2\pi]-\pi/2$ (the $\pi/2$ factor is necessary to find the correct phase for comparison). The extrema of the colorbars are chosen to be the same values as those in the above panels. 
       }
    \label{fig:vr}
\end{figure*}
\begin{figure}
    \centering
    \includegraphics[width=\columnwidth]{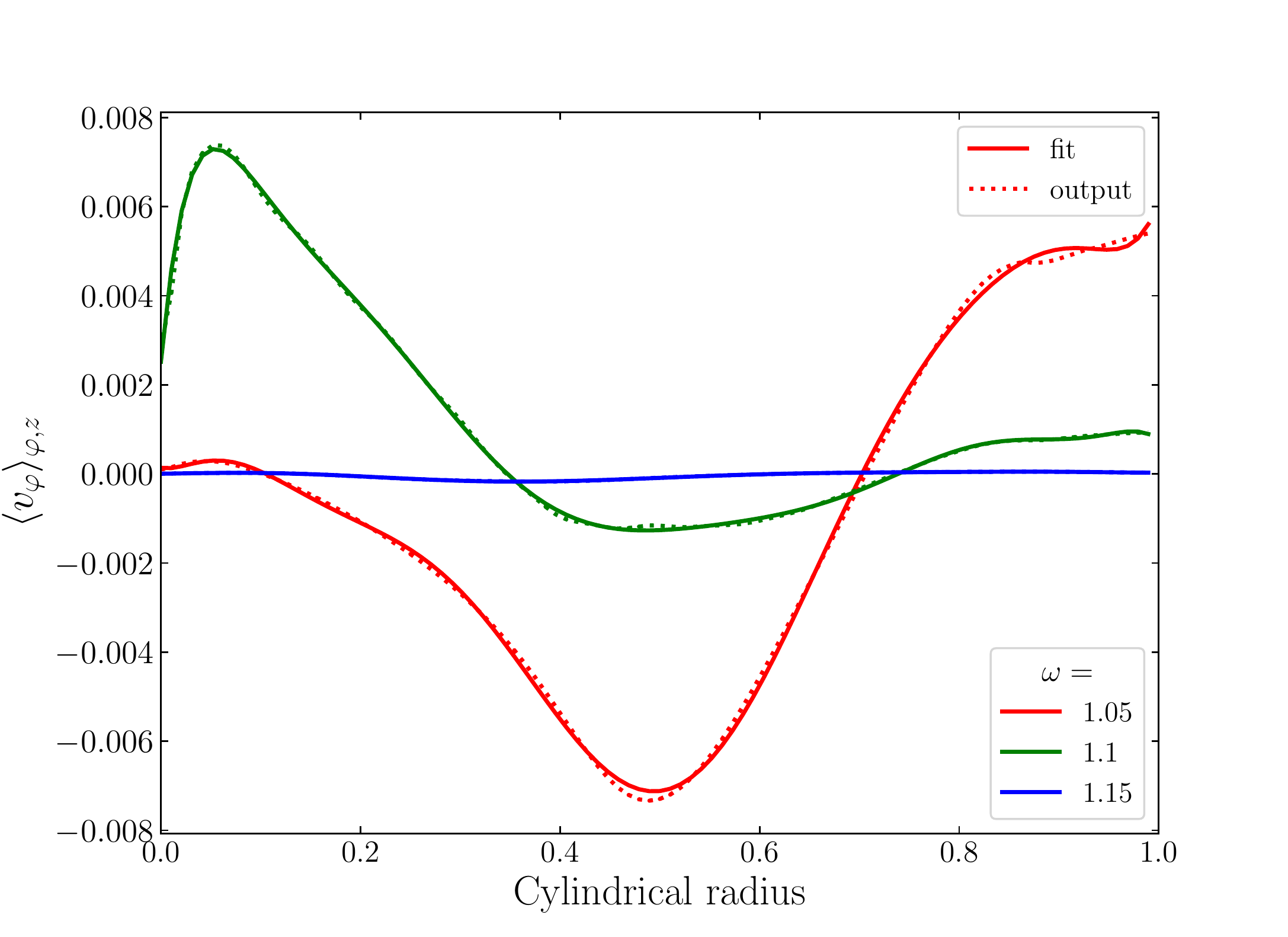}
    \caption{Vertically and azimuthally averaged \cor{wavelike azimuthal velocity}  $\langle u_\varphi\rangle_{\varphi,z}$ ("output") at the end of the non-linear simulations ($\Omega t=5000$) against cylindrical radius for three different forcing frequencies. The fitting curves ("fit") are degree 10 polynomials computed using a least squares \cor{method}.}
    \label{fig:zonflo}
\end{figure}
\subsection{Effects of the zonal flows on tidal dissipation}
In this section, we demonstrate that the development of differential rotation can explain the discrepancies between linear and wavelike/wavelike nonlinear tidal dissipation rates in many of these simulations, as we observe in the rights panels of Figs \ref{fig:kD} and \ref{fig:ang}. At late times in our nonlinear simulations (several thousand rotation times), these simulations reach an approximate steady state for $K$, $D_\nu$, $E_\mathrm{dr}$ and $\delta\Omega/\Omega$, in which zonal flows are fully developed and hardly evolve any more. To further analyse these simulations we extract a $\varphi-$ and $z-$average of the azimuthal component of the velocity $\langle u_\varphi\rangle_{\varphi,z}$ when the simulation reaches such a steady state (as done e.g. in Fig. \ref{fig:zonflo}, showing this quantity for the three frequencies studied earlier). We then investigate the energy exchanges in new linear simulations of tidally-forced inertial waves, where we apply the final zonal flow from nonlinear simulations as a ``background flow". This allows us to determine whether the most important nonlinear effect in our simulations is the generation of differential rotation and the back-reaction of this on the tidally forced waves, as opposed to triadic wave interactions (or parametric instabilities) involving inertial waves, for example. We adopt a similar approach to the linear calculations of \cite{BR2013} with initial cylindrical differential rotation, except that inertial waves are tidally forced in our model and we perform these as an initial value problem using simulations (while they studied free inertial modes as an eigenvalue problem).

In our model, the azimuthal background flow is $\bm U=\bm r\wedge\bm\Omega=s\Omega_s(s)\,\bm e_\varphi$, where the rotation $\Omega_s(s)$ is a function of  the cylindrical radius\footnote{Any meridional circulation is neglected since $\langle u_r\rangle_\varphi$ and $\langle u_\theta\rangle_\varphi$ have been found to be negligible in most of the nonlinear simulations, or at least they are weaker (with such cases to be emphasised in the following), compared to $\langle u_\varphi\rangle_\varphi$.} $s=r\sin\theta$. 
The rotation can be decomposed into a constant mean rotation $\Omega$ plus an $s$-dependent small departure $\delta\Omega_s$, such that $\Omega_s(s)=\Omega+\delta\Omega_s(s)$. Since the numerical simulations are in the co-rotating frame, the correct initial zonal flow to choose is given by $\delta\Omega_s=\langle u_\varphi\rangle_{\varphi,z}/s$. Unlike in \cite{BR2013} where a quadratic function of $s$ is used to describe the cylindrical rotation \cor{(chosen to maintain matrix sparsity as much as possible)}, 
the rotation profile obtained is not a simple function of $s$ in our models (see Fig. \ref{fig:zonflo}). We have made the choice to reconstruct the zonal flow $\langle u_\varphi\rangle_{\varphi,z}$ as a degree $n$ polynomial such that $\langle u_\varphi\rangle_{\varphi,z}=\sum_{i=0}^{n}a_is^i$ (with $n=10-20$) and where $a_i$ are fitted constants. 

\begin{figure}
    \centering
    \includegraphics[width=\columnwidth]{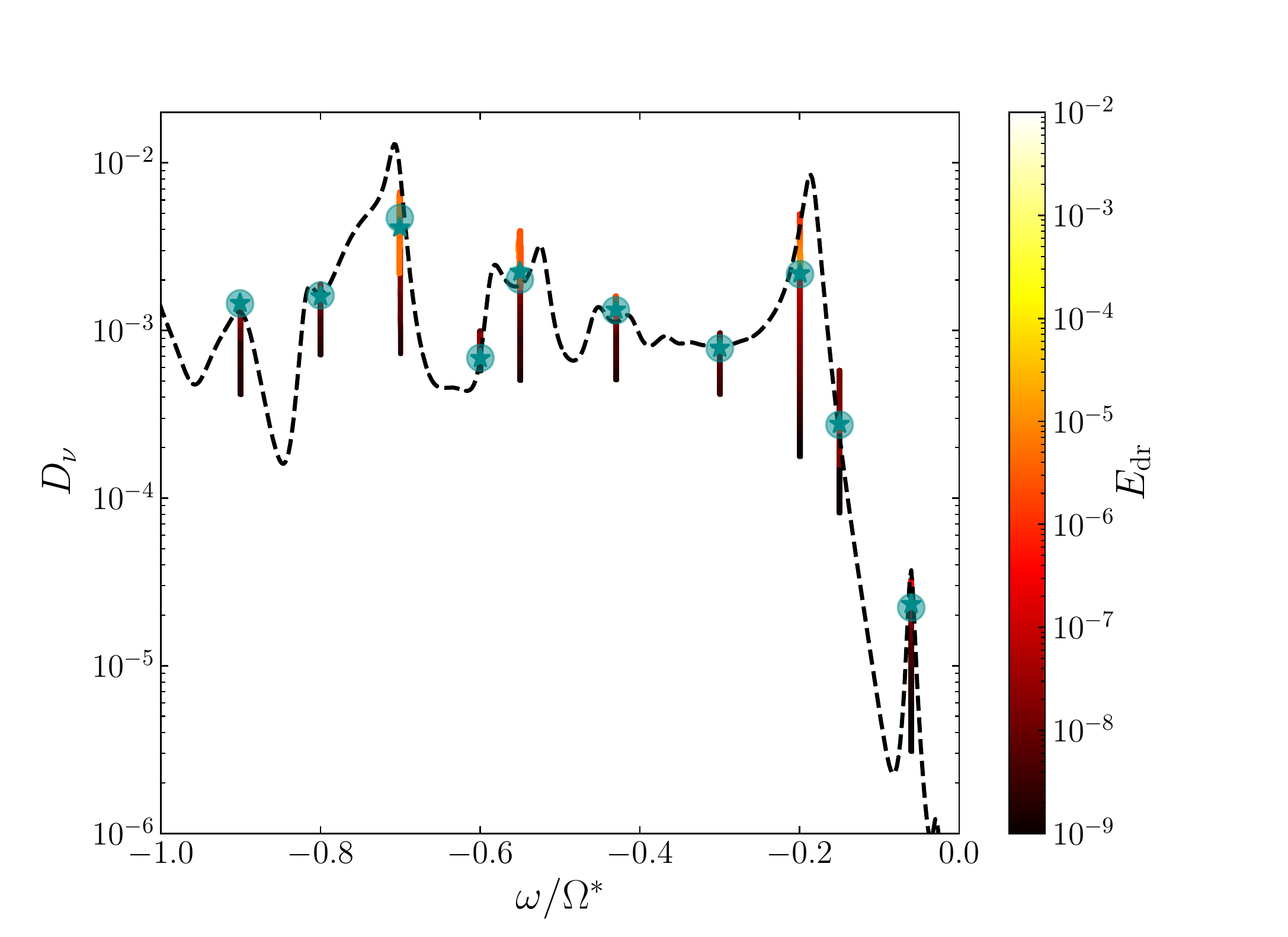}
    \caption{Dissipation $D_\nu$ as a function of the normalised frequency $\omega/\Omega^*$. Nonlinear dissipation with wavelike/wavelike nonlinearities (reddish curves and pale blue bullets indicating the final dissipation) is computed from $\Omega t=20$ to more than 8000 (when an average steady state is reached) for a tidal amplitude $\Ct=10^{-2}$ and initial tidal forcing frequency $\omega$ in the range $[-1,0]$. The nonlinear dissipation is rescaled by a factor $\Ct^2$ to compare with the linear predictions for a uniformly rotating body of frequency-dependent (dashed lines). The colorbar indicates the kinetic energy in the differential rotation $E_\mathrm{dr}$. The tidal power rates $P_\mathrm{t,dr}$ computed from the final steady-state reached in linear simulations with an initial cylindrical zonal flow (corresponding to the one attained in nonlinear simulations) are also added as blue stars.
}
    \label{fig:span_negom}
\end{figure}
\begin{figure*}
\includegraphics[width=0.49\textwidth]{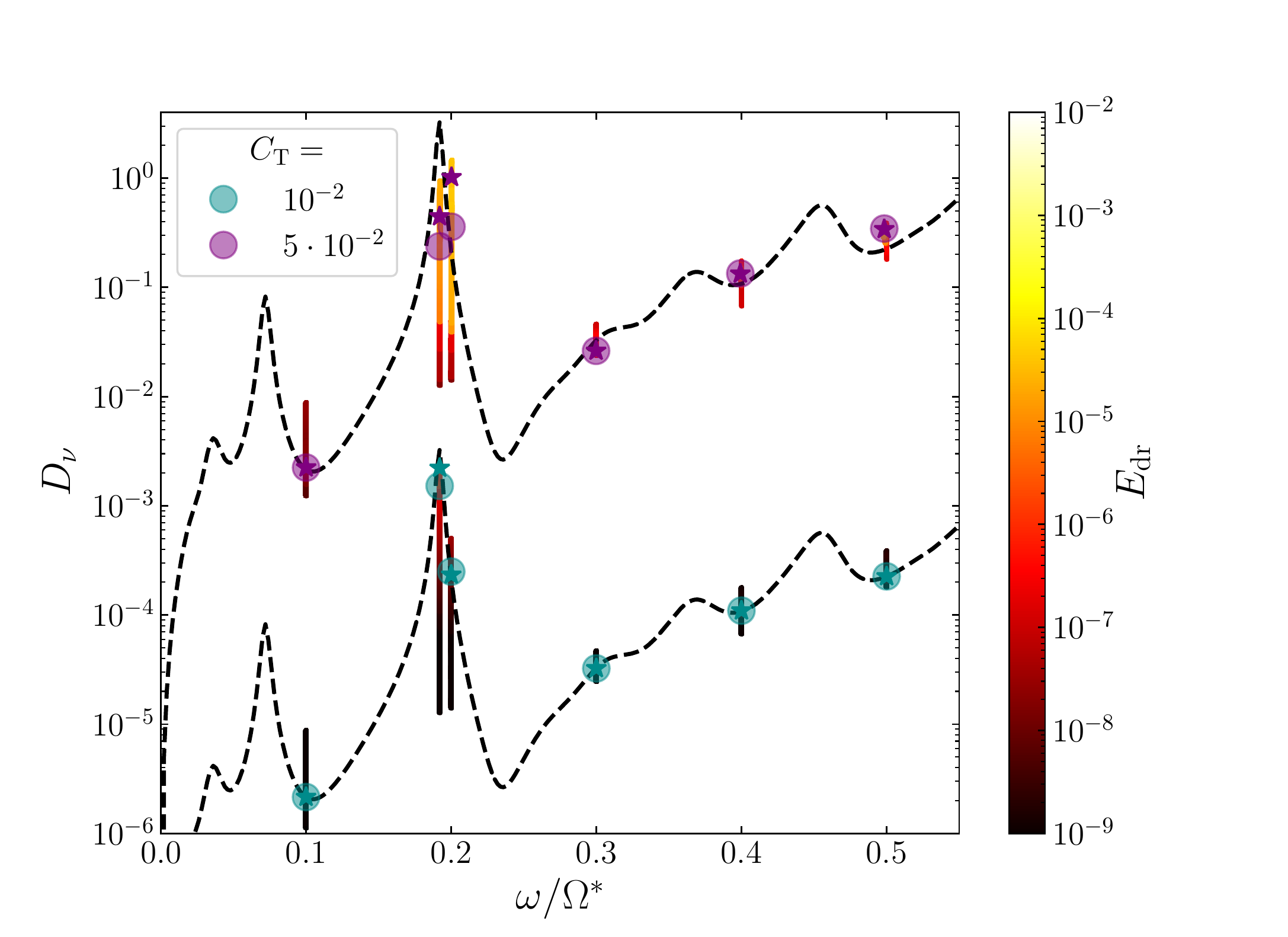}
\includegraphics[width=0.49\textwidth]{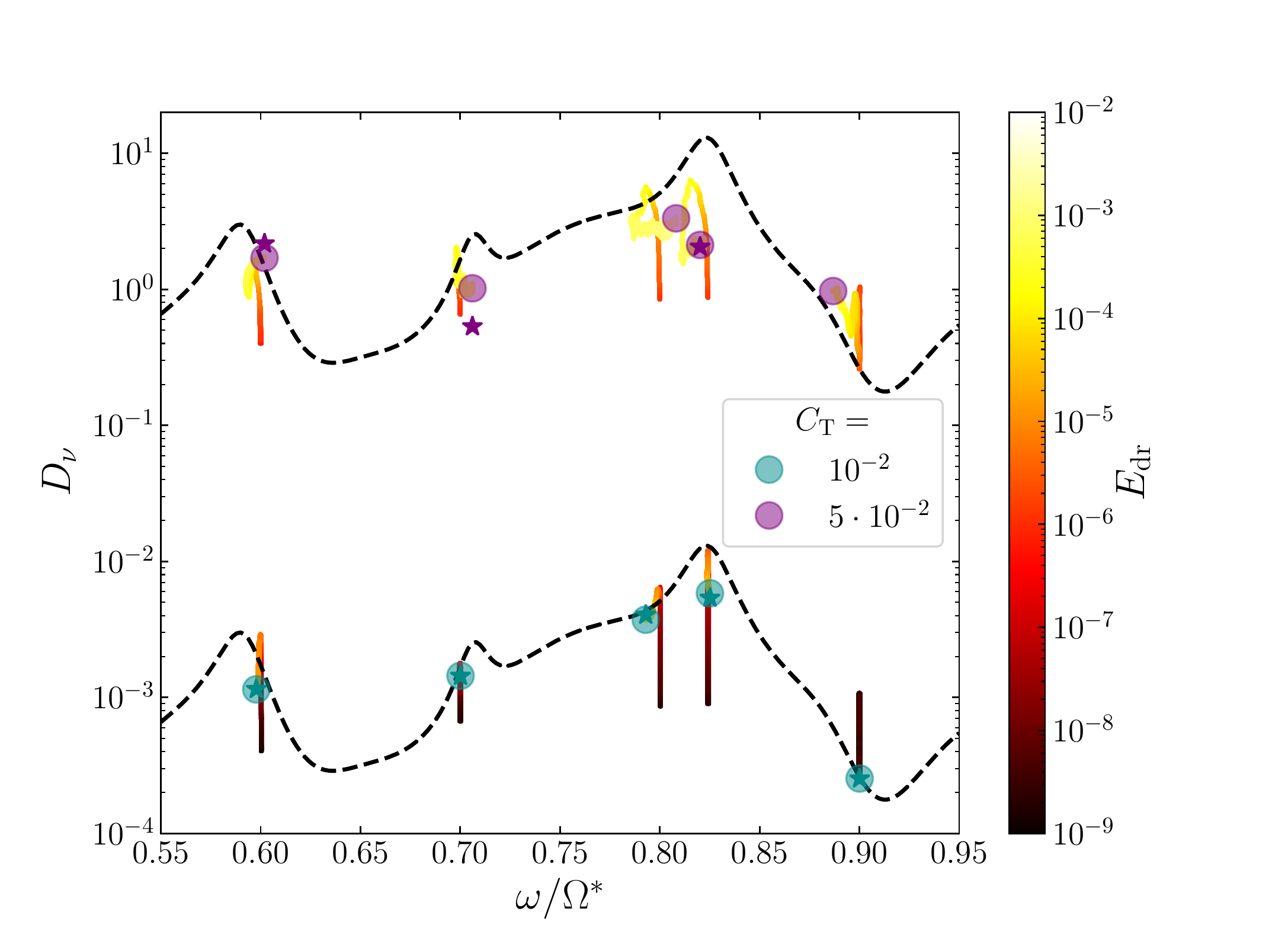}
\caption{Same as in Fig. \ref{fig:span_negom} but for initial tidal frequencies spanning the range $[0,1]$ and for two tidal forcing amplitudes $\Ct$.
The simulations for the highest $\Ct$ are shifted upwards by a factor $10^{3}$ for ease of visualisation (upper curves). 
}
\label{fig:span_spec_diss}
\end{figure*}
In the frame rotating at the rate $\Omega$, the linearised momentum equation for inertial waves excited by the effective tidal forcing $\bm f_\mathrm{t}$, and propagating in our background zonal flow, reads \citep[see also][]{BR2013}: 
\begin{equation}
    (\partial_t+\delta\Omega_s\partial_\varphi)\uw+2\Omega_s\bm e_z\wedge\uw+s\bm e_\varphi(\uw\cdot\bn)\delta\Omega_s=-\frac{\bn p_\mathrm{w}}{\rho}+\nu\Delta\uw+\bm f_\mathrm{t}.
    \label{eq:mom_sheared}
\end{equation}
The terms on the left hand side of Eq. (\ref{eq:mom_sheared}, except for $\partial_t\uw$) come from the linear development of the inertial terms $(\bm U+\uw)\cdot\bn(\bm U+\uw)$ in the co-rotating frame, remembering that $\bm U=s(\Omega+\delta\Omega_s)\bm e_\varphi$
and keeping the first order perturbation only.
Thus, these new terms directly originate from the initial differential rotation imposed here. 

We can also derive an energy balance as we did in Eq. (\ref{eq:bal}), here for linear inertial waves in a cylindrically, differentially rotating background flow (to avoid any confusion, we add the subscript $\mathrm{dr}$):
\begin{equation}
    \partial_tK_\mathrm{dr}= 
    \mI{w-dr}-D_{\nu, \mathrm{dr}}+P_{\mathrm{t},\mathrm{dr}}.
\end{equation}
We have also introduced the energy transfer term (involving Reynolds stresses) between the wavelike and zonal flows:
\begin{equation}
\begin{aligned}
    \mI{w-dr}&=-\langle\rho s u_\varphi(\uw\cdot\bn)\delta\Omega_s\rangle.
    \label{eq:balzon}
\end{aligned}
\end{equation}
Since we also have $\mI{w-dr}=\langle\rho\delta\Omega(\uw\cdot\bn)s u_\varphi\rangle$ using the divergence-free constraint and the boundary conditions on the wavelike flow, a negative $\mI{w-dr}$ means energy is transferred from inertial waves to the zonal flow, while a positive $\mI{w-dr}$ means energy is extracted from the zonal flow towards the waves. 

In the right panel of Fig. \ref{fig:ang}, we have added the tidal power (featured by stars) from the new linear simulations solving Eq. (\ref{eq:mom_sheared}) with MagIC, with the initial zonal shear flow coming from the end of the associated nonlinear simulations with the same tidal forcing frequencies $\omega=1.05,~1.1$ and $1.1$5 (see Fig. \ref{fig:zonflo}). Interestingly, the dissipation rates at the end of the nonlinear simulations (when an overall steady state is reached, given by blue circles) match very well with the linear tidal power terms for all three cases. For these simulations, it means that the lower  rates of nonlinear dissipation compared to linear predictions are fully explained by the setting up of zonal flows inside the shell, and of their effects on the waves.
We emphasise that the presence of the transfer term $\mI{w-dr}$ in the energy balance Eq. (\ref{eq:balzon}) means that all of the energy injected by the tidal flows is not entirely dissipated by viscosity, but can also be transferred to (or extracted from) zonal flows.

\begin{figure*}
    \centering
    \includegraphics[width=\columnwidth]{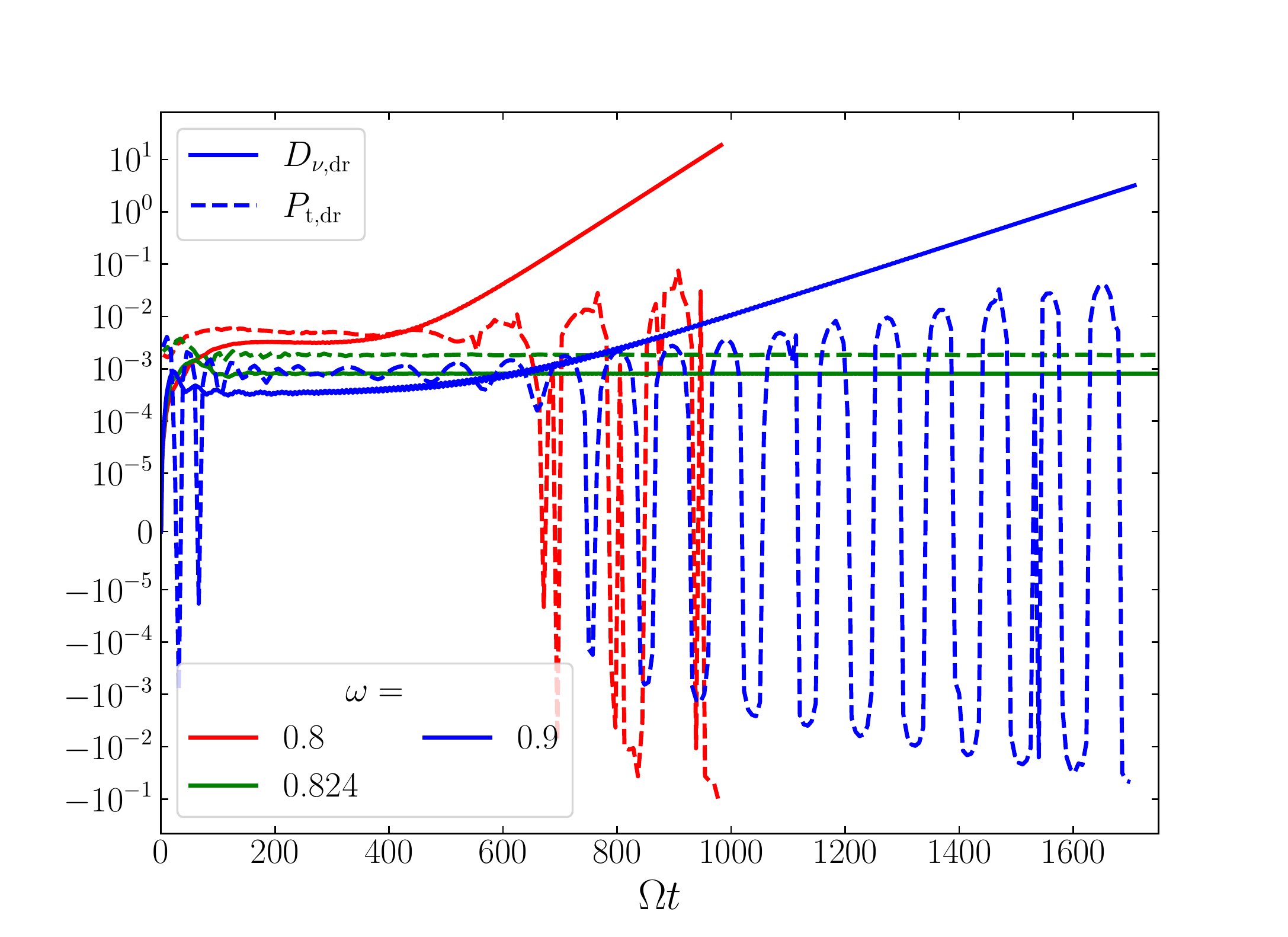}
    \includegraphics[width=\columnwidth]{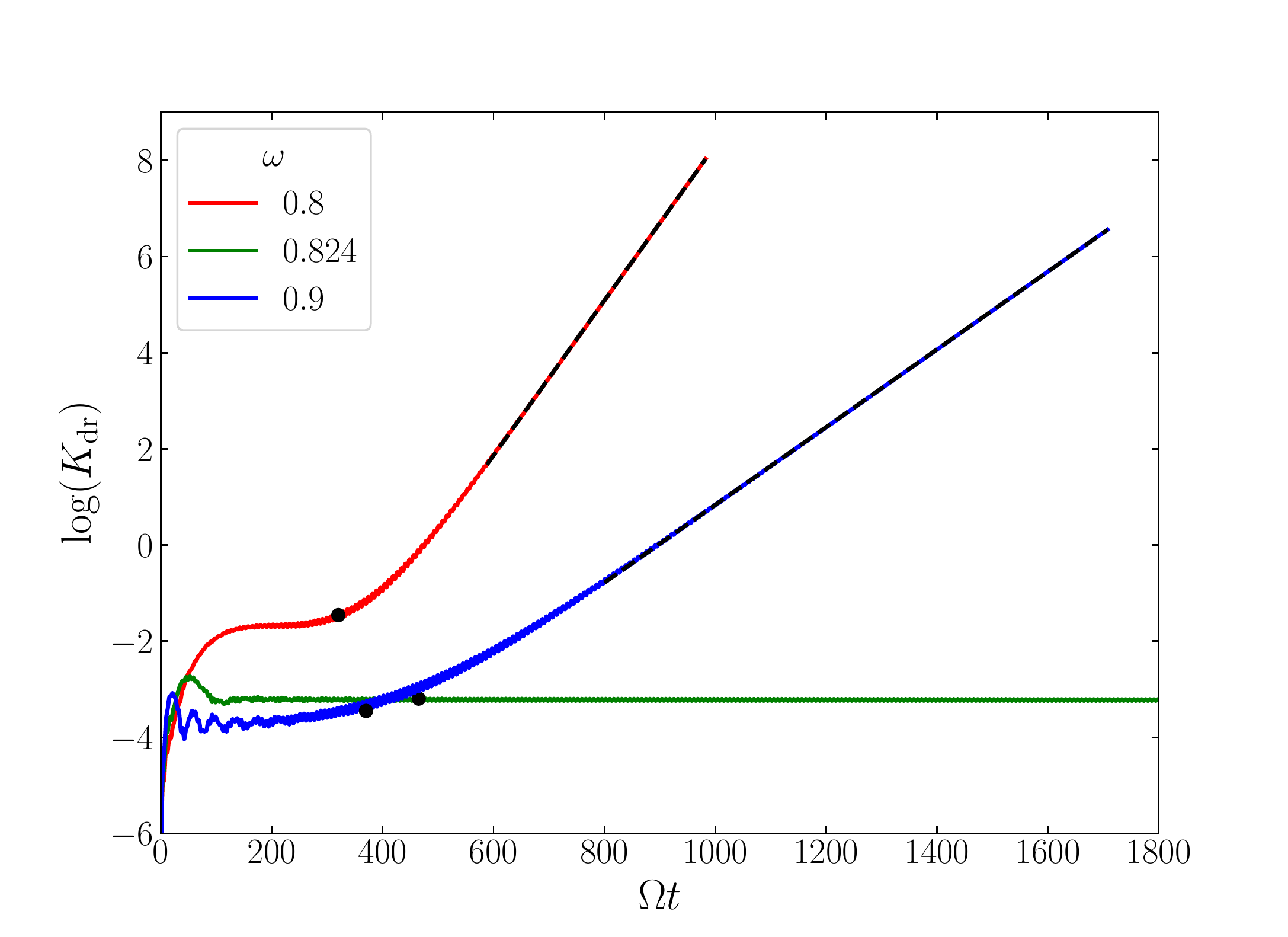}
    \caption{
    \textit{Left:} Tidal power $P_\mathrm{t,dr}$ and dissipation $D_{\nu,\mathrm{dr}}$ versus time $\Omega t$ in three linear simulations with a background cylindrical zonal flow.
    \textit{Right:} Natural logarithm of the kinetic energy $K_\mathrm{dr}$  for the same three simulations. The meridional plane snapshots of the kinetic energy in Fig. \ref{fig:corot} are at the times indicated by the black dots. Black dashed lines at late times are linear fits with slopes $\sim1.6\cdot10^{-2}$ for $\omega=0.8$, and $\sim8.1\cdot10^{-3}$ for $\omega=0.9$.
    }
    \label{fig:logK}
\end{figure*}
We have further explored the range of frequencies for which inertial waves can be excited by the tidal forcing, between -1 and 1 in Figs \ref{fig:span_negom} and \ref{fig:span_spec_diss} (with our choice of time unit, we can excite inertial waves between $-2$ and $2$). For a tidal amplitude of $\Ct=10^{-2}$, there are no significant departures in nonlinear dissipation rates from linear predictions (dashed line), though the simulations having an initial tidal forcing frequency close to a (linear) resonant peak of dissipation have more energy inside the differential rotation and a dissipation rate that differs more from the linear one, as was already observed in Fig. \ref{fig:ang} (right panel). Moreover, the tidal power term $P_\mathrm{t,dr}$ agrees quite nicely in all cases with the final nonlinear dissipation. For a tidal amplitude forcing of $\Ct=5\cdot10^{-2}$, this is no longer systematically true, with simulations having a notable mismatch between $D_\nu$ and $P_\mathrm{t,dr}$ for example, are $\omega=0.192,\ 0.2,\text{ and }0.7$, or even no steady linear tidal power $P_\mathrm{t,dr}$, which still wildly oscillates at late times for $\omega=0.8\text{ and } 0.9$, as we can see in the left panel of Fig.~\ref{fig:logK}. Some of these cases are investigated in further detail in the following.  

\begin{figure*}
    \centering
    \includegraphics[width=\columnwidth]{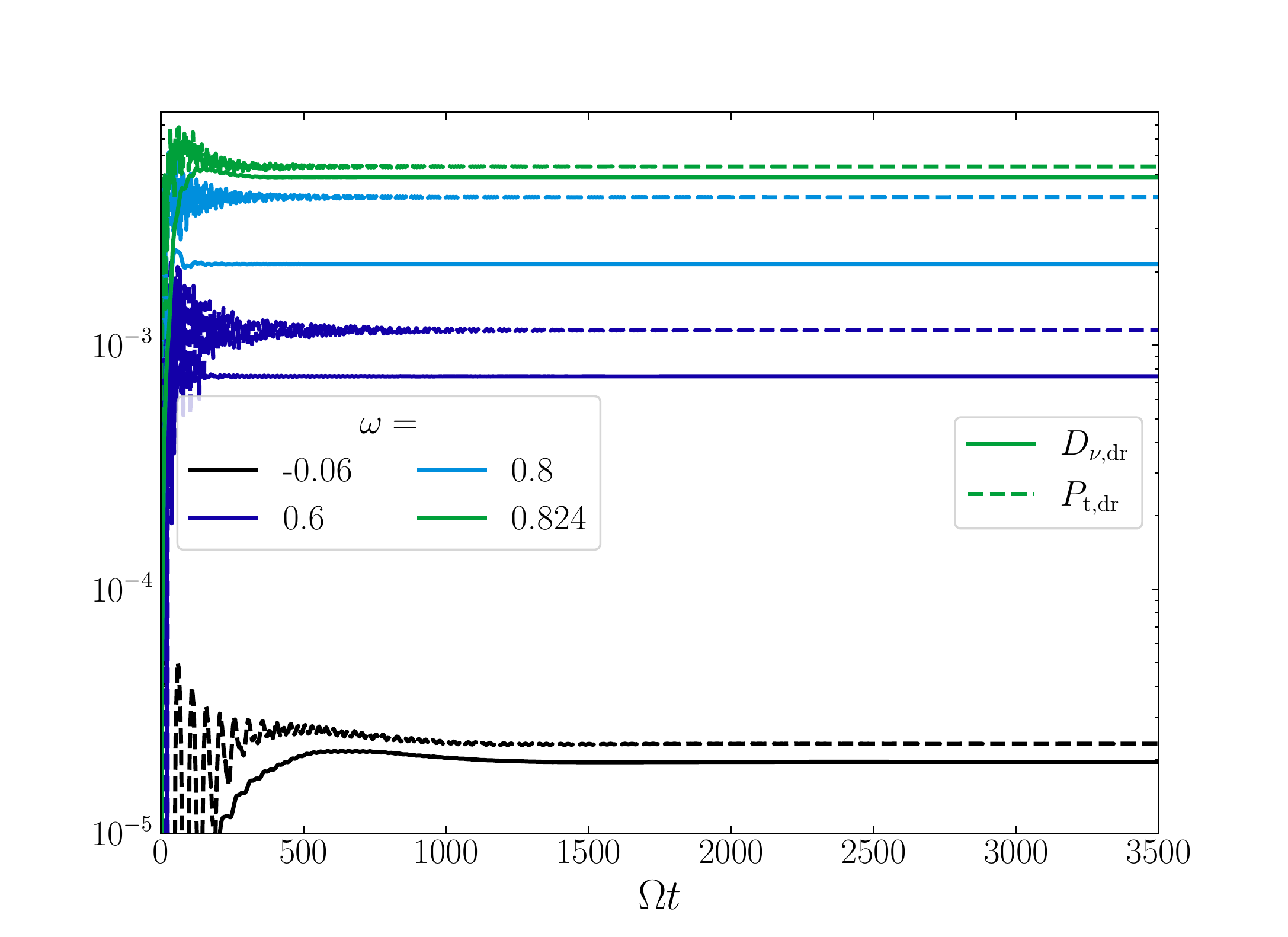}
    \includegraphics[width=\columnwidth]{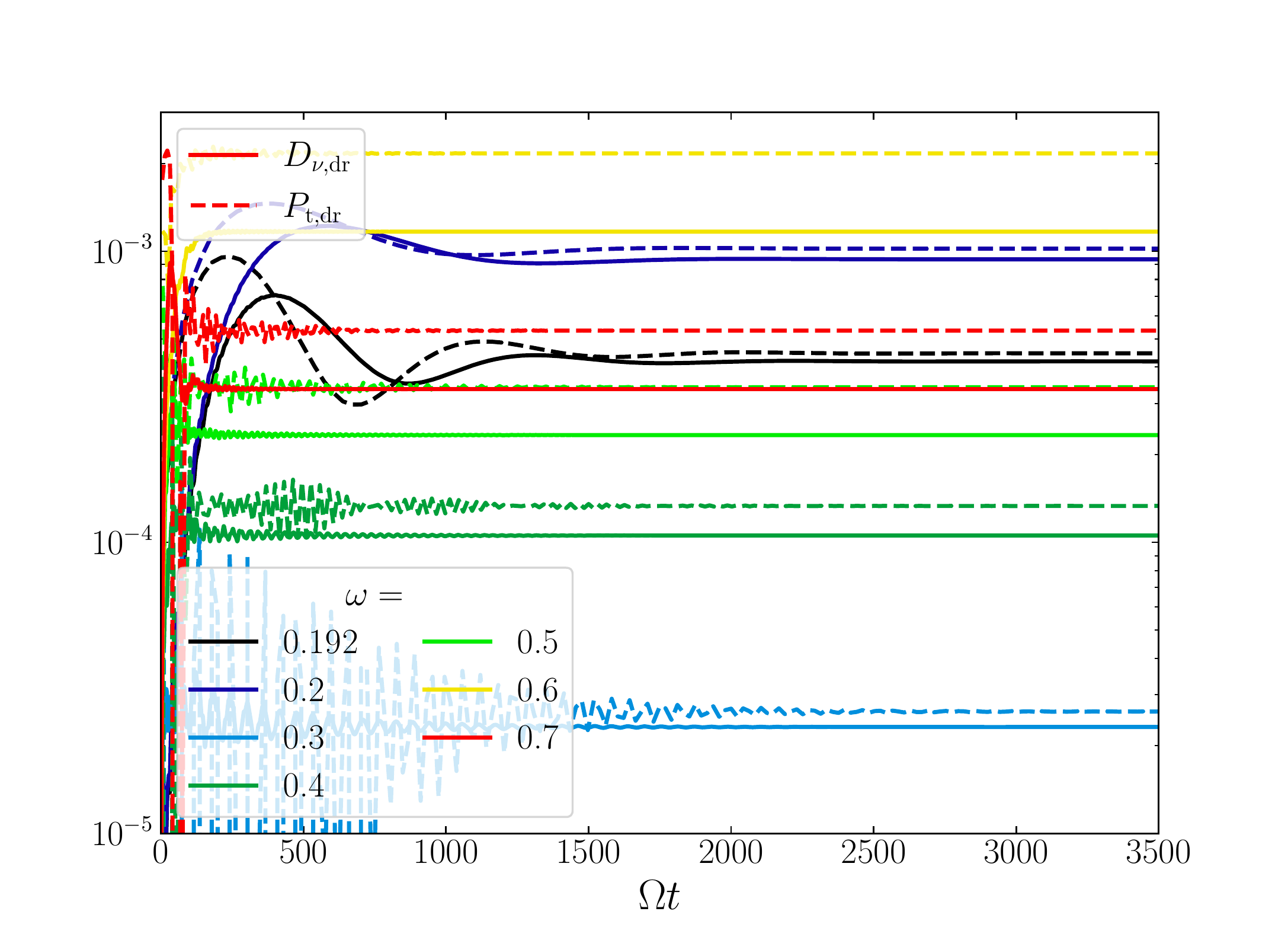}
    \caption{Tidal power $P_\mathrm{t,dr}$ and dissipation $D_{\nu,\mathrm{dr}}$ versus time $\Omega t$ in linear simulations (for tidal forcing frequencies $\omega$  of Figs \ref{fig:span_negom} and \ref{fig:span_spec_diss}) for which the difference between $P_\mathrm{t,dr}$ and $D_{\nu,\mathrm{dr}}$ is larger than $5\%$ (except for $\omega=0.8,\ 0.824,\text{ and } 0.9$ with $\Ct=5\cdot10^{-2}$ shown in the left panel of Fig.~\ref{fig:logK}). 
    \textit{Left:} The tidal forcing amplitude is set to $\Ct=10^{-2}$ \textit{Right:} $\Ct=5\cdot10^{-2}$.
    }
    \label{fig:diss_pt}
\end{figure*}

We show in Fig. \ref{fig:diss_pt} the tidal power $P_\mathrm{t,dr}$ and the dissipation $D_{\nu,\mathrm{dr}}$ for the linear simulations with background zonal flows for which $P_\mathrm{t,dr}$ differs by more than $5\%$ from $D_{\nu,\mathrm{dr}}$, for the two tidal forcing amplitudes $\Ct=10^{-2}$ and $5\cdot10^{-2}$ (left and right panels,  respectively). In all cases presented in these panels, we observe that the amplitude of the tidal power input is greater than the viscous dissipation, confirming that part of the energy injected by tides is redirected to the zonal flows via the energy transfer term $\mI{w-dr}$. The difference between $P_\mathrm{t,dr}$ and  $D_{\nu,\mathrm{dr}}$, i.e.~the value of $\mI{w-dr}$, is even more important for higher tidal forcing amplitudes (right vs left panel), so for stronger zonal flows\footnote{This might be partly because inertial waves are primarily damped near a corotation resonance, as we investigate later in this Section.}. The maximum differences for steady cases are for $\omega=0.824$ (more than $50\%$), as is shown in the left panel of Fig.~\ref{fig:logK}.
In the linear simulations with $\omega=0.8$ and $\omega=0.9$, $P_\mathrm{t,dr}>D_{\nu,\mathrm{dr}}$ is not satisfied at late times, as soon as the kinetic energy and tidal power start to diverge, implying that energy is being extracted from the zonal flow due to the positive sign of $\mI{w,dr}$.

In fact, these two last cases exhibit an instability; the kinetic energy $K_\mathrm{dr}$ grows exponentially soon after the beginning of the simulation, as we can see in the left panel of Fig. \ref{fig:logK}. 
We have verified that this behaviour is not due to unphysical numerical effects by varying the spatial and temporal resolution (as well as by using a different time-stepping scheme). We have also run again these linear simulations 
but with a random noise in the velocity field, but without tidal forcing (namely $\Ct=0$). We found that such a random noise is not able to destabilise these flows and produce an exponential growth of the kinetic energy. We also point out that no instabilities leading to a diverging kinetic energy are found in the associated nonlinear simulations for $\omega=0.8$ and $\omega=0.9$. Nonlinearities may have a stabilising effect here by slightly modifying the zonal flows while this is not possible in these linear simulations.

\begin{figure}
    \centering
    \includegraphics[width=\columnwidth]{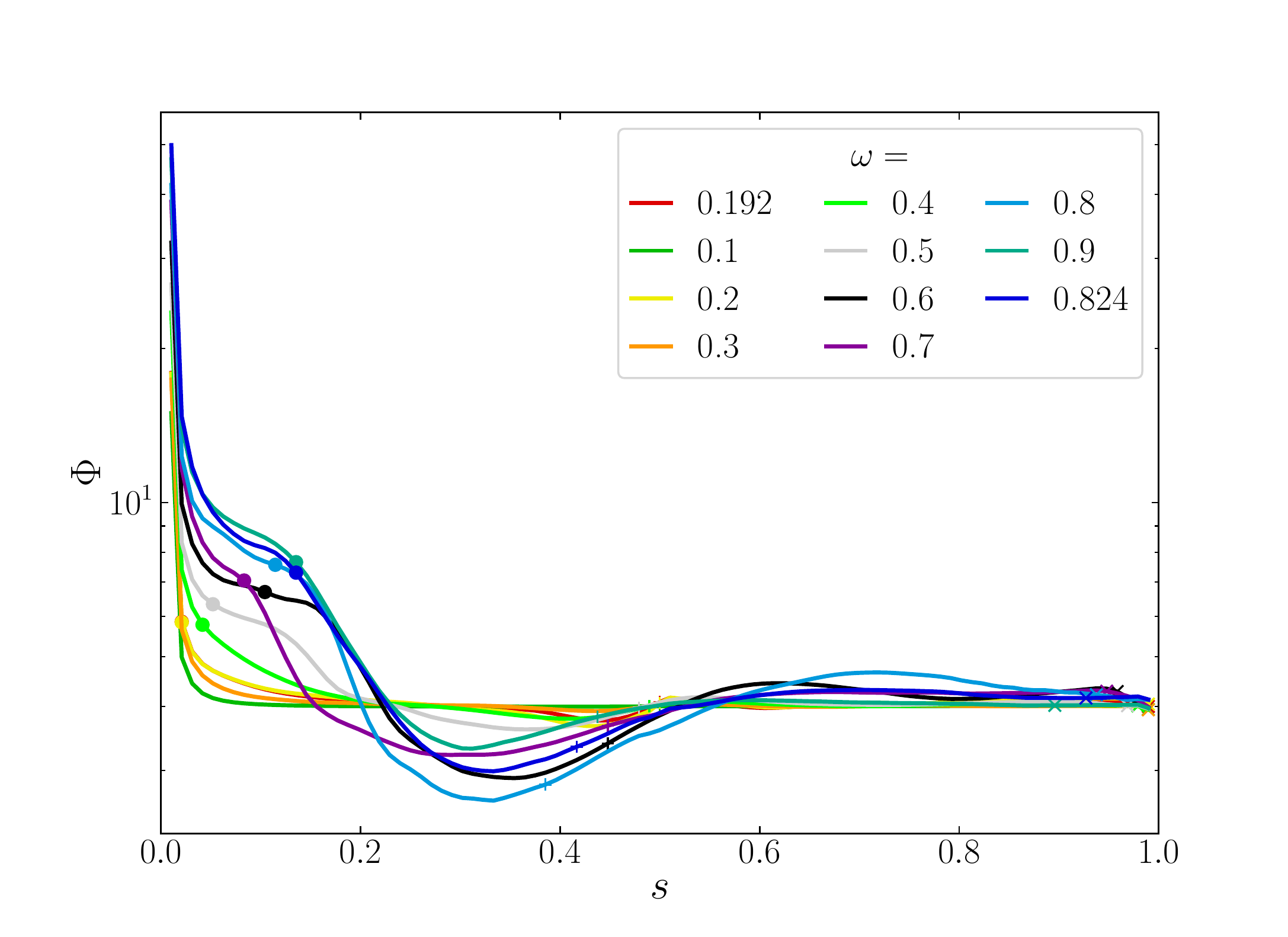}
    \caption{Plot of the Rayleigh discriminant $\Phi$ versus the cylindrical radius $s$ for the same initial forcing frequencies as in Fig. \ref{fig:span_spec_diss} for a tidal forcing amplitude $\Ct=5\cdot10^{-2}$. Bullets indicate corotation resonances (Eq. (\ref{eq:corot})), while ticks indicate inner critical latitudes and crosses outer critical latitudes, all determined by Eq.~(\ref{eq:critlat}).
    }
    \label{fig:rayleigh_crit}
\end{figure}

To understand the nature of this instability, we consider first whether our zonal flows could be centrifugally unstable.
In the inviscid limit, cylindrical differentially rotating flows depending only on cylindrical radius (albeit which is an imperfect approximation of our flow) can be destabilised by axisymmetric perturbations if the specific angular momentum decreases outwards in magnitude from the rotation axis, or in other words if the Rayleigh discriminant, defined here as
\begin{equation}
    \Phi(s)=\frac{1}{s^3}\frac{\mathrm{d}(s^2\Omega_s)^2}{\mathrm{d}s},
    \label{eq:rayleigh_crit}
\end{equation}
is negative inside the shell \citep[the condition is necessary and sufficient from the work of][generalises this result to local non-axisymmetric perturbations]{S1933,BG2005}. In Fig. \ref{fig:rayleigh_crit}, the Rayleigh discriminant $\Phi$ is plotted versus $s$ using the zonal flows extracted from the nonlinear simulations of Fig. \ref{fig:span_spec_diss} for $\Ct=5\cdot10^{-2}$. For these columnar flows, $\Phi$ is always positive inside the shell, meaning that they are centrifugally stable to any inviscid axisymmetric perturbations. 

However, it is unclear whether non-axisymmetric perturbations (like $m=2$ inertial waves) are able to destabilise these zonal flows. 
Considering an inertial mode of complex frequency $\omega=\omega_\mathrm{r}+\mathrm{i}\omega_\mathrm{i}$ of real part $\omega_\mathrm{r}$ and growth (damping) rate $\omega_\mathrm{i}$, if this is positive (negative), the temporal dependence of its velocity scales as  $\exp(-\mathrm{i}\omega t)$, and thus its kinetic energy is proportional to $\exp(2\omega_\mathrm{i}t)$. Hence, the "linear" slopes for $\omega=0.8$ and $\omega=0.9$ in the plot showing $\log(K _\mathrm{dr})$ (right panel of Fig. \ref{fig:logK}) after the early transient phase, would correspond to twice the growth rate of the triggered unstable mode, here $\omega_\mathrm{i}\simeq8\cdot10^{-3}$ for $\omega=0.8$ and $\omega_\mathrm{i}\simeq4\cdot10^{-3}$ for $\omega=0.9$. For comparison, the maximum shear rate $s\partial_s\Omega_s$ for these cases is of the order of $10^{-1}$. The knowledge of unstable eigenmodes in the shell with our particular differential rotation profiles could help us to know whether the growth of the kinetic energy in the linear simulations with $0.8$ and $0.9$ is related to their presence or not. In the analytical study of Dandoy et al. in prep., the authors looked at the destabilisation of a columnar convective vortex located near the rotation axis that is interacting with (tidal) inertial waves. They found that when vortices are centrifugally-unstable under the Rayleigh criterion (Eq. (\ref{eq:rayleigh_crit})), tidal inertial waves can trigger the most unstable (non-axisymmetric) mode of the vortex and its destabilisation. Since the Rayleigh criterion for stability is satisfied here, such a complex analysis is beyond the scope of this paper and should be postponed to a future study. 

\begin{figure*}
    \centering
    \includegraphics[width=\columnwidth]{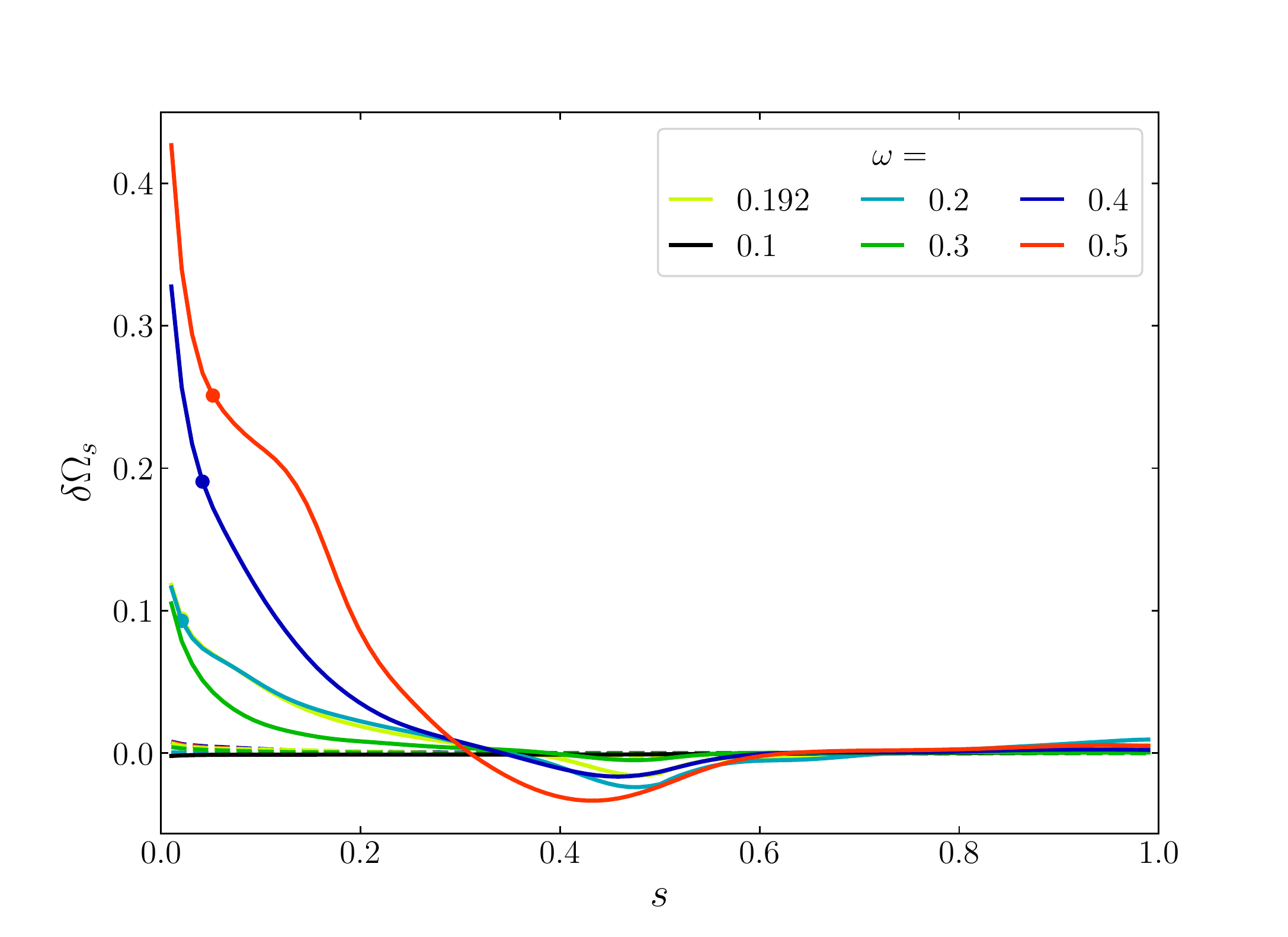}
    \includegraphics[width=\columnwidth]{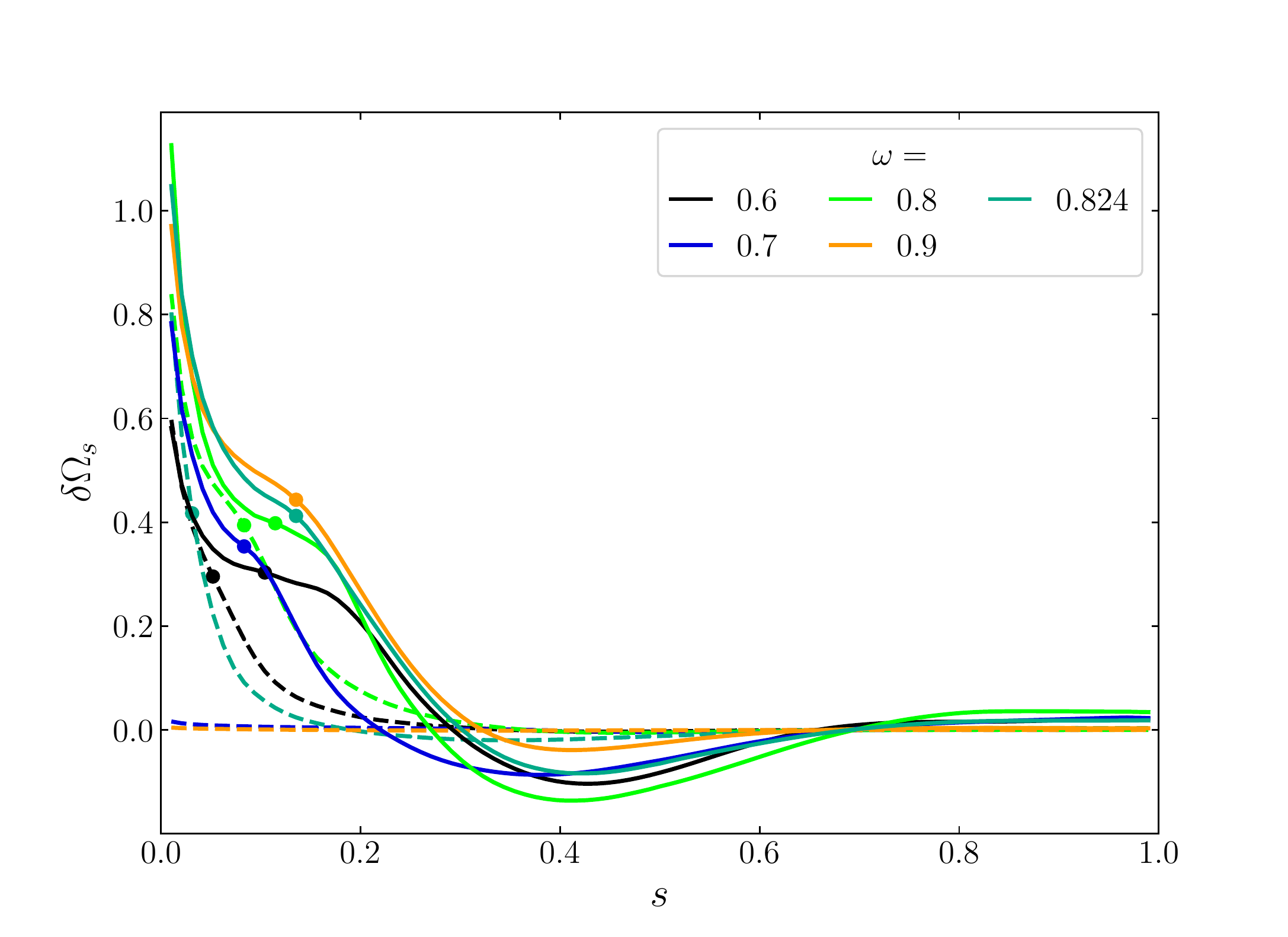}
    \caption{Zonal flows $\delta\Omega_s$ produced in the nonlinear simulations presented in Fig. \ref{fig:span_spec_diss} (left and right panels) in terms of the cylindrical radius $s$ for $\Ct=10^{-2}$ (dashed lines) and $\Ct=5\cdot10^{-2}$ (solid lines). For each zonal flow (of different colour and linestyle), the corotation resonance is indicated by a bullet when it exists (i.e.~ when it satisfies Eq. (\ref{eq:corot})).}
    \label{fig:dOms}
\end{figure*}
Another possible cause of the sharp increase of the kinetic energy observed in some linear simulations with zonal flows is the presence of corotation resonances inside the shell. For the initial $m=2$ inertial wave perturbation of frequency $\omega$ in the corotating frame (i.e. rotating with $\Omega$), a corotation resonance (or ``critical layer") 
happens when the Doppler-shifted frequency of the wave in the zonal flow frame\footnote{A wave with a frequency $\tilde{\omega}$ in the inertial frame has a frequency in the $\Omega$-rotating frame of $\omega=\tilde\omega-m\Omega$, and in the zonal flow frame it is $\sigma=\tilde\omega-m\Omega_s$ which gives Eq. (\ref{eq:sigma}).}
\begin{equation}
\sigma=\omega-m\delta\Omega_s,    
\label{eq:sigma}
\end{equation}
 vanishes, namely at a cylindrical radius $s=\sc$, when:
\begin{equation}
    \delta\Omega_s(\sc)=\omega/2.
    \label{eq:corot}
\end{equation}
One should note that the definition of the Doppler-shifted frequency Eq (\ref{eq:sigma}, and thus of the corotation resonance) differs from the one in \cite{BR2013} for example, due to the different sign convention for the wavelike part of the solution (Eq. (\ref{eq:X})) and the different reference frames used (they worked in an inertial frame). In Fig. \ref{fig:dOms}, we show the different zonal flows $\delta\Omega_s$ produced in nonlinear simulations for the same tidal forcing frequencies and amplitudes as in Figs~\ref{fig:span_spec_diss}. We first notice that some flows exhibit one corotation resonance at a critical cylinder $\sc$ satisfying Eq. (\ref{eq:corot}), usually close to the rotation axis. These corotation resonances are more likely to be present in strong zonal flows, arising when the tidal forcing amplitude is high and the (non-)linear dissipation is strong (e.g.~which can be seen by comparing the left and right panels of Figs \ref{fig:span_spec_diss} and \ref{fig:dOms}). 

When present, the corotation resonance seems to have an effect on the zonal flow profile by perturbing it in most cases, producing a small bump around $s=\sc$. When approaching a corotation resonance, a wave can exchange (deposit or extract) energy (or angular momentum) with the zonal flow, which is in turn modified. It has been analytically and numerically studied in great detail for (inertia-)gravity waves in (rotating) stratified shear flows \cite[e.g.][]{BB1967,G1975,LB1985,BO2010,B2011,AM2013,SL2020} and also for inertial waves in a differentially rotating shell or box \citep[][]{BR2013,GM2016,GB2016,AP2021}. For the simulations for which a corotation resonance exists for both $\Ct=10^{-2}$ and $\Ct=5\cdot 10^{-2}$, the location of the resonance $\sc$ appears to move away from the rotation axis for higher tidal forcing amplitudes. As we can see in \cite{BR2013} for cylindrical differential rotation, free inertial waves are often very strongly damped at the corotation resonance. 
We can thus expect that the successive deposition of energy from inertial waves emitted from the critical latitudes at these corotation points can efficiently contribute to modifying the zonal flow at this location. This process can potentially move the location of the corotation resonance, as has been observed for example in \cite{BO2010} and \cite{SL2020} in their local simulations of gravity waves breaking nonlinearly.

\begin{figure*}
    \centering
    \includegraphics[width=0.33\textwidth]{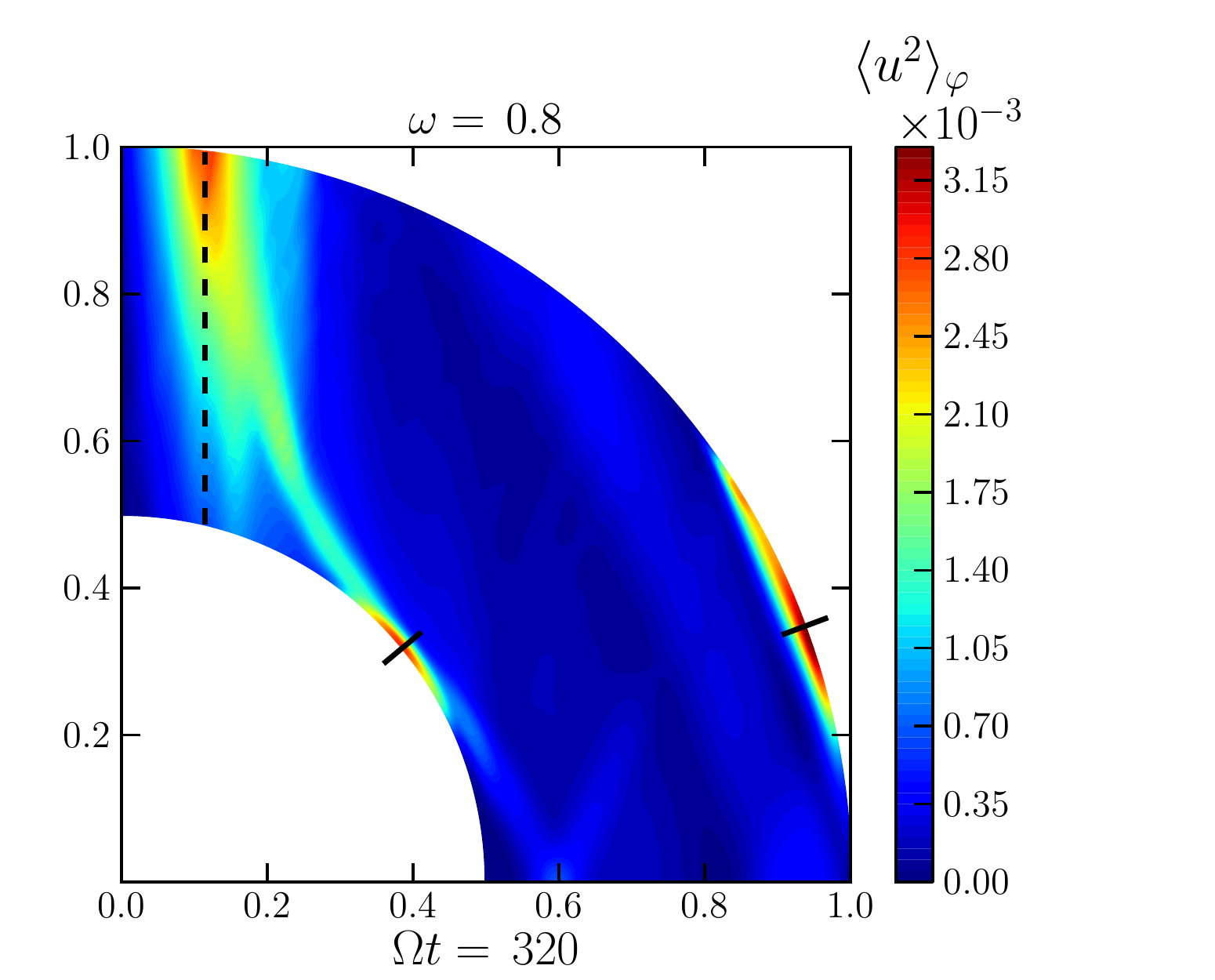}
    \includegraphics[width=0.33\textwidth]{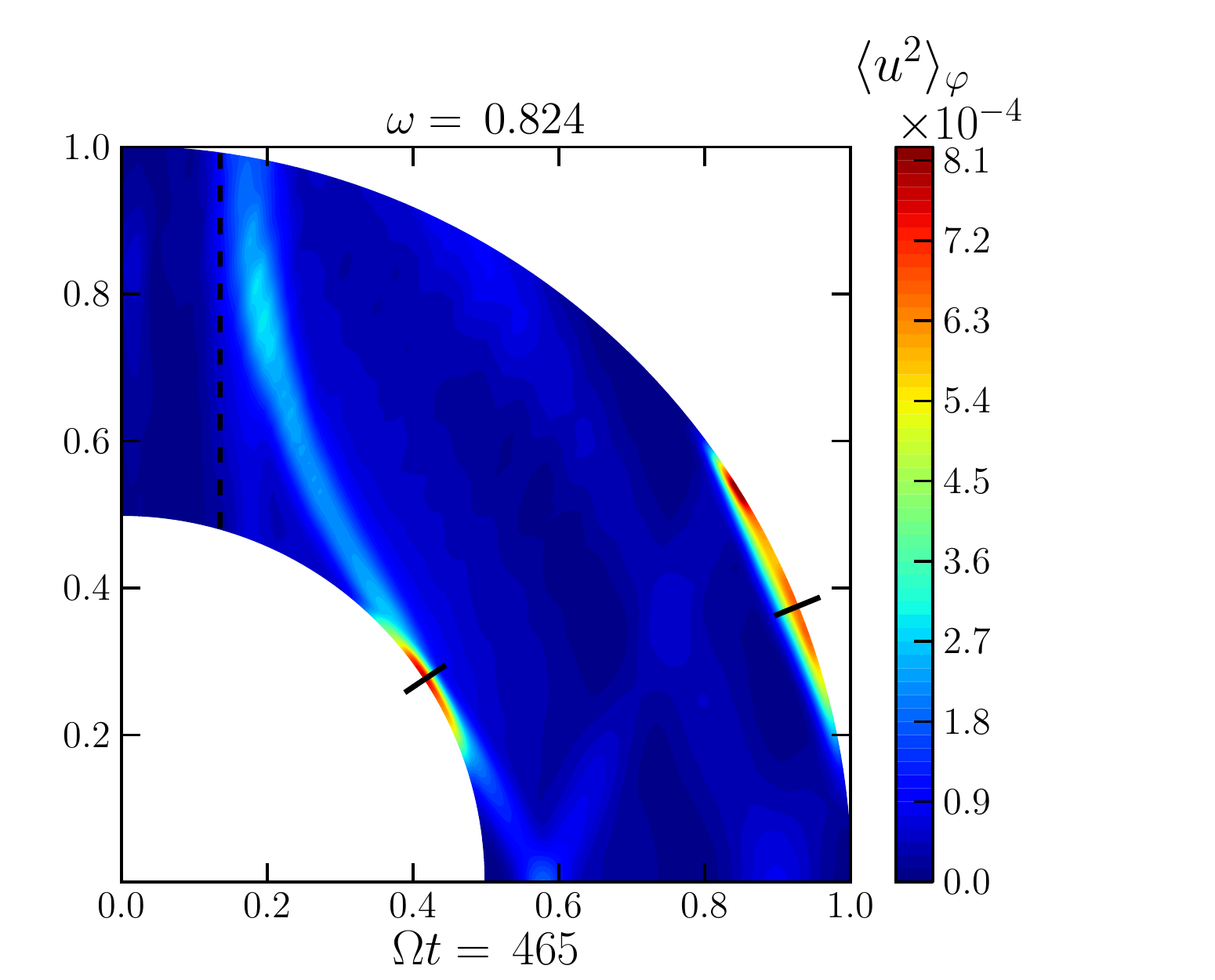}
    \includegraphics[width=0.33\textwidth]{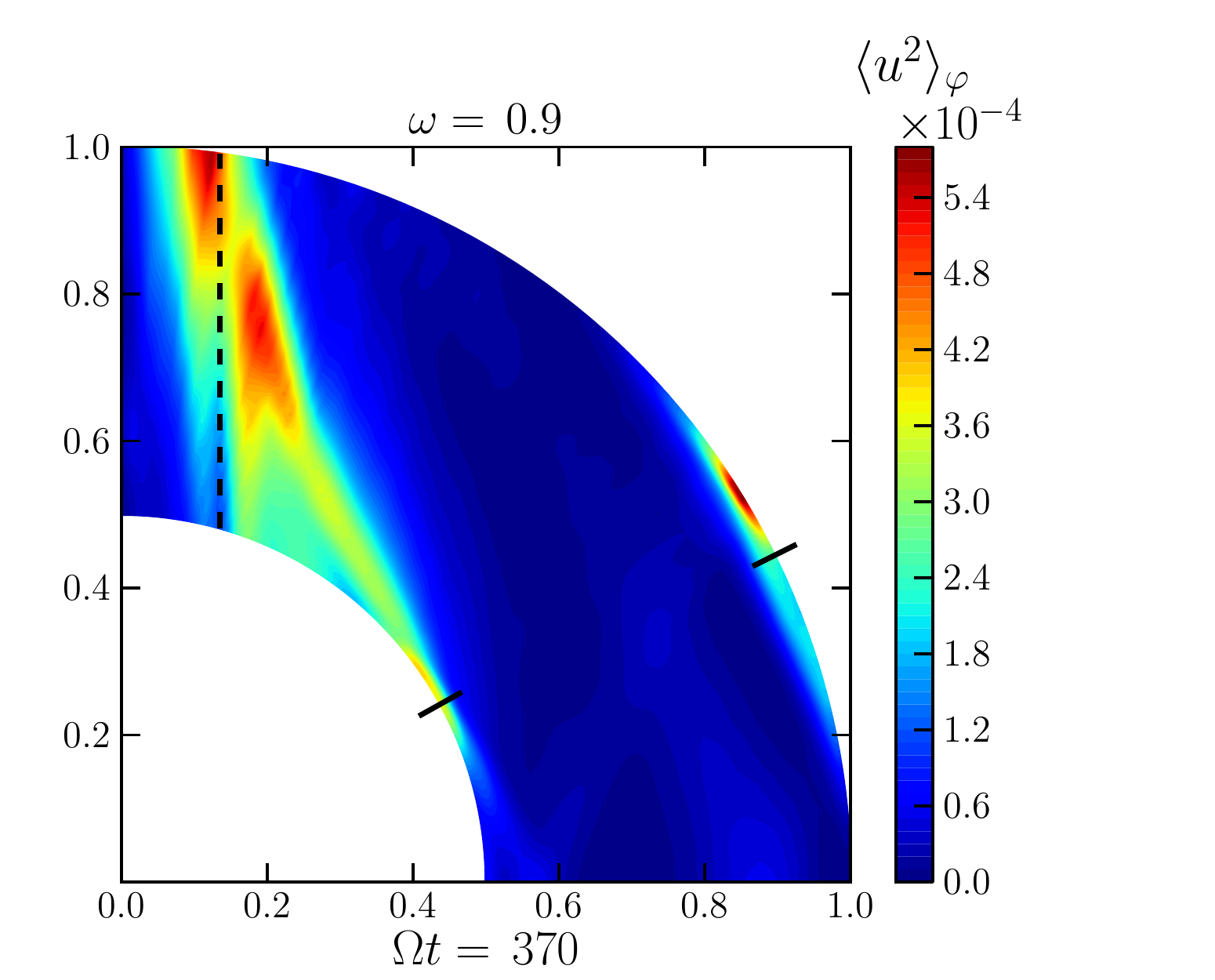}
    \caption{Azimuthal average of the \cor{wavelike} kinetic energy $\langle u^2\rangle_\varphi$ in the meridional plane in linear simulations with initial zonal flows for the same three tidal frequencies as in Fig. \ref{fig:logK} at early times (indicated by black dots on that figure). The critical cylinder corresponding to the corotation resonance is indicated by a vertical dashed line and the critical latitudes $\theta$ defined by Eq. (\ref{eq:critlat}) on the inner and outer shells 
    are featured by the inclined black ticks.}
    \label{fig:corot}
\end{figure*}
\begin{figure*}
    \centering
    \includegraphics[width=0.33\textwidth]{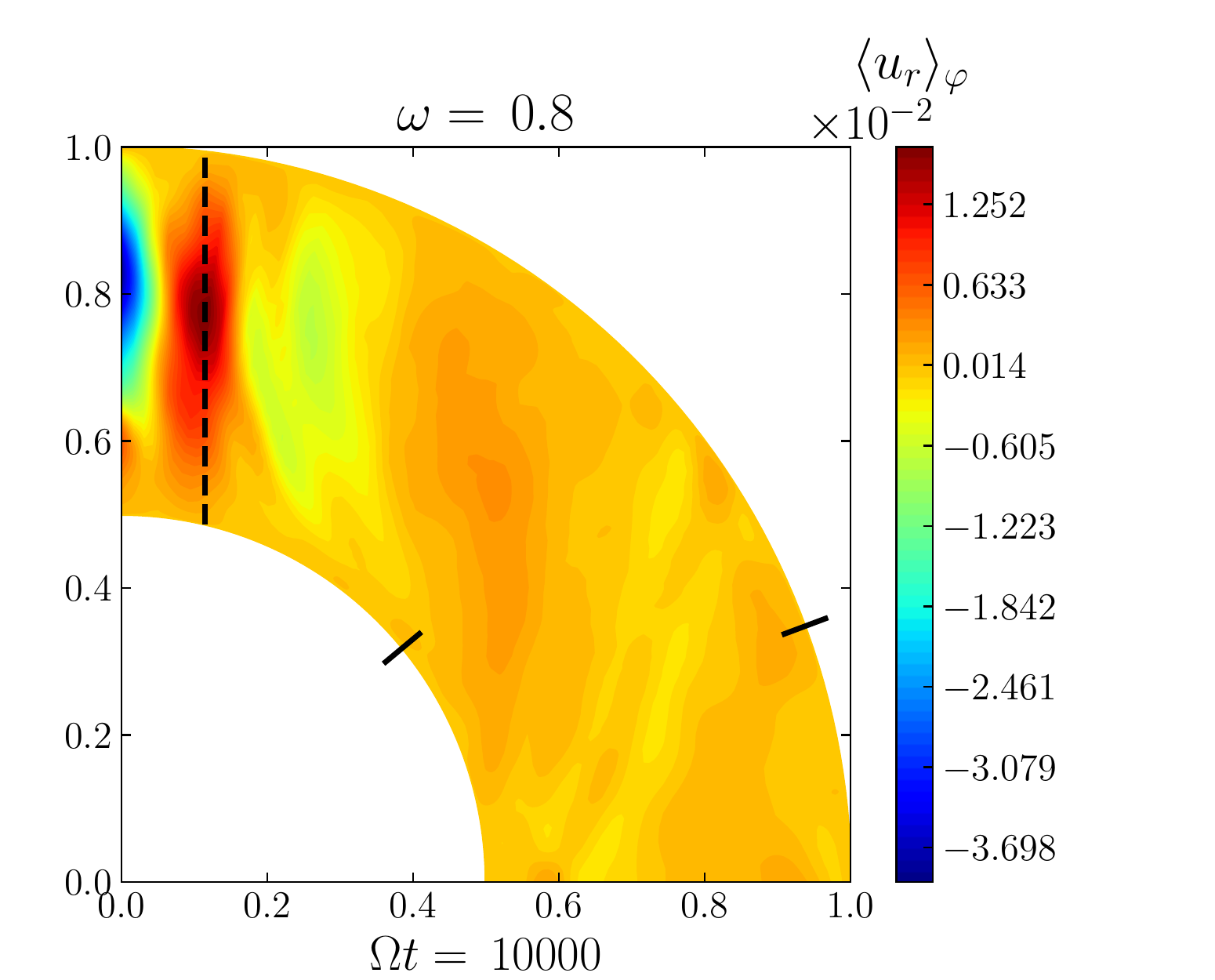}
    \includegraphics[width=0.33\textwidth]{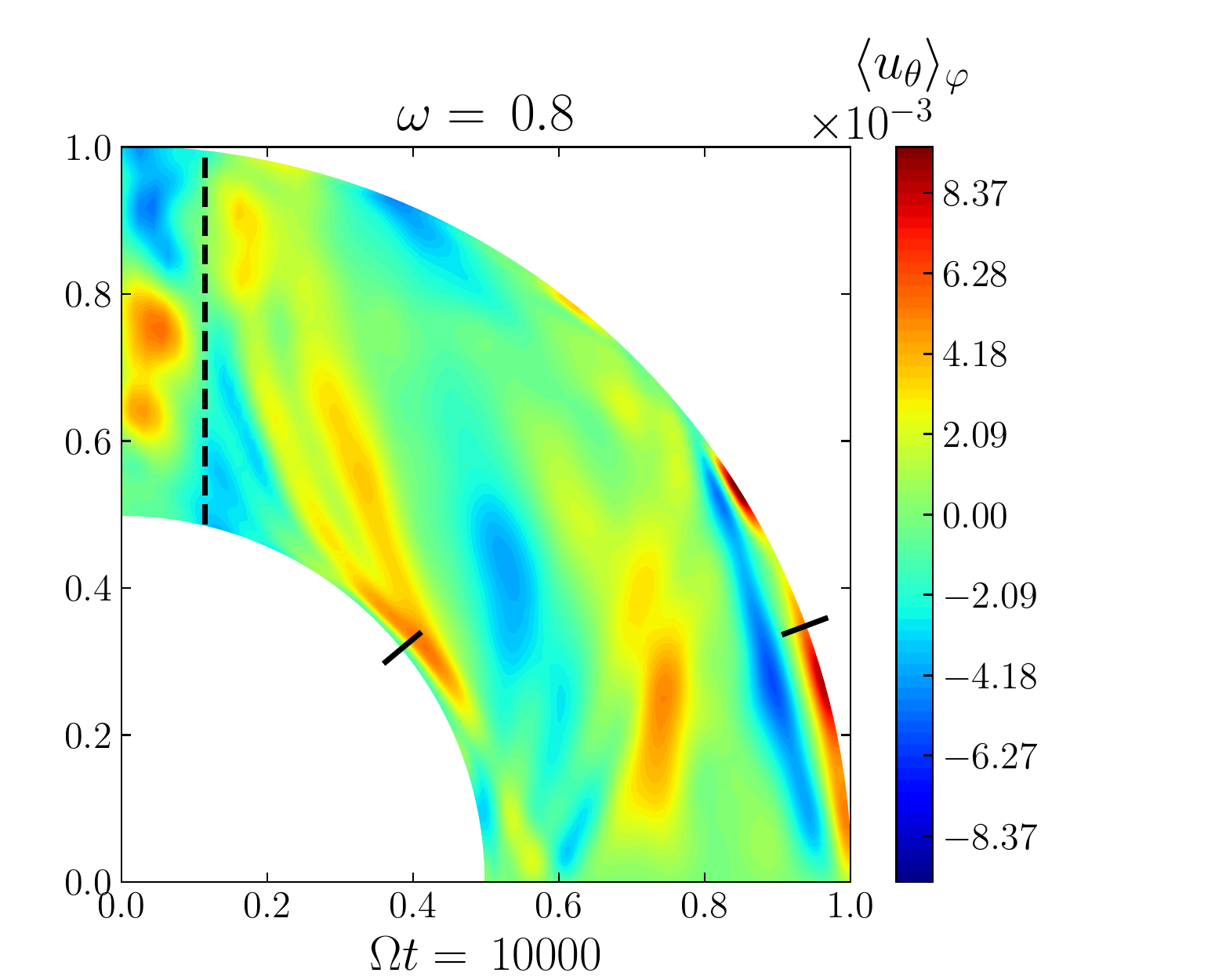}
    \includegraphics[width=0.33\textwidth]{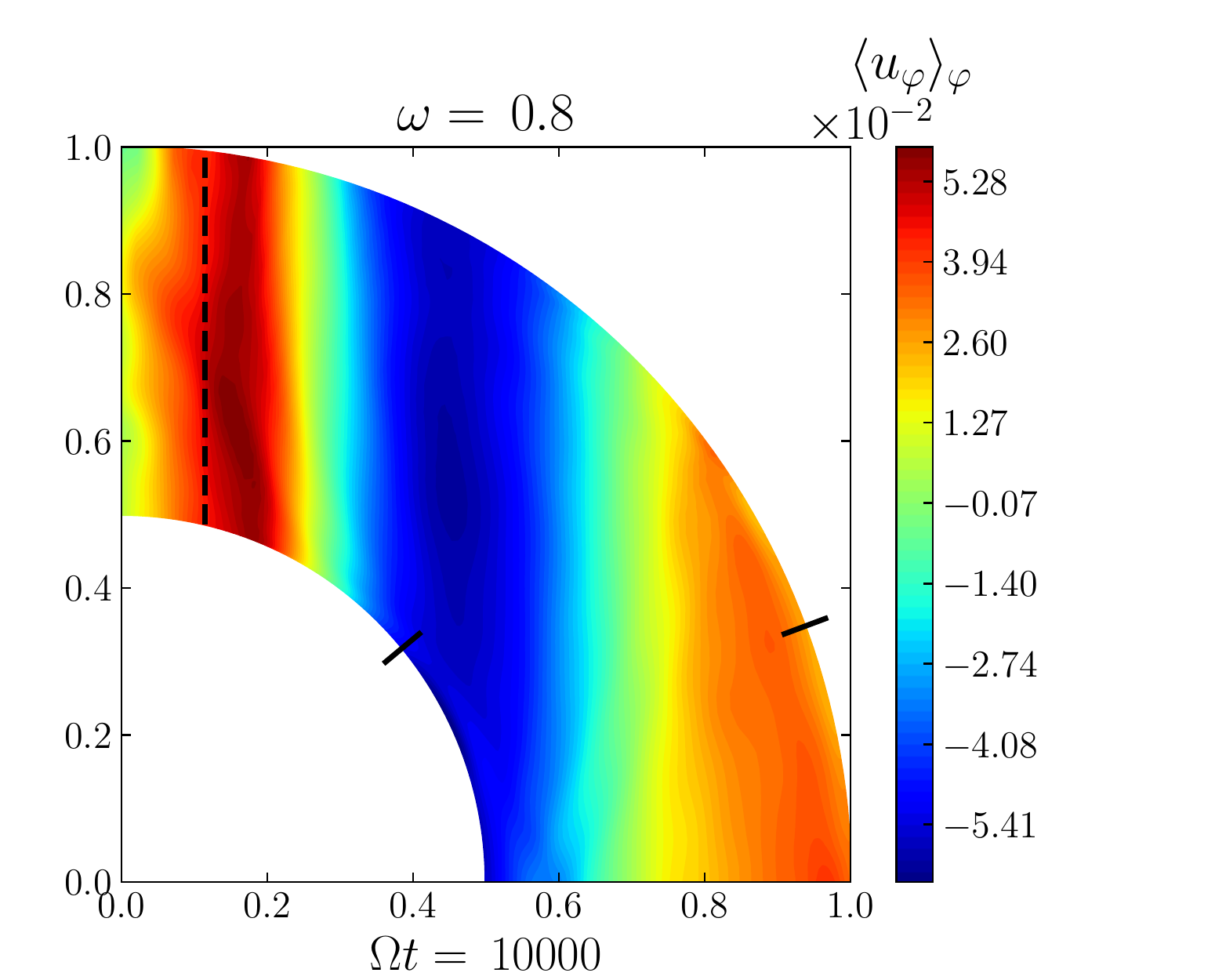}
    \caption{Azimuthal average in one quadrant of the three components of the wavelike velocity $\uw$ at the end of the nonlinear simulation with $\omega=0.8$. As in Fig.~\ref{fig:corot}, the ticks indicate the critical latitudes and the dashed line the corotation resonance.}
    \label{fig:om0p8_flows}
\end{figure*}

The role of corotation resonances is clear from Fig.~\ref{fig:corot}, which shows the azimuthally-averaged kinetic energy for three linear simulations containing corotation resonances close to the poles at early times. In each of the three panels, inertial waves are emitted from the inner and possibly outer critical latitudes, whose expressions (detailed at the end of the section) are modified by differential rotation. 
When approaching the corotation resonance at the critical cylinder, shear layers bend as a result of the differential rotation. In the left and right panels (for which the kinetic energy diverge in these linear simulations at later times, see Fig.~\ref{fig:logK}) the kinetic energy is concentrated along the critical cylinder on both sides, while in the middle panel (for which the kinetic energy reaches a steady state at late times) the critical cylinder seems to act like an absorbing barrier for incoming inertial waves which do not cross the corotation resonance.

Since we observe instabilities inside the simulations (left and right panels of Fig. \ref{fig:corot}), we first investigate the possibility of an instability similar to the ``Papaloizou-Pringle instability'' \citep{PP1984,PP1985,PP1987}, which occurs in Rayleigh-stable (i.e. with $\Phi>0$, just like here) Keplerian disks, and has been observed e.g.~in \cite{BO2016}. For this kind of instability to occur, the inertial wave needs to be sent multiple times into the corotation region to exhibit amplification. Since in our cases, the differential rotation is not strong enough to confine the waves in a portion of the domain\footnote{The condition for turning surfaces to exist is not met for the zonal flows presented here i.e. where $\sigma^2=\Phi$ (see Eq. (\ref{eq:carac})), and hence we only observe what \cite{BR2013} refer to as ``D modes''.} (i.e. for turning points to exist in the differential equation governing inertial wave propagation, where it changes character from hyperbolic/wavelike to elliptic/evanescent), it is the spherical boundaries and the rotation axis that could reflect the waves back towards the corotation region. This has the potential to lead to instability if the waves can ``over-reflect" and be amplified at corotation. 

This phenomenon is also similar to the over-reflection mechanism described in the review of \cite{L1988} in the case of internal gravity waves in shear flows and leading to shear instability \cite[see also e.g.][for Rossby waves]{J1968,LT1978,LB1985,AM2013,BM1994,HH2007}. Several conditions on the ``wave geometry'' are required for such an instability to happen. The bounded domain where the waves propagate must be divided into three regions: two ``propagative'' regions with wavelike solutions where the Richardson number\footnote{It is the ratio of the squared Brunt-Vaïsälä frequency with the squared shear rate.} $\mathrm{Ri}$ is greater than $1/4$, surrounding a region containing the critical layer where $\mathrm{Ri}<1/4$ \citep[a necessary condition for instability,][]{MH1964}. Under suitable boundary conditions for the waves to be reflected back to the critical cylinder and interfere constructively with waves on the other side of corotation (i.e. with wave quantization on the phase speed or frequency), a shear instability can occur after multiple amplifications of the waves at corotation. 

In \citet{AP2021}, an analogous criterion \cor{to the Richardson criterion (also known as the Miles-Howard theorem, and establishing a necessary condition for instability when} $\mathrm{Ri}\cor{\leq}1/4$) but for inertial waves has been derived in the context of a local model of cylindrical differential rotation. They demonstrate that over-reflection and over-transmission are also possible for inertial waves in a similar three layer model, which opens the way to possible shear instabilities (though their study did not investigate these instabilities directly). In Appendix \ref{sec:appendix1}, we demonstrate that this criterion, based on an inviscid analysis of free inertial waves with cylindrical rotation, may predict the stable regime for all zonal flows investigated in this study. In such case, we expect waves to be strongly damped at corotation, which would rule out shear-type instabilities. However, the derivation of this criterion in Appendix \ref{sec:appendix1} and \citet{AP2021} does not take into account viscous dissipation and tidal forcing, which could modify the criterion. For gravito-inertial waves, viscosity is able to destabilise stratified shear flows and allow  instabilities even if the Richardson number is greater than $1/4$ \cite[e.g.][]{ML1988}. For inertial waves, \cite{BR2013} and \cite{GM2016} numerically observed unstable modes triggered below a certain viscosity threshold for shellular and conical differential rotation. Tidal forcing could also modify this criterion, as highlighted in Appendix \ref{sec:appendix1}, but a more detailed analysis is postponed to a future study.
In either case (whether it is a shear or ``Papaloizou-Pringle''-like instability), the mechanism leading to non-axisymmetric instability looks similar: incoming inertial waves launched from the critical latitude(s) (or elsewhere) propagate towards the corotation resonance, partly tunnel through, partly (over-)reflect and are partly damped \citep{NG1987,GN1985,L1988}. The transmitted wave reflects from the rotation axis, and the reflected wave reflects from outer boundary (or core) and back. These waves potentially sustain a positive growth of the kinetic energy where they meet near the corotation resonance (near $s=s_c$ as in Fig.~\ref{fig:corot}). 

An important point to stress about the linear simulations involving these kind of instabilities, is that the meridional circulation observed in the associated nonlinear simulations are not negligible, as assumed in Eq.~(\ref{eq:mom_sheared}). In the snapshots showing the $\varphi-$average of the nonlinear velocity components for $\omega=0.8$ in Fig.~\ref{fig:om0p8_flows}, the azimuthal average of the radial velocity $\langle u_r\rangle_\varphi$ is not negligible anymore compared to the zonal flow $\langle u_\varphi\rangle_\varphi$, especially close to the corotation resonance, though it is still weaker in amplitude. This also demonstrates that  radial (vertical) angular momentum flux is efficiently deposited at the corotation resonance and contributes to the creation of a strong radial meridional flow there\footnote{This also has been derived and observed in the preliminary study of V. Skoutnev  et al., as part of the Kavli Summer Program in Astrophysics 2021 (\url{https://kspa.soe.ucsc.edu/sites/default/files/KSPA_VSkoutnev.pdf}).}. We also observed strong meridional (and especially radial) flows close to the corotation resonance in the nonlinear simulations for $\omega=0.7$ and $\omega=0.9$ (not shown in this study). Whether these meridional flows are responsible for the departure of the nonlinear dissipation from linear predictions in Fig. \ref{fig:span_spec_diss} is an open question. We emphasise however that no strong meridional flows have been observed for the nonlinear simulations for $\omega=0.192$ and $\omega=0.2$ where a departure in the dissipation rate is also observed. In these simulations, the corotation resonance is extremely close to the rotation axis (see Fig.~\ref{fig:dOms}) and does not appear to play an important role in the redistribution and dissipation of inertial wave kinetic energy, as shown in the meridional snapshots and averages of $u_r$ and $u_\varphi$ displayed in Fig.~\ref{fig:om0p2_snap} for $\omega=0.2$. This is presumably because the inner critical latitude is close to the equatorial plane, from which the waves propagate in almost vertical shear layers and deposit their energy preferentially close to the equator. 

We finally compute the location of the critical latitudes which differ from the uniform rotation case when the zonal flows are strong.
The critical latitudes $\theta$ for which rays (inertial wave shear layers) are tangent to the shell's inner core and outer surface satisfy the equation in the first quadrant in the meridional plane
\begin{equation}
    \frac{\mathrm{d}z}{\mathrm{d}s}=-\frac{s}{z},
    \label{eq:caracritlat}
\end{equation}
at $r=\sqrt{s^2+z^2}=\alpha R$ or $R$, respectively. Moreover, following \citet[Eq. (3.8)]{BR2013}, the paths of characteristics for inviscid and free inertial waves are defined here by:
\begin{equation}
    \frac{\mathrm{d}z}{\mathrm{d}s}=\pm\sqrt{\frac{\kappa^2(s)}{\sigma^2(s)}-1},
    \label{eq:carac}
\end{equation}
with $\kappa=\sqrt{\Phi}$ the epicyclic frequency (commonly used in astrophysical disks).
Combining Eqs.~(\ref{eq:caracritlat}) and (\ref{eq:carac}), we obtain the expression for the critical latitudes in the first quadrant:
\begin{equation}
    \sin\theta=\frac{\sigma(s)}{\kappa(s)},
    \label{eq:critlat}
\end{equation}
at $r=\alpha R$ for the inner shell and $r=R$ 
for the outer shell, with $\theta$ measured from the equator, where $z=\alpha R\sin \theta, s=\alpha R \cos\theta$ describes the inner critical latitude (and $z=R\sin \theta, s=R \cos\theta$  the outer critical latitude). 
\begin{figure*}
    \centering
    \includegraphics[width=\columnwidth]{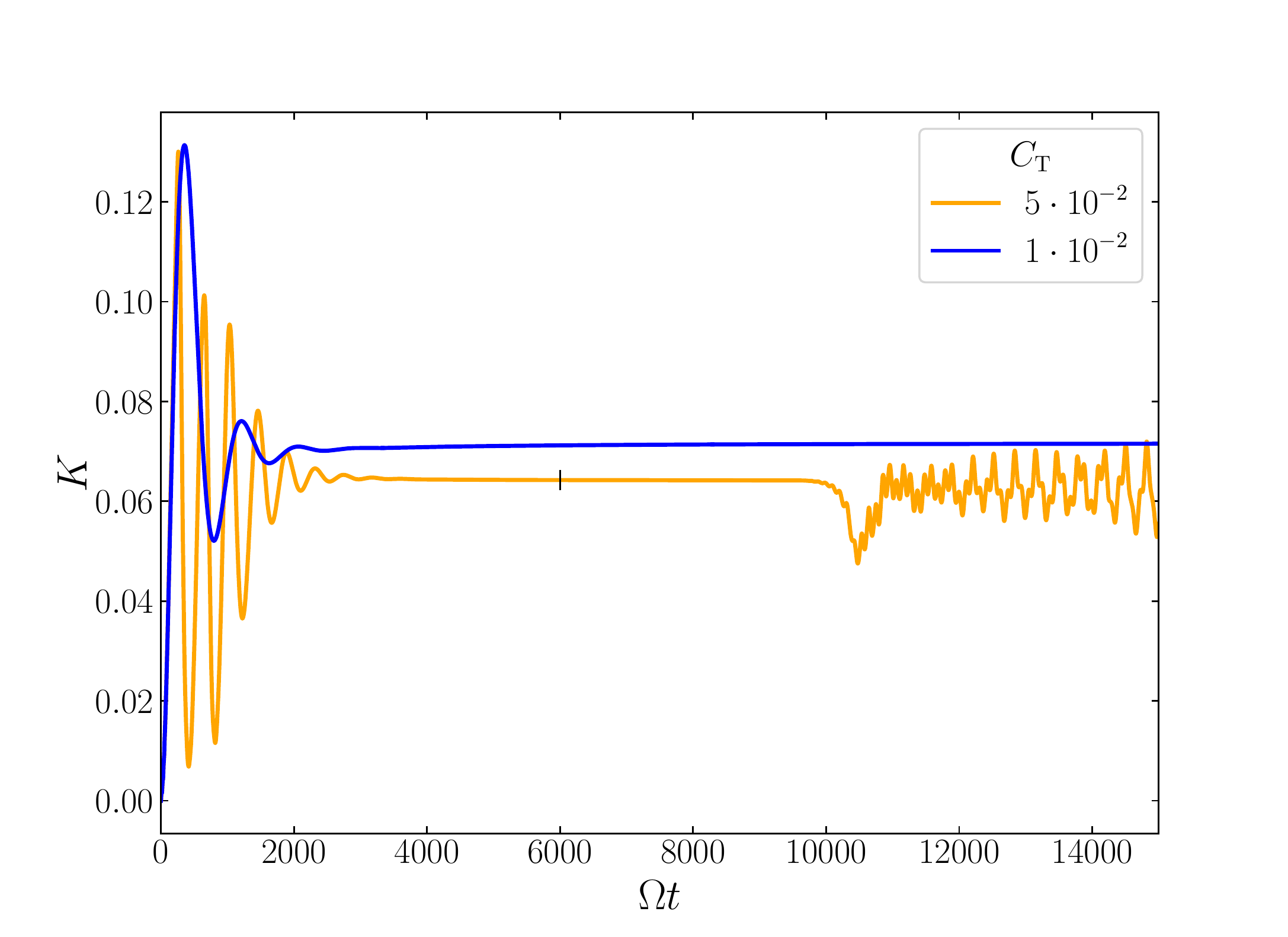}
    \includegraphics[width=\columnwidth]{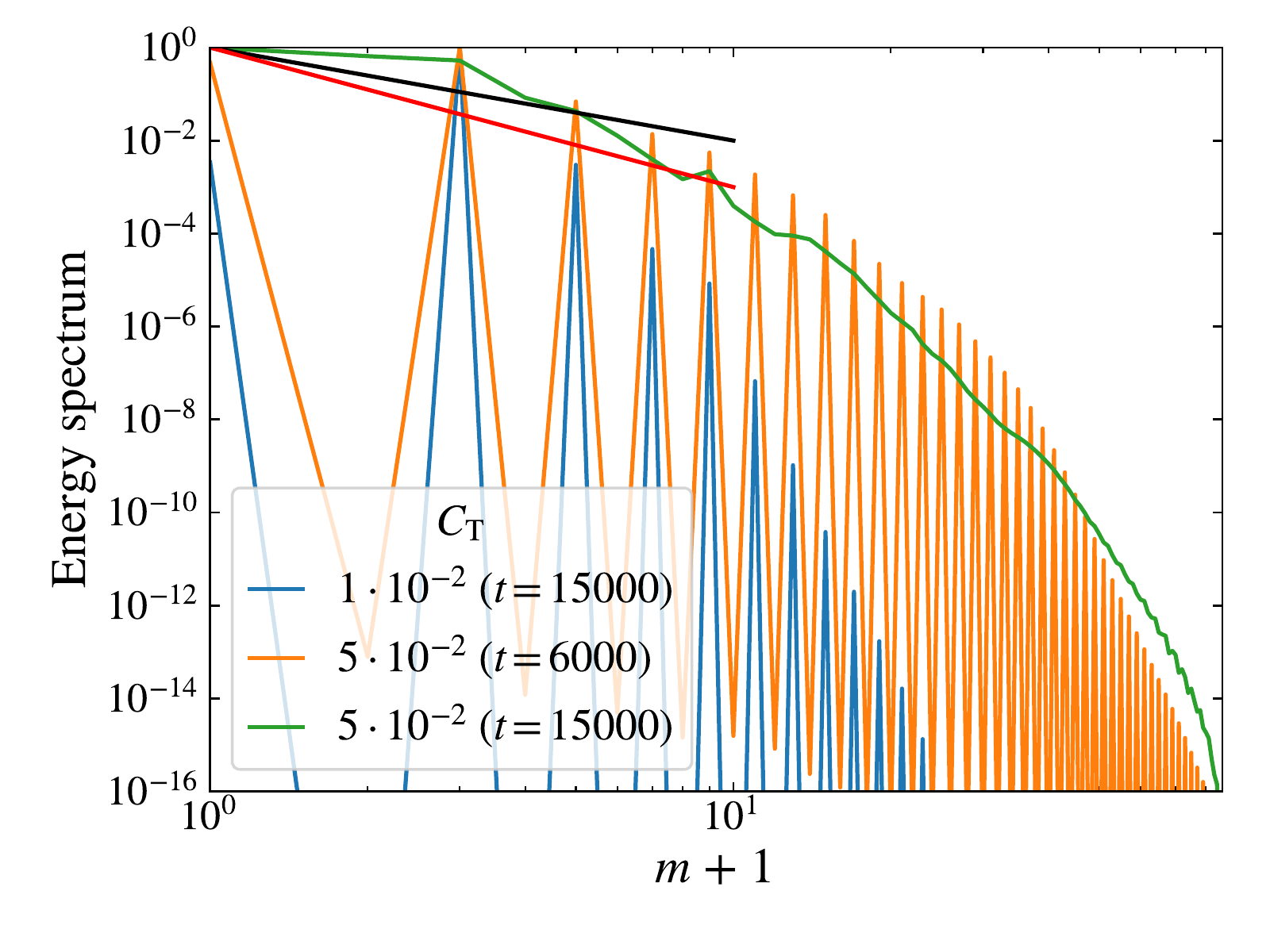}
    \caption{\textit{Left:} Kinetic energy versus time $\Omega t$ in two nonlinear simulations with different forcing amplitudes $\Ct$ with $\omega=0.2$. \textit{Right:} Normalised (to the maximum \cor{value for each simulation}) kinetic energy spectrum in terms of the azimuthal order $m+1$ for $\omega=0.2$ for two tidal amplitudes $\Ct$, at different times. \cor{The $m^{-3}$ (red) and $m^{-2}$ (black) scaling laws, found for example in numerical simulations of \citet[for elliptical instability with strong geostrophic flows]{BL2013} and \citet[for ``inertial wave turbulence"]{LRF2020}, have been plotted for reference, although neither is a particuarly good fit to our data. Note that odd $m$ values of the kinetic energy for the orange and blue curves are of the order of numerical errors or lower (less than $10^{-12}$), so their contributions are negligible and mostly not shown.}
    }
    \label{fig:kinspec}
\end{figure*}
One should note that for weak amplitudes of the zonal flow and shear, $\sigma\simeq\omega$, $\kappa\simeq2\Omega$, i.e.~$\Phi\simeq4$ in our time unit, as we can observe in Fig. \ref{fig:rayleigh_crit} and \ref{fig:dOms} (it concerns notably the outer critical latitudes), and  Eq. (\ref{eq:critlat}) tends towards the classical definition of the critical latitude with uniform rotation $\sin\theta=\omega/2\Omega$.
For strong enough zonal flows, the difference between Eq. (\ref{eq:critlat}) and the latter definition is quite important (note also that inner and outer critical latitudes are the same with uniform rotation). For example, for a tidal forcing frequency $\omega=0.8$, $\theta\simeq23.6\degree$ with uniform rotation, while the inner critical latitude is $\theta\simeq39.6\degree$ and the outer critical latitude is $\theta\simeq20.4\degree$ using Eq. (\ref{eq:critlat}) derived with the zonal flows at this frequency (note that the outer critical latitude is closer to the  critical latitude with uniform rotation for the reason explained above). These inner and outer critical latitudes nicely match the locations of the high amplitudes in kinetic energy in Fig. \ref{fig:corot} (black ticks across inner and outer surfaces show the predicted $\theta$'s).
\begin{figure*}
    \centering
    \includegraphics[width=0.33\textwidth]{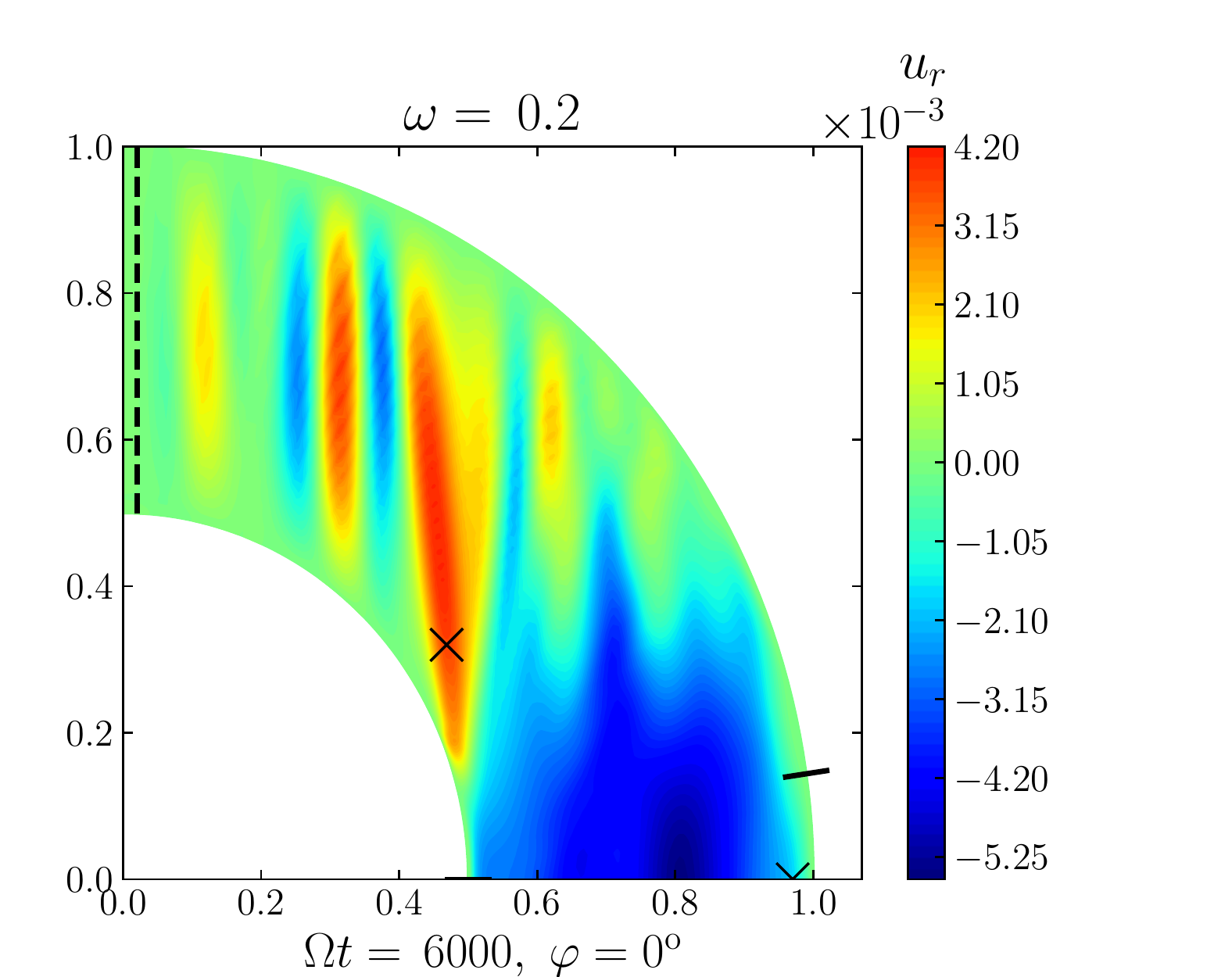}
    \includegraphics[width=0.33\textwidth]{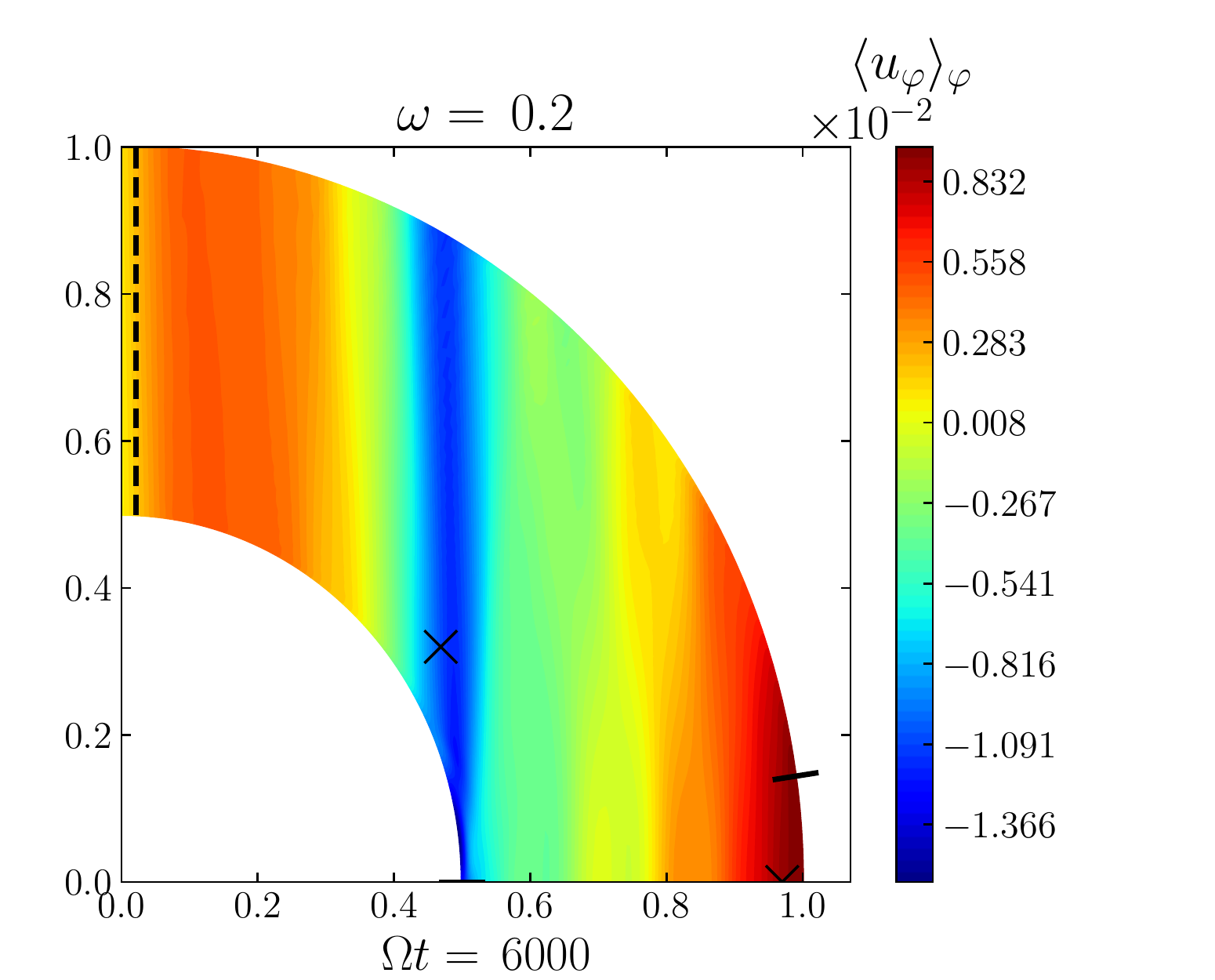}
    \includegraphics[width=0.33\textwidth]{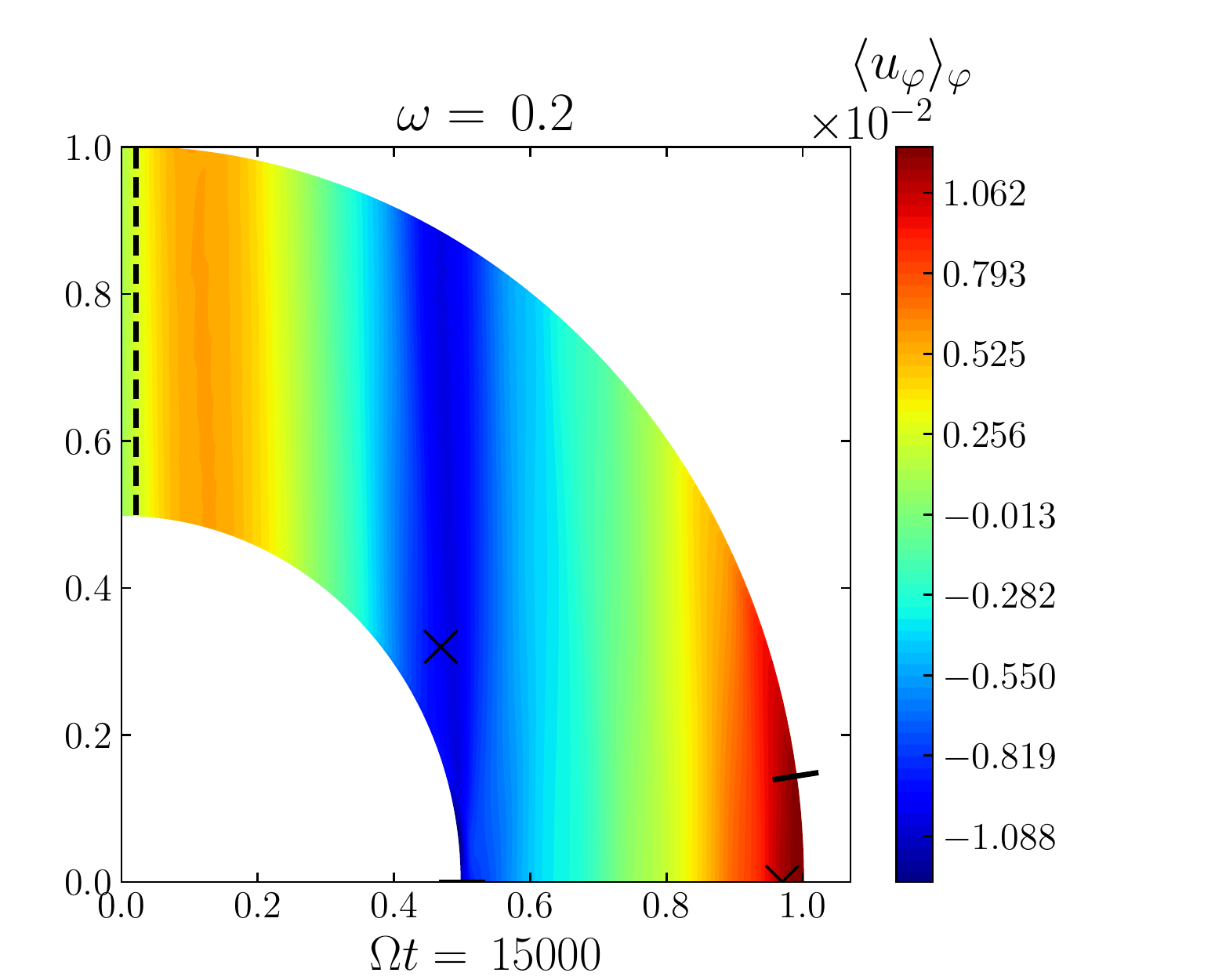}
    \caption{Snapshots of one quadrant for the nonlinear simulation with $\omega=0.2$ and $\Ct=5\cdot10^{-2}$. Critical latitudes and the corotation resonance are indicated as in Fig. \ref{fig:corot} (the inner critical is almost in the equatorial plane); crosses are the locations of the points where the Fourier transform of $u_r$ has been performed (in the left panel of Fig. \ref{fig:fft}). \textit{Left:} $\varphi$-slice of the radial component of the \cor{wavelike} velocity $u_r$ at early time. \textit{Middle:} $\varphi$-average of the azimuthal component of the  \cor{wavelike} velocity $\langle u_\varphi\rangle_\varphi$ at the same time. \textit{Right:} $\varphi$-average of the azimuthal component of the \cor{wavelike}  velocity $\langle u_\varphi\rangle_\varphi$ at a later time. }
    \label{fig:om0p2_snap}
\end{figure*}
\begin{figure*}
    \centering
    \includegraphics[width=\columnwidth]{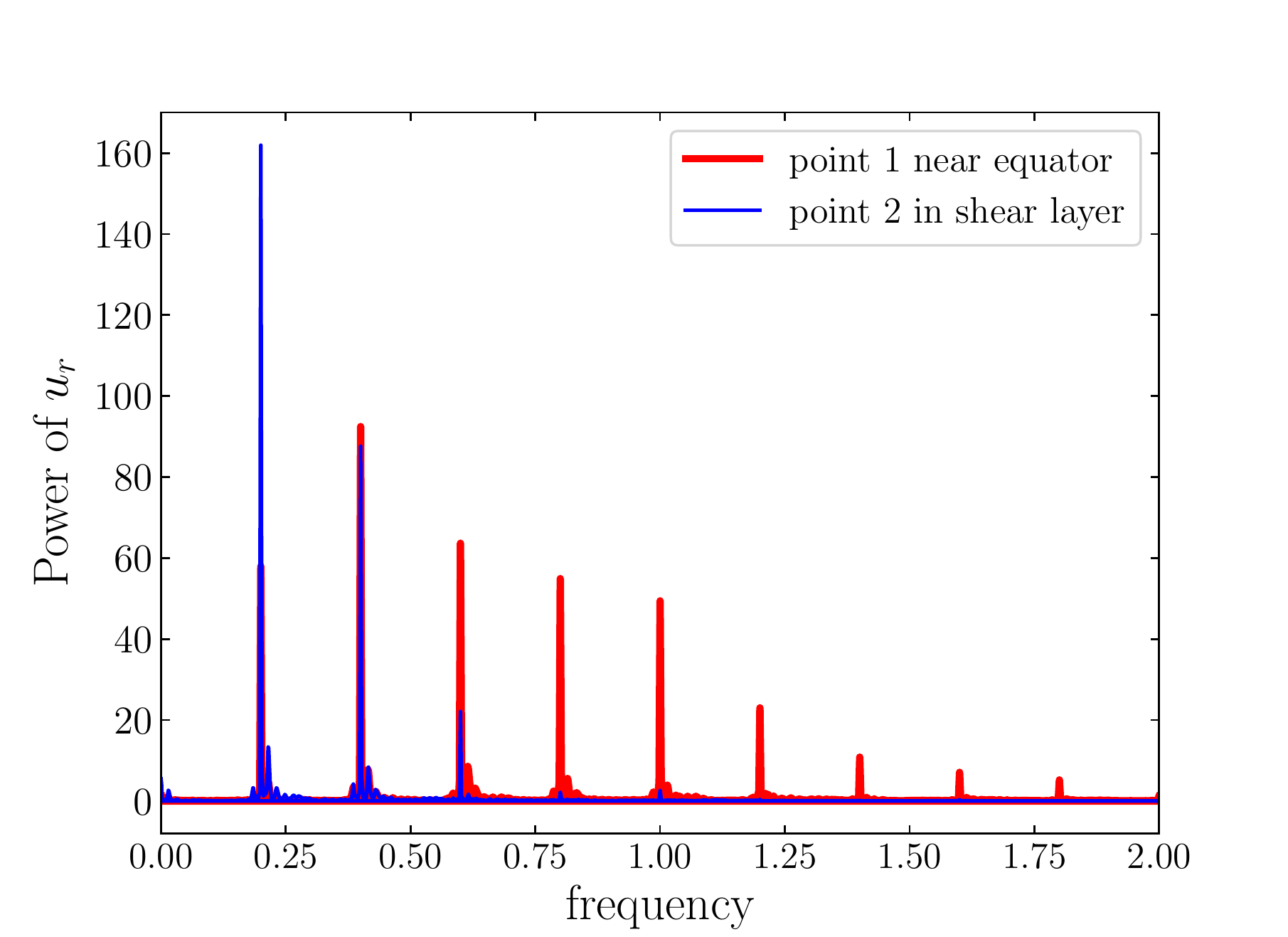}
    \includegraphics[width=\columnwidth]{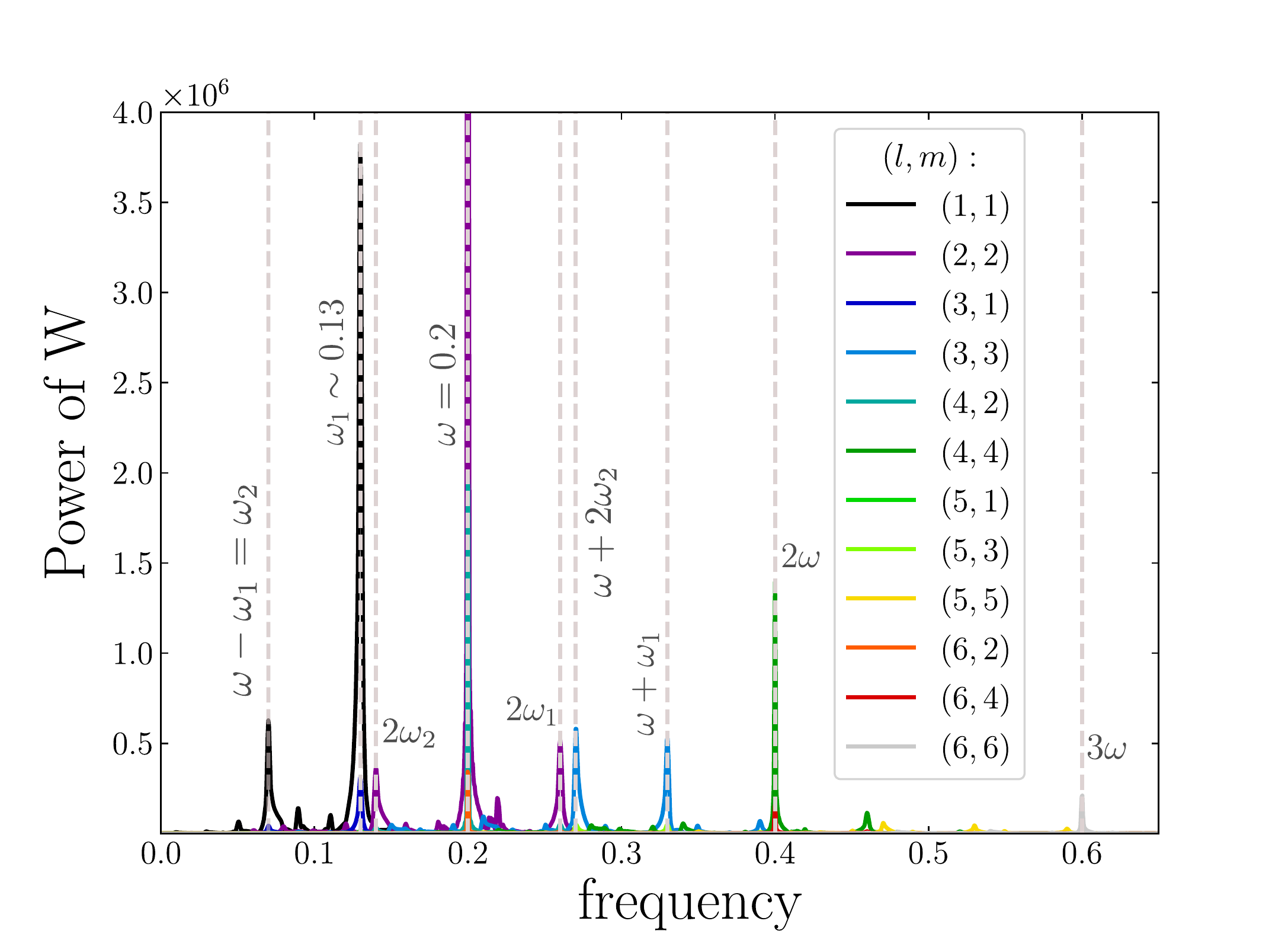}
    \caption{\textit{Left:} Fourier transform of the radial component of the nonlinear wavelike velocity $u_r$ for two points in the shell indicated by crosses in Fig.~\ref{fig:fft}  and for the time range $t\leq6000$.
     \textit{Right:} Fourier transform of the poloidal velocity potential $W$ for the whole time range $t\leq15000$, with azimuthal order $m$ and degree $l$ indicated in colour. Dominant frequencies excited are indicated by dashed grey lines and labelled. 
    }
    \label{fig:fft}
\end{figure*}
\subsection{Parametric instability of inertial waves}
In this section we focus on the nonlinear simulation with $\omega=0.2$ and $\Ct=5\cdot10^{-2}$. After an early transient phase, the kinetic energy of this simulation reaches a steady state, and then an instability sets in producing oscillations with several different periods for $t\gtrsim10000$, as we can see on the left panel of Fig. \ref{fig:kinspec}. To understand what is happening here, we analyse the azimuthal kinetic energy spectrum in the right panel of Fig. \ref{fig:kinspec} as a function of azimuthal order $m$ at different times (before and after the oscillations start), which we also compare with the $\Ct=10^{-2}$ simulation.
When $\Ct=10^{-2}$, $m=2$ (of the initial tidally-forced waves) is the dominant component in the energy spectrum, but when $\Ct=5\cdot 10^{-2}$, the $m=0$ component is of the same order of magnitude (both at early and late times), implying a strong zonal flow, and $m=1$ is also strong at late times. 

We examine in further detail the Fourier transform of the nonlinear radial velocity $u_r$ (after applying a Hamming window function) at two points inside the shell (indicated by crosses in Fig. \ref{fig:om0p2_snap}): one in the equatorial plane where the amplitude of the zonal flow is strong, the other in a shear layer directly emerging from the inner critical latitude. This is done in the left panel of Fig.~\ref{fig:fft} at the early time indicated. As expected, the contribution of the tidal forcing at $\omega=0.2$ is dominant for the point inside the shear layer. Interestingly, superharmonics of the frequency $\omega$ are also excited before the instability is triggered, with particularly high power in the region where the zonal flow is strong (point 1) compared to inside the shear layer (point 2). These oscillations have a frequency $n\omega$ and azimuthal order $m n=2n$ with $n\leq10$, a positive integer which must satisfy $n\omega\leq2$ in our time units, corresponding to the upper limit of propagation of inertial waves. We also observe superharmonics in the simulation for $\Ct=10^{-2}$. Superharmonics have been clearly reported for gravity waves \citep[e.g.][]{BS2020,BP2021,ICB2022}, but have not been observed in many prior studies of inertial waves, though they have been predicted analytically, and observed, for instance in \cite{BT2018}. One reason could be the limited frequency range allowed for inertial waves, imposing $|\omega|\ll2\Omega$ to observe multiple higher harmonics of the main frequency $\omega$. 

In the right panel of Fig.~\ref{fig:fft}, we perform a Fourier transform (applying again a Hamming window) of the poloidal velocity potential $W(l,m)$ at the mid-shell for the whole time range, to explore the properties of the waves (frequency, $m$ and degree $l$) excited after the instability becomes important for $t\geq10000$. In addition to the primary wave with a frequency $\omega$ and its (self-excited) superharmonics, we also observe a strong peak at the frequency $\omega_1\simeq0.13$ with wavenumbers $(l,m)=(1,1)$, along with a number of even $l+m$ modes (due to symmetry; with odd $l+m$ being excited for the toroidal velocity potential). This provides clear evidence for triadic resonances \citep[theorised and  observed for inertial waves in e.g.][]{K1999,BT2018} since all of the excited modes can be recovered using only linear combinations of $\omega$ and $\omega_1$, and their corresponding azimuthal orders and degrees. In particular, we have 
\begin{equation}
    \begin{aligned}
    &\omega_\mathrm{a}\pm\omega_\mathrm{b}=\omega_\mathrm{c},
    &m_\mathrm{a}\pm m_\mathrm{b}=m_\mathrm{c},\ \text{ and }\     l_\mathrm{a}\pm l_\mathrm{b}=l_\mathrm{c},
    \end{aligned}
\end{equation}
with for example a parent primary wave $\omega_\mathrm{a}=\omega$ with $(l_\mathrm{a},m_\mathrm{a})=(2,2)$, interacting with two secondary/daughter waves, one with $\omega_\mathrm{b}=\omega_1$ with $(l_\mathrm{b},m_\mathrm{b})=(1,1)$, and another with $\omega_\mathrm{c}=\omega-\omega_1=\omega_2$ with $(l_\mathrm{c},m_\mathrm{c})=(l_\mathrm{a}-l_\mathrm{b},m_\mathrm{a}-m_\mathrm{b})=(1,1)$. In this case, we have a parametric instability of the primary wave involving these three components, as is shown in the right panel of Fig. \ref{fig:fft}. It should be noted that there are also superharmonics of these daughter waves (for example with frequencies $2\omega_2$ and $2\omega_1$). 

\begin{figure*}
    \centering
    \includegraphics[width=\textwidth]{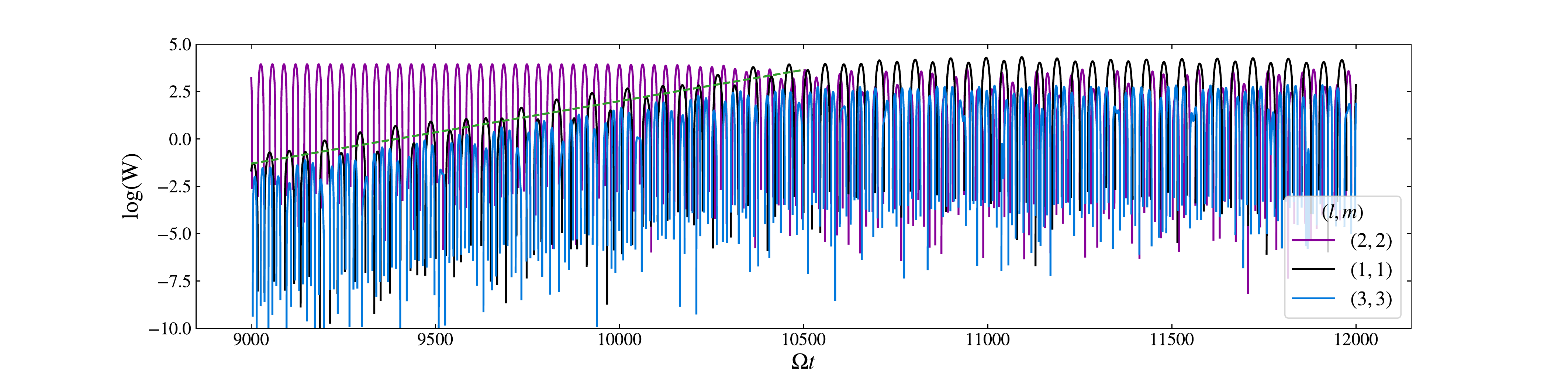}
    \caption{Time series of the poloidal velocity potential $\log(W)$ (base e) for different modes (note that the same $(l,m)$ component can relate to different frequencies as in Fig. \ref{fig:fft}). The dashed green line illustrates the exponential growth of the modes of components $(l,m)=(1,1)$ and $(l,m)=(3,3)$. This indicates that the primary tidal wave is unstable to a slowly growing parametric instability.}
    \label{fig:tserie_modes}
\end{figure*}

The exponential growth of the parametric instability  involving these $m=1$ modes is clear from Fig. \ref{fig:tserie_modes}, which shows the time series of the logarithm of the poloidal velocity potential for $(l,m)=(1,1)$, $(2,2)$ and $(3,3)$. Indeed, the dashed green curved with linear slope $\sim0.033$ indicates that the secondary waves with wavenumbers $(1,1)$ and $(3,3)$ are exponentially growing at the expense of the primary wave by draining its energy, until a saturation level is reached after $\Omega t\gtrsim10500$. Thanks to viscous dissipation, the simulation reaches a quasi-steady state, similarly to \cite{CL2022}, where an inertial mode also triggers wave-wave interactions and parametric instabilities in their local simulations of protoplanetary disks. 

Parametric instabilities typically transfer energy to waves excited close to reflection points, with lower frequencies and shorter wavelengths, as observed in nonlinear simulations of inertial wave attractors in 2D Cartesian geometry \citep{JO2014}. It is possible that the mode of frequency $\omega_1$ is related to this phenomenon. \cor{As the Ekman number is lower (for which nonlinearities are expected to be enhanced), the frequency of the secondary could become closer to exact subharmonics with $\omega/2$ for the fastest growing modes \citep[though this has not always been found in related problems in spherical geometry, e.g.][where $\omega/3$ and $2\omega/3$ have been obtained instead for moderate Ekman numbers, similar to our case]{Linetal2015}}, as obtained in the inviscid theoretical parametric subharmonic instability study in the aforementioned paper, and as also expected for the related elliptical instability \citep[e.g.][which is usually analysed for simple elliptical flows]{K2002,B2016}. This kind of instability was not observed in \fb, perhaps because of the strong zonal flows that dominated their simulations.

It is clear from the energy spectrum in the right panel of Fig. \ref{fig:kinspec}, that kinetic energy is cascaded towards high values of azimuthal wavenumbers $m$ (when comparing blue and orange curves at different tidal forcing amplitudes). This is initially due to superharmonics of the main frequency $\omega$, that correspond with even $m$ (orange), and then to odd values of $m$ (green curve) presumably due to the triadic resonances with secondary waves with $m=1$, which subsequently initiate an inertial wave turbulent cascade \citep[e.g.][]{LRFBLB2017}. 

Finally, we emphasise that undertaking the eigenvalue problem for this differentially rotating shell\footnote{Note that the eigenvalues are known to  drift from the uniformly rotating case \citep[e.g.][]{BT2018} depending on the strength of the shear and also with the Ekman number \citep[e.g.][]{BC2022}.}, could help us to understand which modes are likely to be excited and cause the particular instability we observe.
\subsection{Scaling laws for tidally-generated differential rotation: effects of the viscosity}
\begin{figure}
    \centering
    \includegraphics[width=0.49\textwidth]{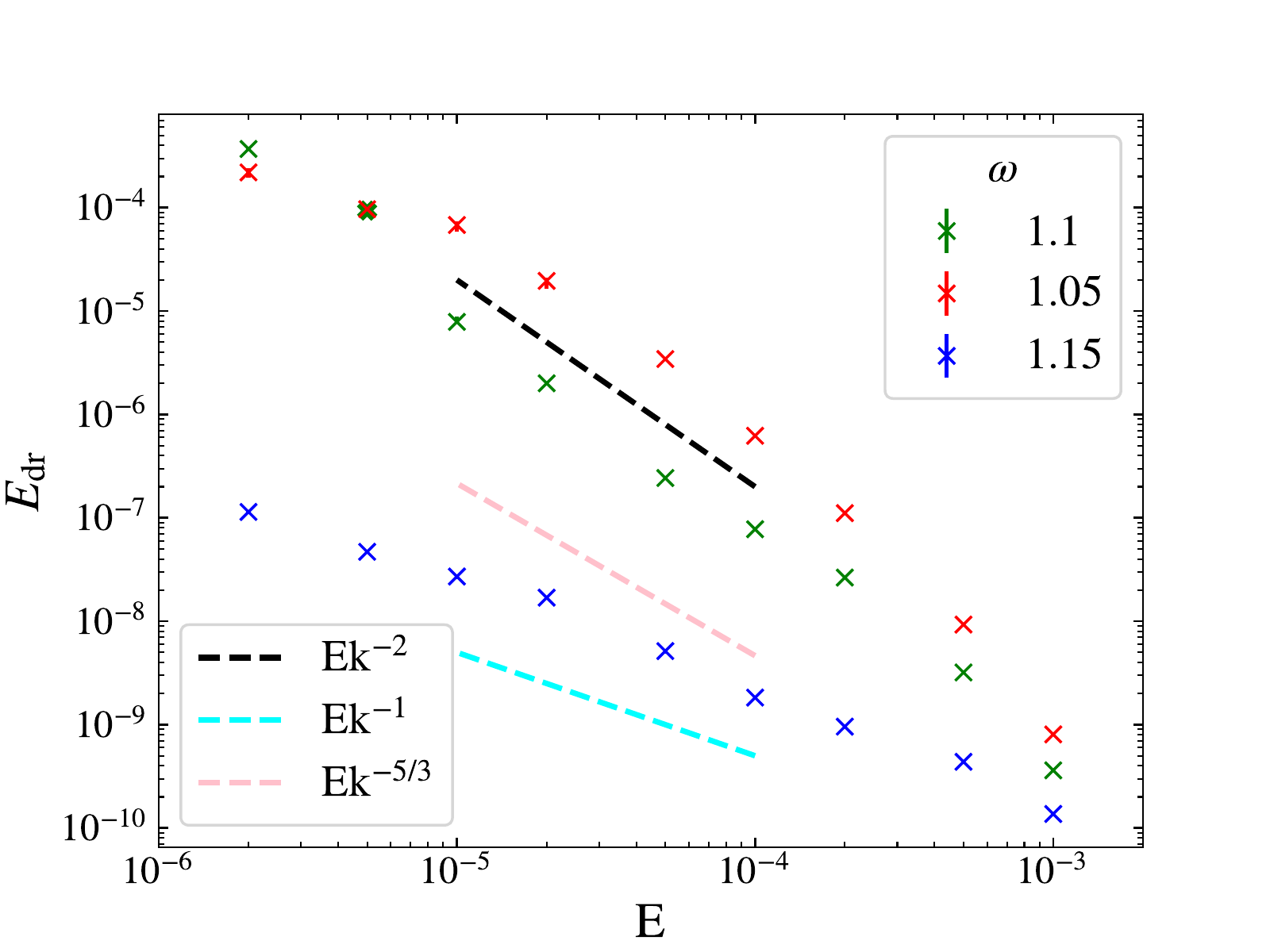}
    \caption{Energy inside the differential rotation $E_\mathrm{dr}$ as a function of the Ekman number $\Ek$ for various nonlinear simulations with three tidal forcing frequencies $\omega$ (with $\Ct\approx9\cdot10^{-3}$).
    Simulations are run until an overall steady state is reached and the values taken are the mean in the last $3000$ rotation times (and errorbars correspond to minimum and maximum values in that range, being non-negligible only for one point). Some scaling laws have been added as dashed lines for comparison.
    }
    \label{fig:edr_ek}
\end{figure}

In this section we analyse how the energy inside the tidally-generated differential rotation, $E_\mathrm{dr}$, varies as we vary the viscosity/Ekman number $\Ek$. In fig. \ref{fig:edr_ek}, we display $E_\mathrm{dr}$ in terms of $\Ek$ when an overall steady state is reached in our nonlinear simulations for the three frequencies studied in Sect. \ref{sec:simu_nl}. For the simulations with $\Ek\leq5\cdot10^{-6}$, the spatial resolution is increased to $l_\mathrm{max}=170$ (so $n_\varphi=512$) and $n_r=161$. 

Looking at $E_\mathrm{dr}$ for $\omega=1.05$ and $1.1$ (for which the linear dissipation is close to a resonant peak), we observe that $E_\mathrm{dr}\propto \Ek^{-2}$, while for $\omega=1.15$ (for which the linear dissipation is at a trough), the scaling is less steep, closer to $E_\mathrm{dr}\propto \Ek^{-1}$. At low Ekman numbers, such as $\Ek\leq5\cdot10^{-6}$, there is a break in the $\Ek^{-2}$ law for $\omega=1.05$. At such a low viscosities, the amplitude of prograde zonal flows starts to become large (for $\Ek=5\cdot10^{-6}$) or even dominant (for $\Ek=2\cdot10^{-6}$) close to the rotation axis for the $\omega=1.05$ and $1.15$ cases, contrary to the ``equatorially" dominant flow we found for the same frequencies in Fig. \ref{fig:vp_nonlinear} for $\Ek=10^{-5}$. This is presumably explained by the development of a corotation resonance close to the pole, since the amplitude of the zonal flow is anti-correlated with the viscosity. For the $\omega=1.1$ case, since the zonal flow is already strong near the pole, the presence of the corotation resonance may have amplified it, which is what we suspect for $\Ek=5\cdot10^{-6}$. 

It is difficult to determine theoretically a universal scaling law, given the strongly frequency-dependent predictions from linear theory. However \cite{T2007} argued for an upper threshold $E_\mathrm{dr}<\Ek^{-2}\ell^{-2}$, which predicts $E_\mathrm{dr}\propto \Ek^{-2}$ as we have observed if $\ell=O(1)$, but is steeper than $\Ek^{-2}$ if we naively use $\ell\propto\Ek^{1/3}$ here (or indeed any positive power of $\Ek$). Following similar lines to Tilgner's analysis, who distinguished a momentum equation for the $m=2$ tidal inertial waves $\bm u_1$ (his Eq. 3) and one for the axisymmetric flow $\bar{\bm u}_2$ generated by their nonlinear self-interaction (Eq. 5), we can derive $\Ek \hat{\bm\varphi}\cdot\bn^2\bar{\bm u}_2\sim \hat{\bm\varphi}\cdot (\bm u_1 \cdot \bn) \bm{u}_1$,
with $\hat{\bm\varphi}$ the unit vector in the azimuthal direction.  
If we assume the tidal waves 
to have a typical (transverse) lengthscale $\ell$ and the zonal flow to have a radial scale $L$, this would suggest the zonal flow magnitude to scale as $\bar{u}_2\sim u_1^2 L^2/(\Ek \ell)$, in terms of the typical velocity magnitude of the waves. Hence $E_\mathrm{dr}\sim \langle \bar{u}_2^2\rangle\sim \langle u_1^4L^4/(\Ek^2 \ell^2)\rangle$, which predicts $E_\mathrm{dr}\propto \Ek^{-2}$ if $u_1$, $L$, $\ell$, and the volume filling fraction (VFF) of the waves combine to produce an $O(1)$ number in the above scaling. 
On the other hand, if we assume $u_1\sim \Ek^{-1/6}, L\sim\ell\sim \Ek^{1/3}$ and VFF also scales as $\Ek^{1/3}$, we find the less steep scaling $E_\mathrm{dr}\propto \Ek^{-5/3}$ (reducing to $E_\mathrm{dr}\propto \Ek^{-2}$ under the same assumptions if VFF is $O(1)$). If we instead focus on flows generated near the critical latitude, and consider $u_1\sim \Ek^{-1/5}, L\sim\ell\sim \Ek^{2/5}$ and VFF scaling like $\Ek^{2/5}$ we would obtain  $E_\mathrm{dr}\propto \Ek^{-8/5}$.
Different assumptions would lead to alternative scalings, but these are hard to justify a priori.

The Ekman number scalings we have observed are very different from the zonal flows generated by tidal forcing in a full sphere in the experiments (with a deformable no-slip boundary) of, for example, \citet[][]{ML2010}, where the zonal velocity scales as $\epsilon^2 \Ek^{-3/10}$, or for those produced by libration-driven inertial waves, as studied in \cite{CV2021} and \cite{LN2021}, where it instead scales as $\epsilon^2 \Ek^0$ or $\epsilon^2 \Ek^{-1/10}$, respectively (though these also result from self-interaction, so the zonal velocity also scales as $\epsilon^2$). 
We underline that the best fitting laws here seem quite different from the ones emerging in \fb~ in $\Ek^{-3/2}$ and $\Ek^{-1/2}$, presumably because of the non-wavelike nonlinearities involved in their work. It is difficult to predict whether the scaling laws we have observed would still hold for solar-like and Jupiter-like values of the Ekman number, particularly because the corotation resonance at $\delta\Omega_s=\omega/2$ near the rotation axis starts to trigger instabilities for $\Ek=10^{-6}$ for the three frequencies studied in this section. These instabilities appear to flatten the dependence, and could even cause $E_\mathrm{dr}$ to become independent of $\Ek$ below a critical value. Further work is required to understand these results theoretically and to explore a wider range of parameters.
\section{Importance of nonlinear effects for exoplanetary systems}
\label{sec:astro}
\begin{figure}
    \centering
    \includegraphics[width=\columnwidth]{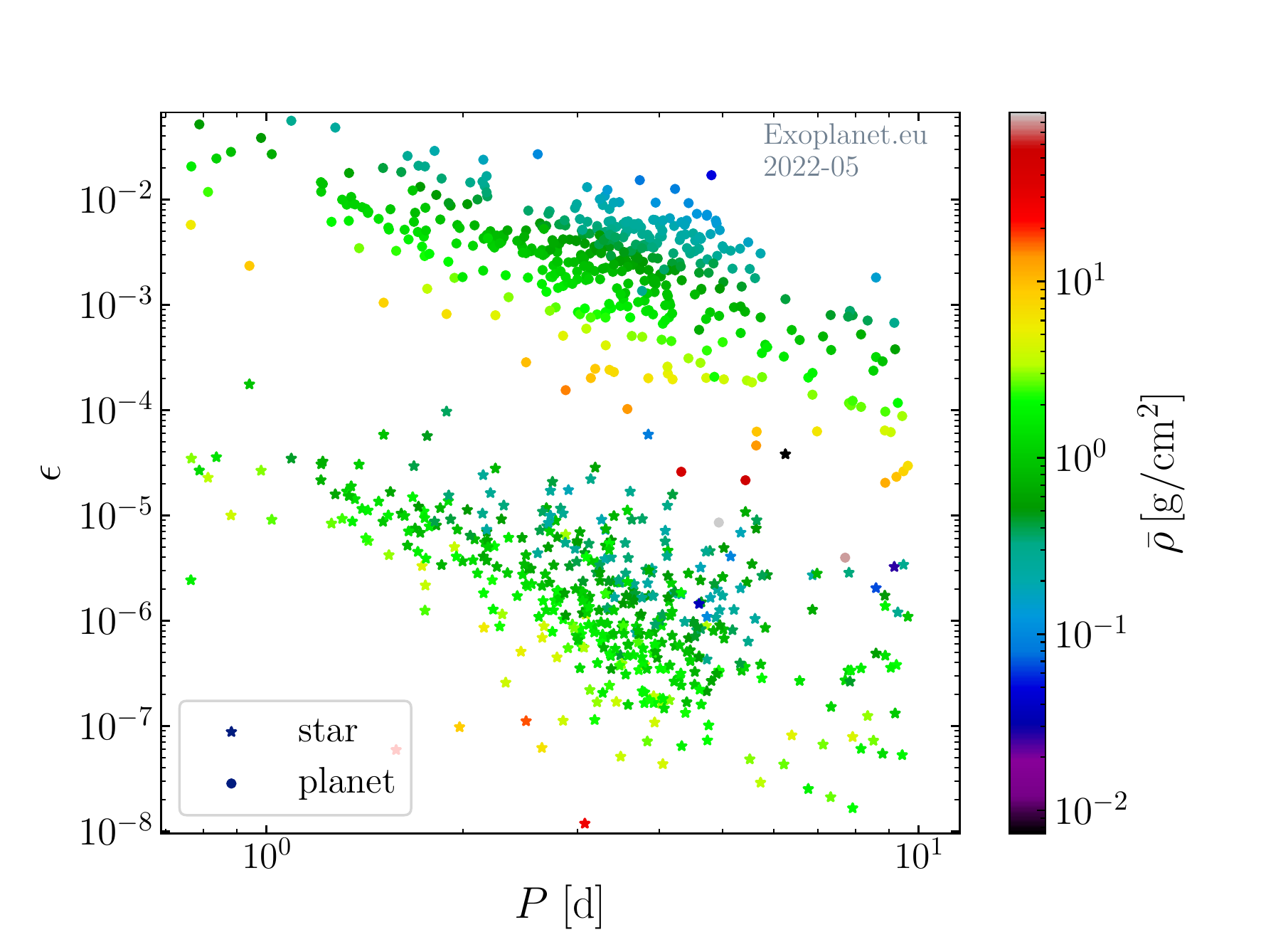}
    \caption{Stellar and planetary tidal amplitudes $\epsilon$ in terms of the orbital period $P$ of the planet in days, and the mean density $\bar{\rho}=M_1/(4/3\pi R^3)$ of the star or planet in colour. Only compact systems satisfying $P<10$ d, and $M_\mathrm{p}/M_\star>10^{-4}$ (bodies with $\rho>100\,\mathrm{g/cm^3}$ have also been removed) have been selected (a significant fraction of these are Hot Jupiter systems), and radii, masses, periods and semi-major axes have been taken from the online database \url{http://exoplanet.eu/}. 
    }
    \label{fig:eps}
\end{figure}

In parallel with this theoretical modelling, it is important to consider which astrophysical systems are likely to be affected by these nonlinear effects. 
Using the time and length scales defined in Sect. \ref{sec:bal}, the importance of 
wavelike nonlinearities can be quantified by:
\begin{equation}
    \frac{(\uw\cdot\bn)\uw}{\partial_t\uw}\sim \frac{u_\mathrm{w}}{\omega \ell}\sim \Ct E^{-\alpha-\beta},
    \label{eq:scal}
\end{equation}
where $-\alpha-\beta=-1/2$ or $-5/12$ for shear layers, and $-\alpha-\beta=-3/5$ for the critical latitudes. Note that as hypothesised in Paper I, wave breaking is expected when the RHS is $O(1)$, which is more likely to happen at the critical latitudes for small $\Ek$, perhaps preventing the propagation of shear layers from these locations \citep[see also][]{GL2009}.
We confirm that nonlinear effects are important for both shear layers and critical latitudes according to Eq. (\ref{eq:scal}) in most of our simulations (i.e. for $\Ek\lesssim10^{-5}$ and $\Ct\gtrsim10^{-2}$), in that the RHS is $\gtrsim O(1)$.

An estimation of Eq. (\ref{eq:scal}) in compact exoplanetary systems can be made using typical values for the stellar or planetary tidal amplitude $\epsilon$ (depending on whether the primary body in which tidal flows are investigated is the star or the planet), which we display in Fig. \ref{fig:eps}. We recall that $\Ct$ and $\epsilon$ differ only by $k_2$, which is around $1/2$ for giant gaseous planets, like Jupiter and Saturn, and also possibly for Hot Jupiters \citep[e.g.][]{GF2022}. For a $n=1$ non-rotating polytrope, a similar value is found, but can vary with rotation to between $0.3$ and $1.2$ \citep{DL2022}. The planetary tidal amplitude parameter varies in Fig. \ref{fig:eps} between $10^{-6}$ and $10^{-1}$, with maximum values reached for the Hot Jupiters WASP-12 b and WASP-19 b at $\epsilon\approx5\cdot10^{-2}$. Using a typical value for the kinematic Ekman number of about $\Ek=10^{-18}$, and $\Ct\approx3\epsilon/2$, nonlinear effects for wavelike tidal flows are predicted to matter for every gaseous planet according to Eq. (\ref{eq:scal}) for any of our choices of exponent $-\alpha-\beta$. 

The stellar tidal amplitude parameter varies between $10^{-8}$ and $10^{-4}$, with a maximum value for the host star WASP-19, in which $\epsilon\approx2\cdot10^{-4}$. Assuming a Sun-like  kinematic Ekman number of about $\Ek=10^{-12}$ (probably even lower at the base of the convective envelope), nonlinear effects are significant for only a fraction of low-mass stars (preferentially those with $\epsilon\gtrsim10^{-5}$), depending on the value of $-\alpha-\beta$.
We point out that the results in our paper may be more relevant to fast rotating bodies like Hot Jupiters, which possibly exhibit cylindrical-like differential rotation in their upper atmospheres \citep[e.g.][]{GW2013}, than to host stars for which we generally expect a more conical-like differential rotation in the convective envelope \citep[e.g.][]{BB2018}, like in the Sun. Although extrapolating $E_\mathrm{dr}$ for much lower Ekman numbers in Fig. \ref{fig:edr_ek} suggests strong shears and thus a much higher impact on the tidal dissipation (than observed e.g. in Fig. \ref{fig:span_spec_diss}), these statements must be tempered by the unknown action of turbulent convective motions on tidal inertial waves, which potentially drive even stronger zonal flows, and the presence of magnetism and fluid instabilities that could mitigate them. 

If we assume that the action of convection on tidal flows can be modelled by an eddy viscosity that behaves just like a microscopic viscosity, as done for instance to study Rossby and inertial waves in the Sun in \citet{GF2020} and \citet{FG2022}, and which is on the order of $10^{-4}-10^{-5}$ from mixing length theory,
major differences in outcome are to be expected for the scaling law Eq. (\ref{eq:scal}), and tidal dissipation estimates from our simulations may be more directly applicable.
With a turbulent eddy viscosity on the order of $10^{-5}$, like in most of our simulations, our numerical results first suggest that for all low-mass host stars and Hot Jupiters with $\epsilon\lesssim 10^{-2}$, (see Figs \ref{fig:ang}, \ref{fig:span_negom} and \ref{fig:span_spec_diss} for $\Ct=10^{-2}$), nonlinear effects have a limited effect on the tidal dissipation, modifying the frequency-dependent (and presumably frequency-averaged) tidal dissipation rates by a factor of unity only. However, for Hot Jupiters having $\epsilon\gtrsim 10^{-2}$ (depending on their Love number), and especially for ultra Hot Jupiters like WASP-12 b, WASP-19 b, WASP-121 b, or HIP 65A b having $\Ct\gtrsim5\cdot{10^{-2}}$, nonlinearities could have a much greater impact on tidal dissipation rates and the generation of zonal flows, depending on the forcing frequencies in these systems, the exact value of the Ekman number inside the convective envelope, and also the aspect ratio of the shell. Further exploration, and scanning of the frequency range allowed for inertial waves, could allow us to determine if the frequency-averaged nonlinear dissipation is modified in comparison to its linear analogue \citep{O2013}.  
\section{Conclusions and discussion}
\label{sec:end}
In this paper, we have investigated the impact of nonlinearities on tidal flows in an incompressible and adiabatic spherical shell, which models the convective envelope of a low-mass star or a giant gaseous planet in a compact system. In our study, we have performed a suite of 3D hydrodynamical direct numerical simulations in spherical shell geometry for a range of tidal frequencies and amplitudes, and fluid viscosities (Ekman numbers), building upon the prior study of \cite[Paper I]{FB2014}. We show in particular that nonlinear tidal effects are likely to be important in the convective envelopes of planets and stars in Hot Jupiter systems.

Unlike in the nonlinear simulations of Paper I where the flow was forced from the outer surface through an imposed radial velocity, here we use a tidal body forcing to excite inertial waves, resulting from the residual action of the Coriolis acceleration on the equilibrium (non-wavelike) tide. We consider this to be a more realistic way to tidally excite inertial waves, and it has the advantage that it allows us to clearly identify and assess the contribution of wavelike ($\uw\cdot\bn\uw$), mixed ($\uw\cdot\bn\unw+\unw\cdot\bn\uw$), and non-wavelike ($\unw\cdot\bn\unw$) nonlinear terms in the energy and momentum balances. We demonstrate that the mixed nonlinearities generate a nonphysical radial flux through the boundaries due to our spherical (and not tidally elliptically deformed) boundaries, which is responsible for the unrealistic angular momentum evolution leading to the de-synchronisation of the body observed for some frequencies in \fb. By removing these nonlinear terms, which is justified by physical scaling arguments for astrophysical parameter regimes, 
the angular momentum is instead conserved in this model. Our model thus probes the instantaneous energy transfers between tidal inertial waves and zonal flows on times much shorter than long tidal evolutionary timescales.

Like in \fb~ and \cite{T2007}, we report the development of strong axisymmetric zonal flows describing cylindrical differential rotation in the shell. These flows are generated by the nonlinear self-interaction of tidal inertial waves in various locations in the shell: inside shear layers, and developing around critical latitudes, at the points of reflection between the wave beams and the boundaries or the rotation axis, and finally near a corotation resonance. We observe the formation of corotation resonances (also called critical layers) in our simulations, namely critical cylinders where the angular velocity matches the angular pattern speed of a wave \citep[as in the 2D linear calculations of][]{BR2013}, that form preferentially close to the rotation axis if the tidal forcing is strong or the viscosity is small. This is explained by the fact that the zonal flows in these cases are stronger, since the energy inside the differential rotation varies as $E_\mathrm{dr}\propto\epsilon^4\Ek^{-\gamma}$, where $\epsilon$ is the tidal amplitude and $\gamma$ is a positive number (typically $\gamma\in[\cor{1,2}]$, though we find indications that $\gamma\to 0$ as $\Ek\to 0$) depending on the tidal frequency. When present, these corotation resonances strongly modify the zonal flow profile and the tidal angular momentum exchanges in the system. 

We also find that the nonlinear tidal dissipation rates depart from linear predictions, with a large discrepancy when the tidal frequency is close to a resonant peak of enhanced dissipation according to linear theory. These cases are also correlated with stronger energy inside the differential rotation. However, unlike \fb, the discrepancy between linear and nonlinear tidal dissipation rates is in general somewhat less important in our simulations (we typically find less than one order of magnitude differences between the two). We demonstrate, by injecting the zonal flows resulting from nonlinear simulations into new tidally forced linear simulations as a ``background flow", that the development of cylindrical differential rotation is responsible for this discrepancy between nonlinear and linear dissipation rates in many of our simulations. This is especially the case for moderate values of the tidal forcing $\Ct=10^{-2}$ and Ekman number $\Ek=10^{-5}$. 

For higher values of the tidal forcing amplitude, or lower values of the Ekman number, we observe the emergence of complex wave/wave and wave/zonal flow interactions that we try to characterise. In particular, we identify parametric instabilities involving triadic resonances between inertial waves, which are reminiscent of the elliptical instability \citep[e.g.][albeit for a primary tidal wave with a more complicated flow than one with simple elliptical streamlines]{K2002,B2016,LRFBLB2017}. We also identify the appearance of corotation resonances, which may trigger (secondary) shear instabilities in some cases, and which lead to strong absorption of waves in other cases. These interactions contribute significantly to change tidal dissipation rates, either by absorbing or amplifying waves at corotation resonances \citep[see also][]{AP2021}, or through generating daughter waves which dissipate on smaller lengthscales \citep[e.g.][]{JO2014}, and thus redistribute energy inside the shell. Analytical investigation of nonlinear three-wave interactions in \cite{K1999} and \citet{BT2018} notably emphasise three different outcomes of nonlinear self-interaction of inertial waves that have all been observed in our model: a geostrophic $m=0$ mode, which is a quasi-axisymmetric zonal flow, superharmonics, as we observe for low frequencies compared to $2\Omega$, and triadic resonances between two daughter waves and a primary tidal wave. 
 
We have demonstrated that nonlinear effects are likely to play an important role on tidal flows, and in modifying tidal dissipation rates, particularly
when considering realistic values of atomic viscosities. If turbulent convective motions can be modelled as an effective viscosity on tidal inertial waves, and are thus the dominant contribution for the viscous dissipation term, the latter statement may only be true for the envelopes of ultra short-period Hot Jupiters. However, these statements have to be qualified since various different key processes have not been investigated in this study, including the influence of magnetism and proper turbulent convective motions, that could shape different amplitudes and profiles of the differentially-rotating background flow, the effect of compressibility, and also the size of the convective shell. The latter is likely to be important because the linear tidal dissipation prediction with uniform rotation depends strongly on this parameter, with a frequency-averaged dissipation depending on  $\alpha^5/(1-\alpha^5)$. Although this implies tidal dissipation is higher for a thin convective shell, \cite{BR2013} have shown that even for a tiny core, differential rotation is able to retain shear layers, which potentially make a huge difference in tidal dissipation rates. There remains much to be explored in this problem.
\section*{Acknowledgements}
This research has been supported by STFC grants ST/R00059X/1, ST/S000275/1 and ST/W000873/1. Simulations were undertaken using the MagIC software on ARC4, part of the High Performance Computing facilities at the University of Leeds, and using the DiRAC Data Intensive service at Leicester, operated by the University of Leicester IT Services, which forms part of the STFC DiRAC HPC Facility (\href{www.dirac.ac.uk}{www.dirac.ac.uk}). The equipment was funded by BEIS capital funding via STFC capital grants ST/K000373/1 and ST/R002363/1 and STFC DiRAC Operations grant ST/R001014/1. DiRAC is part of the National e-Infrastructure. We are grateful to T. Gastine for his help and valuable advice on the use of MagIC. We would like to thank M. Rieutord, R. V. Valdetarro and C. Baruteau for allowing us to use the linear spectral code LSB. We also thank J. Park and S. Mathis for fruitful exchanges about the paper, \cor{and the referee for a prompt, helpful and constructive report}. This research has made use of data obtained from or tools provided by the portal \url{exoplanet.eu} of The Extrasolar Planets Encyclopaedia, and of NASA’s Astrophysics Data System Bibliographic Services.
\section*{Data Availability}
The data underlying this article will be shared on reasonable request to the corresponding author.
\bibliographystyle{mnras}
\bibliography{biblio.bib} 
\appendix
\section{Inertial waves near corotation}
\label{sec:appendix1}
We recall that the linear momentum Equation (\ref{eq:mom_sheared}) for inertial waves with a background zonal flow is analogous to Equation (2.1) in \cite{BR2013} with cylindrical differential rotation. The differences are that they omit tidal forcing and work in an inertial frame. By assuming a plane wave in the vertical and azimuthal directions, we can derive a similar second-order ordinary differential equation (ODE) for the free (unforced) pressure perturbation $p_\mathrm{w}$ in the inviscid limit (as in their Appendix A). Under the short-wavelength approximation, and in the vicinity of the corotation resonance at the critical cylinder $\sc$, this ODE reads\footnote{Eq. (\ref{eq:corot_eq}) can also be derived from \citet[Eq. (3.5)]{BR2013} by taking the leading order term in $s-\sc$ at the corotation resonance at the critical cylinder $\sc$.}
\begin{equation}
    \frac{\partial^2p_\mathrm{w}}{\partial s^2}+\frac{\alpha_c^2(1+\Roc)}{\Roc^2(s-\sc)^2}p_\mathrm{w}=0,
    \label{eq:corot_eq}
\end{equation}
with $\Roc=\cfrac{\sc\partial_s\Omega_s(\sc)}{2\Omega_s(\sc)}$, which is a kind of Rossby number, and the ratio of the vertical to azimuthal wavenumbers $\alpha_c=\sc k_z/m$. We should point out that in our simulations the singularity at the critical cylinder in Eq.~(\ref{eq:corot_eq}) is regularised by viscosity, which is not taken into account here.
Eq. (\ref{eq:corot_eq}) is analogous to the second-order equation derived in \citet[Eq. (68)]{AP2021}
for free inertial waves near a corotation resonance with cylindrical differential rotation in a local model (the different sign in the numerator results from the opposite local orientation of the box). Two regimes emerge depending on the value of 
\begin{equation}
\mathcal{R}=\frac{\alpha_\mathrm{c}^2(1+\Roc)}{\Roc^2},
\label{eq:crit}
\end{equation}
in Eq. (\ref{eq:corot_eq}), which plays an equivalent role to the Richardson number for internal gravity waves. Similarly to the Miles-Howard theorem, if $\mathcal{R}>1/4$, solutions of Eq. (\ref{eq:corot_eq}) take the form of wavelike solutions, which are preferentially strongly damped when reaching the critical cylinder $\sc$, while if $\mathcal{R}<1/4$, Eq. (\ref{eq:corot_eq}) admits hyperbolic solutions, meaning that waves can be amplified. The inequality $\mathcal{R}<1/4$ is also a necessary but not sufficient condition to trigger shear instabilities, that may depend on the shear flow profile, the boundary conditions, and the properties of the incident waves.

The main difficulty to evaluate $\mathcal{R}$ in our simulations is to estimate the vertical wavenumber $k_z$, which should be taken close enough to $\sc$ but in a region where Eq. (\ref{eq:corot_eq}) admits wavelike solutions (i.e. when $\mathcal{R}>1/4$). The wavenumber transverse to a shear layer scales as $\Ek^{-1/3}$ \cite[e.g.][]{RV2018}, thus the vertical wavenumber $k_z$ close to the inner critical latitude $\theta$ scales as $\Ek^{-1/3}\sin\theta$. Of course, this is only valid close to the inner critical latitude and does not account for the bending of the shear layer as it approaches the corotation resonance, where a WKBJ analysis \citep[like in][]{BR2013,GM2016,AP2021} predicts the vertical wavenumber to go to zero, or the transverse wavenumber $k_s$ to go to infinity (in which case the phase and the group velocities go to zero). 
This estimate is based on the shear layer scaling and may no longer apply close to $\sc$, since the wave beam is enlarged there, as seen in Fig.~\ref{fig:corot}. 
We have, however, evaluated $\mathcal{R}$ given by Eq.~(\ref{eq:crit}) for all of the ``background" zonal flows in our linear simulations. In all cases, 
we find $\mathcal{R}>1/4$, leading us to two opposite conclusions:
\begin{description}
    \item[$-$] either our zonal flows are stable under the $\mathcal{R}$-criterion, and the instabilities we have observed for $\omega=0.8$ and $0.9$ have nothing to do with local shear instabilities, and are possibly more similar to the `Papaloizou-Pringle instability''.
    \item[$-$] the criterion on $\mathcal{R}$ discussed here is no longer valid with viscous dissipation and tidal forcing, and must be modified.
\end{description}
Regarding our second hypothesis, the presence of tidal forcing presumably modifies Eq. (\ref{eq:corot_eq}) close to $\sc$, based on the Appendix of \citet[Eq. (A.9) is the equivalent ODE for forced inertial waves]{AP2021}. 
This term could thus modify the expression for $\mathcal{R}$ in Eq. (\ref{eq:crit}), and hence also change the stability criterion.

\bsp	
\label{lastpage}
\end{document}